\documentclass[twocolumn,a4paper,11pt]{article}
\usepackage{blindtext}
\usepackage[utf8]{inputenc}
\usepackage{times}
\usepackage{microtype}

\usepackage{lscape}

\usepackage{epstopdf}

\usepackage{natbib}
\bibpunct{(}{)}{;}{a}{,}{,}
\usepackage{subfigure}
\usepackage{multirow}
\usepackage{multicol}
\usepackage{float}
\usepackage{soul}
\usepackage{xcolor}
\graphicspath{{./images/}}

\usepackage{amssymb}
\usepackage{amsmath}	
\usepackage{mathtools}
\usepackage{amsthm}
\usepackage{bm}
\usepackage{url}
\usepackage{parskip}
\usepackage{pbox}
\usepackage{nicefrac}

\usepackage[titletoc]{appendix}
\usepackage{booktabs}
\usepackage{boldline}
\usepackage{colortbl}
\DeclareMathOperator{\LogNormal}{LogNormal}
\DeclareMathOperator{\GP}{\mathcal{GP}}
\DeclareMathOperator{\Normal}{Normal}
\usepackage{titlesec}

\titleformat{\section}
  {\normalfont\fontsize{11}{13}\bfseries}{\thesection}{0.5em}{}
\titleformat{\subsection}
  {\normalfont\fontsize{11}{13}\bfseries}{\thesubsection}{0.5em}{}
\titleformat{\subsubsection}
  {\normalfont\fontsize{11}{13}\bfseries}{\thesubsubsection}{0.5em}{}

\providecommand{\keywords}[1]
{
  \small	
  \textbf{\textit{Keywords---}} #1
}

\usepackage{listings}
\usepackage{caption}
\usepackage[colorlinks=true,citecolor=black,linkcolor=black,urlcolor=black,breaklinks=true]{hyperref}

\title{Practical Hilbert space approximate Bayesian Gaussian processes for probabilistic programming}

\author{Gabriel Riutort-Mayol$^{1*}$, Paul-Christian Bürkner$^2$, Michael R.\ Andersen$^{3}$,\\
  Arno Solin$^{2}$, Aki Vehtari$^{2}$}

\date{ \small
$^1$ FISABIO-Public Health, Spain 
\break
$^2$ Excellence Cluster for Simulation Technology, University of Stuttgart, Germany \\
\emph{Most of work was done while at Dept. of Computer Science at Aalto University}
\break
$^2$ Department of Computer Science, Aalto University, Finland
\break
$^3$ Department of Applied Mathematics and Computer Science, Technical University of Denmark, Denmark
\break
$^*$ Corresponding author, Email: gabriuma@gmail.com
}

\begin{document}

\maketitle

\begin{abstract}
Gaussian processes are powerful non-parametric probabilistic models for stochastic functions. However, the direct implementation entails a complexity that is computationally intractable when the number of observations is large, especially when estimated with fully Bayesian methods such as Markov chain Monte Carlo. In this paper, we focus on a low-rank approximate Bayesian Gaussian processes, based on a basis function approximation via Laplace eigenfunctions for stationary covariance functions. The main contribution of this paper is a detailed analysis of the performance, and practical recommendations for how to select the number of basis functions and the boundary factor. Intuitive visualizations and recommendations, make it easier for users to improve approximation accuracy and computational performance.  We also propose diagnostics for checking that the number of basis functions and the boundary factor are adequate given the data. The approach is simple and exhibits an attractive computational complexity due to its linear structure, and it is easy to implement in probabilistic programming frameworks. Several illustrative examples of the performance and applicability of the method in the probabilistic programming language Stan are presented together with the underlying Stan model code.
\end{abstract}


\section{Introduction}\label{sec_bf_intro}

Gaussian processes (GPs) are flexible statistical models for specifying probability distributions over multi-dimensional non-linear functions \citep{rasmussen2006gaussian,neal1997monte}. Their name stems from the fact that any finite set of function values is jointly distributed as a multivariate Gaussian distribution. GPs are defined by a mean and a covariance function. The covariance function encodes our prior assumptions about the functional relationship, such as continuity, smoothness, periodicity and scale properties. GPs not only allow for non-linear effects but can also implicitly handle interactions between input variables (covariates). Different types of covariance functions can be combined for further increased flexibility. Due to their generality and flexibility, GPs are of broad interest across machine learning and statistics \citep{rasmussen2006gaussian,neal1997monte}. Among others, they find application in the fields of spatial epidemiology \citep{diggle2013statistical,carlin2014hierarchical}, robotics and control \citep{deisenroth2015gaussian}, signal processing \citep{sarkka2013spatiotemporal}, neuroimaging \citep{andersen2017} as well as Bayesian optimization and probabilistic numerics \citep{roberts2010bayesian,briol2015probabilistic,hennig2015probabilistic}.

The key element of a GP is the covariance function that defines the dependence structure between function values at different inputs. However, computing the posterior distribution of a GP comes with a computational issue because of the need to invert the covariance matrix. Given $n$ observations in the data, the computational complexity and memory requirements of computing the posterior distribution for a GP in general scale as $O(n^3)$ and $O(n^2)$, respectively. This limits their application to rather small data sets of a few tens of thousands observations at most. The problem becomes more severe when performing full Bayesian inference via sampling methods, where in each sampling step we need $O(n^3)$ computations when inverting the Gram matrix of the covariance function, usually through Cholesky factorization. To alleviate these computational demands, several approximate methods have been proposed. 

Sparse GPs are based on low-rank approximations of the covariance matrix. The low-rank approximation with $m \ll n$ {\it inducing points} implies reduced memory requirements of $O(nm)$ and corresponding computational complexity of $O(nm^2)$. A unifying view on sparse GPs based on approximate generative methods is provided by \citet{quinonero2005unifying}, while a general review is provided by \citet{rasmussen2006gaussian}. \citet{Burt+Rasmussen+vanderWilk:2019} show that for regression with normally distributed covariates in $D$ dimensions and using the squared exponential covariance function, $M=O(\log^DN)$ is sufficient for an accurate approximation.
An alternative class of low-rank approximations is based on forming a basis function approximation with $m \ll n$ basis functions. The basis functions are usually presented explicitly, but can also be used to form a low-rank covariance matrix approximation. Common basis function approximations rest on the spectral analysis and series expansions of GPs \citep{loeve1977probability,trees1968detection,adler1981geometry,cramer2013stationary}. Sparse spectrum GPs are based on a sparse approximation to the frequency domain representation of a GP \citep{lazaro2010sparse,quia2010sparse,gal2015improving}. Recently, \citet{hensman2017variational} presented a variational Fourier feature approximation for GPs that was derived for the Mat{\'e}rn class of kernels. Another related method for approximating kernels relies on random Fourier features  \citep{rahimi2008random,rahimi2009weighted}. Certain spline smoothing basis functions are equivalent to GPs with certain covariance functions \citep{wahba1990spline,Furrer+Nychka:2007}. Recent related work based on a spectral representation of GPs as an infinite series expansion with the Karhunen-Loève representation \citep[see, e.g.,][]{grenander1981abstract} is presented by \citet{JSSv090i10}.
Yet another approach is to present Gaussian process using precision matrix, which is the inverse of the covariance matrix. If the precision matrix is sparse, computation taking the benefit of that sparsity can scale much better than $O(n^3)$. See, for example, review by \citet{Lindgren+Havard:2022:spde}.

In this paper, we focus on a recent framework for fast and accurate inference for fully Bayesian GPs using basis function approximations based on approximation via Laplace eigenfunctions for stationary covariance functions proposed by \citet{solin2018hilbert}. Using a basis function expansion, a GP is approximated with a linear model which makes inference considerably faster. The linear model structure makes GPs easy to implement as building blocks in more complicated models in modular probabilistic programming frameworks, where there is a big benefit if the approximation specific computation is simple. Furthermore, a linear representation of a GP makes it easier to be used as latent function in non-Gaussian observational models allowing for more modelling flexibility. The basis function approximation via Laplace eigenfunctions can be made arbitrary accurate and the trade-off between computational complexity and approximation accuracy can easily be controlled.

The Laplace eigenfunctions can be computed analytically and they are independent of the particular choice of the covariance function including the hyperparameters. While the pre-computation cost of the basis functions is $O(m^2n)$, the computational cost of learning the covariance function parameters is $O(mn+m)$ in every step of the optimizer or sampler. This is a big advantage in terms of speed for iterative algorithms such as Markov chain Monte Carlo (MCMC). Another advantage is the reduced memory requirements of automatic differentiation methods used in modern probabilistic programming frameworks, such as Stan \citep{carpenter2017stan} and others. This is because the memory requirements of automatic differentiation scale with the size of the autodiff expression tree which in direct implementations is simpler for basis function than covariance matrix-based approaches. The basis function approach also provides an easy way to apply a non-centered parameterization of GPs, which reduces the posterior dependency between parameters representing the estimated function and the hyperparameters of the covariance function, which further improves MCMC efficiency.

While \citet{solin2018hilbert} have fully developed the mathematical theory behind this specific approximation of GPs, further work is needed for its practical implementation in probabilistic programming frameworks. In this paper, the interactions among the key factors of the method such as the number of basis functions, domain of the prediction space, and properties of the true functional relationship between covariates and response variable, are investigated and analyzed in detail in relation to the computational performance and accuracy of the method. Practical recommendations are given for the values of the key factors based on simple diagnostic values and intuitive graphical summaries that encode the recognized relationships. Our recommendations help users to choose valid and optimized values for these factors, improving computational performance without sacrificing modeling accuracy. The proposed diagnostic can be used to check whether the chosen values for the number of basis functions and the domain of the prediction space are adequate to model the data well. On that basis, we also develop an iterative procedure to achieve accuracte approximation performance with minimal computational costs.

We have implemented the approach in the probabilistic programming language Stan \citep{carpenter2017stan} as well as subsequently in the \textit{brms} package \citep{burkner2017brms} of the R software \citep{R2019R}. Several illustrative examples of the performance and applicability of the method are shown using both simulated and real datasets. All examples are accompanied by the corresponding Stan code. Although there are several GP specific software packages available to date, for example, GPML \citep{rasmussen2010gpml},  GPstuff \citep{vanhatalo2013gpstuff}, GPy \citep{gpy2014}, and GPflow \citep{GPflow2017}, each provide efficient implementations only for a restricted range of GP-based models. In this paper, we do not focus on the fastest possible inference for a small set of specific GP models, but instead we are interested in how GPs can be easily used as modular components in probabilistic programming frameworks. 

The remainder of the paper is structured as follows. In Section~\ref{ch4_gp}, we introduce GPs, covariance functions and spectral density functions. In Section~\ref{sec_method}, the reduced-rank approximation to GPs proposed by \citet{solin2018hilbert} is described. In Section~\ref{sec_accuracy}, the accuracy of these approximations under several conditions is studied using both analytical and numerical methods.  Practical diagnostics are developed there as well. Several case studies in which we fit exact and approximate GPs to real and simulated data are provided in Section~\ref{sec_cases}. A brief conclusion of the work is made in Section~\ref{sec_conclusion}. Appendix~\ref{app_approx_covfun} includes a brief presentation of the mathematical details behind the Hilbert space approximation of a stationary covariance function, and Appendix \ref{sec_periodic} presents a low-rank representation of a GP for the particular case of a periodic covariance function. Online supplemental material with more case studies illustrating the performance and applicability of the method can be found online at \url{https://github.com/gabriuma/basis_functions_approach_to_GP} in the subfolder \url{Paper/online_supplemental_material}.

\section{Gaussian process as a prior}\label{ch4_gp}

A GP is a stochastic process which defines the distribution of a collection of random variables indexed by a continuous variable, that is, $\left\lbrace f(t): t \in \mathcal{T}\right\rbrace$ for some index set $\mathcal{T}$. GPs have the defining property that the marginal distribution of any finite subset of random variables, $\left\lbrace f(t_1), f(t_2), \hdots, f(t_N) \right\rbrace$, is a multivariate Gaussian distribution.

In this work, GPs will take the role of a prior distribution over function spaces for non-parametric latent functions in a Bayesian setting. Consider a data set $\mathcal{D} = \left\lbrace (\bm{x}_n, y_n) \right\rbrace_{n=1}^N$, where $y_n$ is modelled conditionally as $p(y_n \mid f(\bm{x}_n),\phi)$, where $p$ is some parametric distribution with parameters  $\phi$, and $f$ is an unknown function with a GP prior, which depends on an input $\bm{x}_n\in {\rm I\!R}^D$. This generalizes readily to more complex models depending on several unknown functions, for example such as $p(y_n \mid f(\bm{x}_n),g(\bm{x}_n))$ or multilevel models. Our goal is to obtain the posterior distribution for the value of the function $\tilde{f}=f(\tilde{\bm{x}})$  evaluated at a new input point $\tilde{\bm{x}}$.

We assume a GP prior for $f \sim \GP(\mu(\bm{x}), k(\bm{x}, \bm{x}'))$, where $\mu: {\rm I\!R}^D \to {\rm I\!R}$ and $k: {\rm I\!R}^D \times {\rm I\!R}^D \to {\rm I\!R}$ are the mean and covariance functions, respectively,
\begin{align*}
 	\mu(\bm{x}) &= \mathbb{E}\!\left[f(\bm{x})\right],\\ 
 	k(\bm{x}, \bm{x}') &= \mathbb{E}\!\left[\left( f(\bm{x}) - \mu(\bm{x}) \right)\left( f(\bm{x}') - \mu(\bm{x}') \right)\right].
\end{align*} 
The mean and covariance functions completely characterize the GP prior, and control the a priori behavior of the function $f$. Let $\bm{f}=\left\lbrace f(\bm{x}_n) \right\rbrace_{n=1}^N$, then the resulting prior distribution for $\bm{f}$ is a multivariate Gaussian distribution $\bm{f} \sim \Normal(\bm{\mu}, \bm{K})$, where $\bm{\mu} = \left\lbrace \mu(\bm{x}_n) \right\rbrace_{n=1}^N$ is the mean and $\bm{K}$ the covariance matrix, where $K_{i,j}=k(\bm{x}_i,\bm{x}_j)$. In the following, we focus on zero-mean Gaussian processes, that is set $\mu(\bm{x}) = 0$. The covariance function $k(\bm{x}, \bm{x}')$ might depend on a set of hyperparameters, $\bm{\theta}$, but we will not write this dependency explicitly to ease the notation. The joint distribution of $\bm{f}$ and a new $\tilde{f}$ is also a multivariate Gaussian as,
\begin{align*}
p(\bm{f}, \tilde{f})=\Normal \left( \left[ \begin{array}{cc}
\bm{f} \\ 
f^*
\end{array} \right] \,\middle|\, \bm{0},\left[ \begin{array}{cc}
\bm{K}_{\bm{f},\bm{f}} & \bm{k}_{\bm{f},\tilde{f}} \\ 
\bm{k}_{\tilde{f},\bm{f}} & k_{\tilde{f},\tilde{f}}
\end{array} \right] \right),
\end{align*}
where $\bm{k}_{\bm{f},\tilde{f}}$ is the covariance between $\bm{f}$ and $\tilde{f}$, and $k_{\tilde{f},\tilde{f}}$ is the prior variance of $\tilde{f}$. 

If $p(y_n \mid f(\bm{x}_n),\phi)=\Normal(y_n \mid f(\bm{x}_n),\sigma)$ then $\bm{f}$ can be integrated out analytically (with a computational cost of $O(n^3)$ for exact GPs and $O(nm^2)$ for sparse GPs). If $p(y_n \mid f(\bm{x}_n),g(\bm{x}_n))=\Normal(y_n \mid f(\bm{x}_n),g(\bm{x}_n))$ or $p(y_n \mid f(\bm{x}_n),\phi)$ is non-Gaussian, the marginalization does not have a closed-form solution. Furthermore, if a prior distribution is imposed on $\phi$ and $\bm{\theta}$ to form a joint posterior for $\phi$, $\bm{\theta}$ and $\bm{f}$, approximate inference such as Markov chain Monte Carlo \citep[MCMC; ][]{brooks_2011}, Laplace approximation \citep{williams1998bayesian,rasmussen2006gaussian}, expectation propagation \citep{minka2001expectation}, or variational inference methods \citep{gibbs2000variational,csato2000efficient} are required. In this paper, we focus on the use of MCMC for integrating over the joint posterior. MCMC is usually not the fastest approach, but it is flexible and allows accurate inference and uncertainty estimates for general models in probabilistic programming settings. We consider the computational costs of GPs specifically from this point of view.

\subsection{Covariance functions and spectral density}\label{ch4_sec_cov}

The covariance function is the crucial ingredient in a GP as it encodes our prior assumptions about the function, and characterizes the correlations between function values at different locations in the input space. A covariance function needs to be symmetric and positive semi-definite \citep{rasmussen2006gaussian}. A stationary covariance function is a function of $\bm{\tau}=\bm{x}-\bm{x}' \in {\rm I\!R}^D$, such that it can be written $k(\bm{x},\bm{x}') = k(\bm{\tau})$, which means that the covariance is invariant to translations. Isotropic covariance functions depend only on the input points through the norm of the difference, $k(\bm{x},\bm{x}') = k(|\bm{x}-\bm{x}'|) = k(r), r\in {\rm I\!R}$, which means that the covariance is both translation and rotation invariant. The most commonly used distance between observations is the L2-norm $(|\bm{x}-\bm{x}'|_{L2})$, also known as Euclidean distance, although other types of distances can be considered. 

The Mat\'ern class of isotropic covariance functions is given by, 
\begin{align*}
k_{\nu}(r)&=\alpha \, \frac{2^{1-\nu}}{\Gamma(\nu)}\left(\frac{\sqrt{2\nu}r}{\ell}\right)^{\!\nu} \! K_{\nu} \left(\frac{\sqrt{2\nu}r}{\ell}\right),
\end{align*}
where $\nu > 0$ is the order the kernel, $K_{\nu}$ the modified Bessel function of the second kind, and the $\ell > 0$ and $\alpha > 0$ are the length-scale and magnitude (marginal variance), respectively, of the kernel. The particular case where $\nu=\infty$, $\nu=3/2$ and $\nu=5/2$ are probably the most commonly used kernels \citep{rasmussen2006gaussian}, 
\begin{align*}
k_{\infty}(r)&=\alpha \exp \left( -\frac{1}{2} \frac{r^2}{\ell^2}\right),  \\
k_{\frac{3}{2}}(r)&=\alpha\left(1+\frac{\sqrt{3}r}{\ell}\right) \exp\! \left( -\frac{\sqrt{3}r}{\ell}\right),   \\
k_{\frac{5}{2}}(r)&=\alpha\left(1+\frac{\sqrt{5}r}{\ell} + \frac{5r^2}{3\ell^2}\right) \exp\! \left( -\frac{\sqrt{5}r}{\ell}\right). 
\end{align*}
The former is commonly known as the squared exponential or exponentiated quadratic covariance function. As an example, assuming the Euclidean distance between observations, $r=|\bm{x}-\bm{x}'|_{L2}=\sqrt{\sum_{i=1}^{D}(x_i-x_i')^2}$, the kernel $K_{\nu}$ written above takes the form
\begin{align*}
k_{\infty}(r) = \alpha & \exp \left( -\frac{1}{2} \sum_{i=1}^{D}\frac{(x_i-x_i')^2}{\ell_i^2}\right).
\end{align*}
Notice that the previous expressions $k_{\infty}(r)$ has been easily generalized to using a multidimensional length-scale $\bm{\ell}\in {\rm I\!R}^D$. Using individual length-scales for each dimension turns an isotropic covariance function into a non-isotropic covariance function. That is, for a non-isotropic covariance function, the smoothness may vary across different input dimensions. 

Stationary covariance functions can be represented in terms of their spectral densities \citep[see, e.g.,][]{rasmussen2006gaussian}. In this sense, the covariance function of a stationary process can be represented as the Fourier transform of a positive finite measure \citep[\textit{Bochner's theorem}; see, e.g., ][]{akhiezer1993theory}. If this measure has a density, it is known as the spectral density of the covariance function, and the covariance function and the spectral density are Fourier duals, known as the \textit{Wiener-Khintchine theorem} \citep{rasmussen2006gaussian}. The spectral density functions associated with the Mat\'ern class of covariance functions are given by
\begin{align*}
S_{\nu}(\bm{\omega})&= \alpha \, \frac{2^D\pi^{D/2}\Gamma(\nu+D/2)(2\nu)^{\nu}}{\Gamma(\nu)\, \ell^{2\nu}}\left(\frac{2\nu}{\ell^2}+4\pi^2\bm{\omega}^\intercal \bm{\omega} \right)^{-(\nu+D/2)}
\end{align*}
in $D$ dimensions, where vector $\bm{\omega}\in {\rm I\!R^D}$ is in the frequency domain, and $\ell$ and $\alpha$ are the length-scale and magnitude (marginal variance), respectively, of the kernel. The particular cases, where $\nu=\infty$, $\nu=1/2$, $\nu=3/2$ and $\nu=5/2$, take the form
\begin{align}
S_{\infty}(\bm{\omega})&= \alpha\,(\sqrt{2\pi})^D  \ell^D \exp(-\frac{1}{2}\ell^2 \bm{\omega}^\intercal \bm{\omega}), \label{eq_specdens_inf}  \\
S_{\frac{3}{2}}(\bm{\omega})&= \alpha\,\frac{2^D\pi^{D/2}\Gamma(\frac{D+3}{2}){3}^{3/2}}{\frac{1}{2}\sqrt{\pi}\ell^3}\left(\frac{3}{\ell^2}+\bm{\omega}^\intercal \bm{\omega} \right)^{-\frac{D+3}{2}} \label{eq_specdens_32},     \\
S_{\frac{5}{2}}(\bm{\omega})&= \alpha\,\frac{2^D\pi^{D/2}\Gamma(\frac{D+5}{2}){5}^{5/2}}{\frac{3}{4}\sqrt{\pi}\ell^5}\left(\frac{5}{\ell^2}+\bm{\omega}^\intercal \bm{\omega} \right)^{-\frac{D+5}{2}}. \label{eq_specdens_52}   
\end{align}
For instance, with input dimensionality $D=3$ and $\bm{\omega}=(\omega_1,\omega_2,\omega_3)^\intercal$, the spectral densities written above take the form
\begin{align*}
S_{\infty}(\bm{\omega})&= \alpha \, (2\pi)^{3/2}  \prod_{i=1}^{3} \ell_i  \exp\!\left(-\frac{1}{2} \sum_{i=1}^{3} \ell_i^2 \omega_i^2 \right),   \\
S_{\frac{3}{2}}(\bm{\omega})&= \alpha \, 32\,\pi {3}^{3/2}\prod_{i=1}^{3}\ell_i\left(3+\sum_{i=1}^{3}\ell_i^2 \omega_i^2 \right)^{-3},   \\
S_{\frac{5}{2}}(\bm{\omega})&= \alpha \, \frac{64}{3}\,\pi {5}^{5/2}\prod_{i=1}^{3}\ell_i\left(5+\sum_{i=1}^{3}\ell_i^2 \omega_i^2 \right)^{-4}.
\end{align*}
where individual length-scales $\ell_i$ for each frequency dimension $\omega_i$ have been used.

\section{Hilbert space approximate Gaussian process model}\label{sec_method}

The approximate GP method, developed by \citet{solin2018hilbert} and further analysed in this paper, is based on considering the covariance operator of a stationary covariance function as a pseudo-differential operator constructed as a series of Laplace operators. Then, the pseudo-differential operator is approximated with \emph{Hilbert space} methods on a compact subset $\Omega \subset {\rm I\!R}^D$ subject to boundary conditions. For brevity, we will refer to these approximate Gaussian processes as HSGPs. Below, we will present the main results around HSGPs relevant for practical applications. More details on the theoretical background are provided by \citet{solin2018hilbert}. Our starting point for presenting the method is the definition of the covariance function as a series expansion of eigenvalues and eigenfunctions of the Laplacian operator. The mathematical details of this approximation are briefly presented in Appendix~\ref{app_approx_covfun}.

\subsection{Unidimensional GPs} \label{sec_method_uni}

We begin by focusing on the case of a unidimensional input space (i.e., on GPs with just a single covariate) such that $\Omega \in [-L,L] \subset {\rm I\!R}$, where $L$ is some positive real number to which we also refer as boundary condition. As $\Omega$ describes the interval in which the approximations are valid, $L$ plays a critical role in the accuracy of HSGPs. We will come back to this issue in Section~\ref{sec_accuracy}.

Within $\Omega$, we can write any stationary covariance function with input values $x,x' \in \Omega$ as
\begin{equation}\label{eq_approxcov}
k(x,x') = \sum_{j=1}^\infty S_{\theta}(\sqrt{\lambda_j}) \phi_j(x) \phi_j(x'),
\end{equation} 
where $S_{\theta}$ is the spectral density of the stationary covariance function $k$ (see Section \ref{ch4_sec_cov}) and $\theta$ is the set of hyperparameters of $k$ \citep{rasmussen2006gaussian}. The terms $\{\lambda_j\}_{j=1}^{\infty}$ and $\{\phi_j(x)\}_{j=1}^{\infty}$ are the sets of eigenvalues and eigenvectors, respectively, of the Laplacian operator in the given domain $\Omega$. Namely, they satisfy the following eigenvalue problem in $\Omega$ when applying the Dirichlet boundary condition (other boundary conditions could be used as well)
\begin{align}\label{eq_eigenproblem}
\begin{split}
-\nabla^2 \phi_j(x)&=\lambda_j \phi_j(x), \hspace{1cm}  x \in \Omega \\ 
\phi_j(x)&= 0, \hspace{1.85cm}   x \notin \Omega.
\end{split}
\end{align} 
The eigenvalues $\lambda_j>0$ are real and positive because the Laplacian is a positive definite Hermitian operator, and the eigenfunctions $\phi_j$ for the eigenvalue problem in eq.~\eqref{eq_eigenproblem} are sinusoidal functions. The solution to the eigenvalue problem is independent of  the specific choice of covariance function and is given by
\begin{align}
\lambda_j&=\left(\frac{j\pi}{2L}\right)^{\!2}, \label{eq_eigenvalue}\\
\phi_j(x)&=\sqrt{\frac{1}{L}}\, \sin\!\!\left(\sqrt{\lambda_j}(x+L)\right). \label{eq_eigenfunction}
\end{align}

If we truncate the sum in eq.~\eqref{eq_approxcov} to the first $m$ terms, the approximate covariance function becomes
\begin{equation}
k(x,x') \approx \sum_{j=1}^m S_{\theta}(\sqrt{\lambda_j}) \phi_j(x) \phi_j(x') = \bm{\phi}(x)^\intercal \Delta \bm{\phi}(x'), \nonumber
\end{equation}
where $\bm{\phi}(x)=\{\phi_j(x)\}_{j=1}^{m} \in {\rm I\!R}^{m}$ is the column vector of basis functions, and $\Delta  \in {\rm I\!R}^{m\times m}$ is a diagonal matrix of the spectral density evaluated at the square root of the eigenvalues, that is, $S_{\theta}(\sqrt{\lambda_j})$,
\begin{align}
\Delta =  \begin{bmatrix}
    S_{\theta}(\sqrt{\lambda_1}) & & \\
    & \ddots & \nonumber \\
    & & S_{\theta}(\sqrt{\lambda_m}) \\
  \end{bmatrix}.
\end{align}

Thus, the Gram matrix $\bm{K}$ for the covariance function $k$ for a set of observations $i=1,\ldots,n$ and corresponding input values $\{x_i\}_{i=1}^{n}$ can be represented as
\begin{equation}
 \bm{K} = \Phi \Delta \Phi^\intercal, \nonumber
\end{equation}
where $\Phi \in {\rm I\!R}^{n\times m}$ is the matrix of eigenfunctions $\phi_j(x_i)$
\begin{align}
\Phi =  \left[ {\begin{array}{ccc}
   \phi_1(x_1) & \cdots & \phi_m(x_1)  \\
    \vdots &\ddots & \vdots  \nonumber \\ 
    \phi_1(x_n) & \cdots & \phi_m(x_n) \\
  \end{array} } \right].
\end{align}
As a result, the model for $f$ can be written as
\begin{equation}
\bm{f} \sim \Normal(\bm{\mu},\Phi \Delta \Phi^\intercal). \nonumber
\end{equation}
This equivalently leads to a linear representation of $f$ via
\begin{equation}\label{eq_approxf}
f(x) \approx \sum_{j=1}^m \left( S_{\theta}(\sqrt{\lambda_j})\right)^{\frac{1}{2}} \phi_j(x) \beta_j,
\end{equation}
where $\beta_j \sim \Normal(0,1)$. Thus, the function $f$ is approximated with a finite basis function expansion (using the eigenfunctions $\phi_j$ of the Laplace operator), scaled by the square root of spectral density values. A key property of this approximation is that the eigenfunctions $\phi_j$ do not depend on the hyperparameters of the covariance function $\theta$. Instead, the only dependence of the model on $\theta$ is through the spectral density $S_{\theta}$. The eigenvalues $\lambda_j$ are monotonically increasing with $j$ and $S_{\theta}$ goes rapidly to zero for bounded covariance functions. Therefore, eq.~\eqref{eq_approxf} can be expected to be a good approximation for a finite number of $m$ terms in the series as long as the inputs values $x_i$ are not too close to the boundaries $-L$ and $L$ of $\Omega$. The computational cost of evaluating the log posterior density of univariate HSGPs scales as $O(nm + m)$, where $n$ is the number of observations and $m$ the number of basis functions.

The parameterization in eq.~\eqref{eq_approxf} is naturally in the non-centered parameterization form with independent prior distribution on $\beta_j$, which can make the posterior inference easier \citep[see, e.g., ][]{Betancourt+Girolami:2019}. Furthermore, all dependencies on the covariance function and the hyperparameters is through the prior distribution of the regression weights $\beta_j$. The posterior distribution of the parameters $p(\bm{\beta}|\bm{y})$ is a distribution over a $m$-dimensional space, where $m$ is much smaller than the number of observations $n$. Therefore, the parameter space is greatly reduced and this makes inference faster, especially when sampling methods are used.

\subsection{Generalization to multidimensional GPs} \label{sec_method_multi}

The results from the previous section can be generalized to a multidimensional input space with compact support, $\Omega=[-L_1,L_1] \times \dots \times [-L_D,L_D]$ and Dirichlet boundary conditions. 
In a $D$-dimensional input space, the total number of eigenfunctions and eigenvalues in the approximation is equal to the number of $D$-tuples, that is, possible combinations of univariate eigenfunctions over all dimensions. The number of $D$-tuples is given by 
\begin{align} \label{eq_m_multi}
m^{\ast} = \prod_{d=1}^{D} m_d,
\end{align}
where $m_d$ is the number of basis function for the dimension $d$. Let $\mathbb{S}\in {\rm I\!N}^{m^{\ast} \times D}$ be the matrix of all those $D$-tuples. For example, suppose we have $D=3$ dimensions and use $m_{1}=2$, $m_{2}=2$ and $m_{3}=3$ eigenfunctions and eigenvalues for the first, second and third dimension, respectively. Then, the number of multivariate eigenfunctions and eigenvalues is $m^{\ast} = m_{1} \cdot m_{2} \cdot m_{3} = 12$ and the matrix $\mathbb{S}\in {\rm I\!N}^{12 \times 3}$ is given by
%
%
\begin{align}\small
\mathbb{S}=
\left[ {\begin{array}{cccccccccccc}
1 & 1 & 1 & 1 & 1 & 1 & 2 & 2 & 2 & 2 & 2 & 2 \\
1 & 1 & 1 & 2 & 2 & 2 & 1 & 1 & 1 & 2 & 2 & 2 \\
1 & 2 & 3 & 1 & 2 & 3 & 1 & 2 & 3 & 1 & 2 & 3
\end{array} } \right]^\intercal.
\end{align} 

Each multivariate eigenfunction $\phi^{\ast}_j:\Omega \rightarrow {\rm I\!R}$ corresponds to the product of the univariate eigenfunctions whose indices corresponds to the elements of the $D$-tuple $\mathbb{S}_{j\cdotp}$, and each multivariate eigenvalue $\bm{\lambda}^{\ast}_j$ is a $D$-vector with elements that are the univariate eigenvalues whose indices correspond to the elements of the $D$-tuple $\mathbb{S}_{j\bm{\cdotp}}$. Thus, for $\bm{x}=\{x_d\}_{d=1}^D \in \Omega$ and $j=1,2,\ldots,m^{\ast}$, we have 
\begin{align}
\bm{\lambda}^{\ast}_j &= \left\{ \lambda_{\mathbb{S}_{jd}} \right\}_{d=1}^D =  \left\{ \left(\frac{\pi \mathbb{S}_{jd}}{2L_d}\right)^{\!2} \right\}_{d=1}^D, \label{eq_eigenvalue_multi} \\
\phi^{\ast}_j(\bm{x}) &= \prod_{d=1}^{D} \phi_{\mathbb{S}_{jd}}(x_d) = \prod_{d=1}^{D} \sqrt{\frac{1}{L_d}} \sin\!\left(\sqrt{\lambda_{\mathbb{S}_{jd}}}(x_d+L_d)\right). \label{eq_eigenfunction_multi}
\end{align}
The approximate covariance function is then represented as
\begin{equation}\label{eq_approxcov_multi}
k(\bm{x},\bm{x}') \approx \sum_{j=1}^{m^{\ast}} 
S^{\ast}_{\theta}\left(\sqrt{\bm{\lambda}^{\ast}_j}\right)
\phi^{\ast}_j(\bm{x}) \phi^{\ast}_j(\bm{x}'),
\end{equation}
where $S^{\ast}_{\theta}$ is the spectral density of the $D$-dimensional covariance function (see Section \ref{ch4_sec_cov}) as a function of $\sqrt{\bm{\lambda}^{\ast}_j}$ that denotes the element-wise square root of the vector $\bm{\lambda}^{\ast}_j$. We can now write the approximate series expansion of the multivariate function $f$ as
\begin{equation}\label{eq_approxf_multi}
f(\bm{x}) \approx \sum_{j=1}^{m^{\ast}} 
\left( S^{\ast}_{\theta} \left(\sqrt{\bm{\lambda}^{\ast}_j} \right)\right)^{\! \frac{1}{2}} \phi^{\ast}_j(\bm{x}) \beta_j, 
\end{equation}
where, again, $\beta_j \sim \Normal(0,1)$. The computational cost of evaluating the log posterior density of multivariate HSGPs scales as $O(n m^{\ast} + m^{\ast})$, where $n$ is the number of observations and $m^{\ast}$ is the number of multivariate basis functions. Although this still implies linear scaling in $n$, the approximation is more costly than in the univariate case, as $m^{\ast}$ is the product of the number of univariate basis functions over the input dimensions and grows exponentially with respect to the number of  dimensions.

\subsection{Linear representation of a periodic squared exponential covariance function} \label{sec_method_periodic}

A GP model with a periodic covariance function does no fit in the framework of the HSGP approximation covered in this study as a periodic covariance function has not a spectral representation, but it has also a low-rank representation. In Appendix~\ref{sec_periodic}, we briefly present the approximate linear representation of a periodic squared exponential covariance function as developed by \citet{solin2014explicit}, analyze the accuracy of this approximation and, finally, derive the GP model with this approximate periodic squared exponential covariance function.

\section{The accuracy of the approximation}\label{sec_accuracy}

The accuracy and speed of the HSGP model depends on several interrelated factors, most notably on the number of basis functions and on the boundary condition of the Laplace eigenfunctions. Furthermore, appropriate values for these factors will depend on the degree of non-linearity of the function to be estimated, which is in turn characterized by the length-scale of the covariance function. In this section, we analyze the effects of the number of basis functions and the boundary condition on the approximation accuracy. We present recommendations on how they should be chosen and diagnostics to check the accuracy of the obtained approximation. 

Ultimately, these recommendations are based on the relationships among the number of basis functions $m$, the boundary condition $L$, and the length-scale $\ell$, which depend on the particular choice of the kernel function. In this work we investigate these relationships for the squared exponential and the Mat{\'e}rn ($\nu=3/2$ and $\nu=5/2$) covariance functions in the present section, and for the periodic squared exponential covariance function in Appendix~\ref{sec_periodic}. For other kernels, the relationships will be slightly different depending on the smoothness or wigglyness of the covariance function.

\subsection{Dependency on the number of basis functions and the boundary condition} \label{subsec_dependency}

As explained in Section~\ref{sec_method}, the approximation of the covariance function is a series expansion of eigenfunctions and eigenvalues of the Laplace operator in a given domain $\Omega$, for instance in a one-dimensional input space $\Omega=[-L,L]\subset {\rm I\!R}$
\begin{equation*}
k(\tau) = \sum_{j=1}^{\infty} S_{\theta}(\sqrt{\lambda_j}) \phi_j(\tau) \phi_j(0), 
\end{equation*}
where $L$ describes the boundary condition, $j$ is the index for the eigenfunctions and eigenvalues, and $\tau=x-x'$ is the difference between two covariate values $x$ and $x'$ in $\Omega$. The eigenvalues $\lambda_j$ and eigenfunctions $\phi_j$ are given in equations (\ref{eq_eigenvalue}) and (\ref{eq_eigenfunction}) for the unidimensional case and in equations (\ref{eq_eigenvalue_multi}) and (\ref{eq_eigenfunction_multi}) for the multidimensional case. The number of basis functions can be truncated at some finite positive value $m$ such that the total variation difference between the exact and approximate covariance functions is less than a predefined threshold $\varepsilon > 0$:
\begin{equation}\label{eq_diff_covs}
 \int | k(\tau) - \sum_{j=1}^m S_{\theta}(\sqrt{\lambda_j}) \phi_j(\tau) \phi_j(0)|  \,\mathrm{d}\tau  < \varepsilon.
\end{equation}

This inequality can be satisfied for arbritrary small $\epsilon$ provided that $L$ and $m$ are sufficiently large \cite[Theorem 1 and 4]{solin2018hilbert}. 
The specific number of basis functions $m$ needed depends on the degree of non-linearity of the function to be estimated, that is on its length-scale $\ell$, which constitutes a hyperparameter of the GP. The approximation also depends on the boundary $L$ (see equations (\ref{eq_eigenvalue}), (\ref{eq_eigenfunction}), (\ref{eq_eigenvalue_multi}) and (\ref{eq_eigenfunction_multi})), which will affect its accuracy especially near the boundaries. As we will see later on, $L$ will also influence the number of basis functions required in the approximation.

In this work, we choose $L$ such that the domain $\Omega = \left[-L, L\right]$ contains all the inputs points $x_i$. Without loss of generality, we can assume all data points are contained in a symmetric interval around zero. Let $S = \max_{i}|x_i|$, then it follows that $x_i \in \left[-S, S\right]$ for all $i$. We now define $L$ as
\begin{equation}\label{eq_boundary}
L=c \cdot S,
\end{equation} 
where $S > 0$ represents the half-range of the input space, and $c \geq 1$ is the proportional extension factor. In the following, we will refer to $c$ as the boundary factor of the approximation. The boundary factor can also be regarded as the boundary $L$ normalized by the half-range $S$ of the input space.

\begin{figure*}[h]
\centering
\subfigure{\includegraphics[scale=0.29, trim = 1mm 5mm 3mm 0mm, clip]{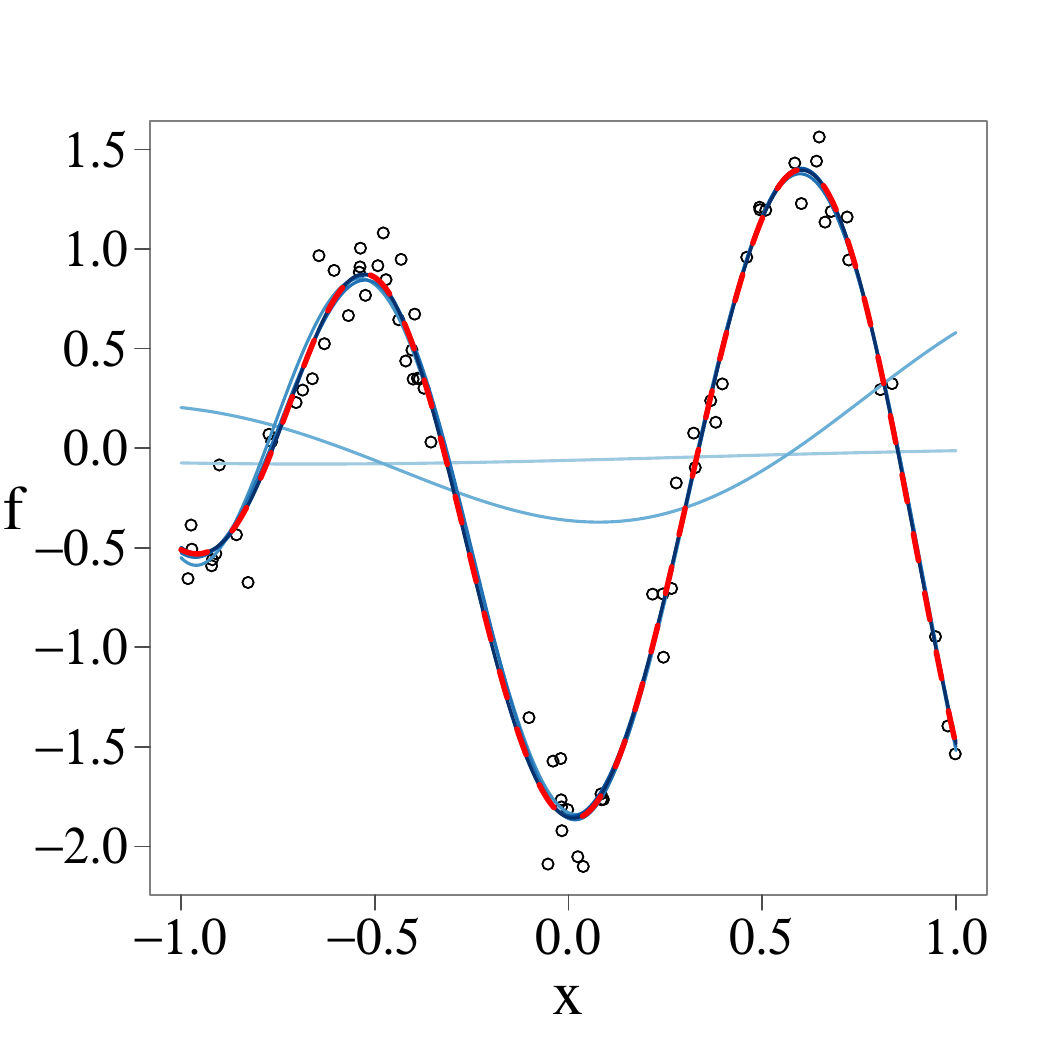}}
\subfigure{\includegraphics[scale=0.29, trim = 0mm 5mm 3mm 0mm, clip]{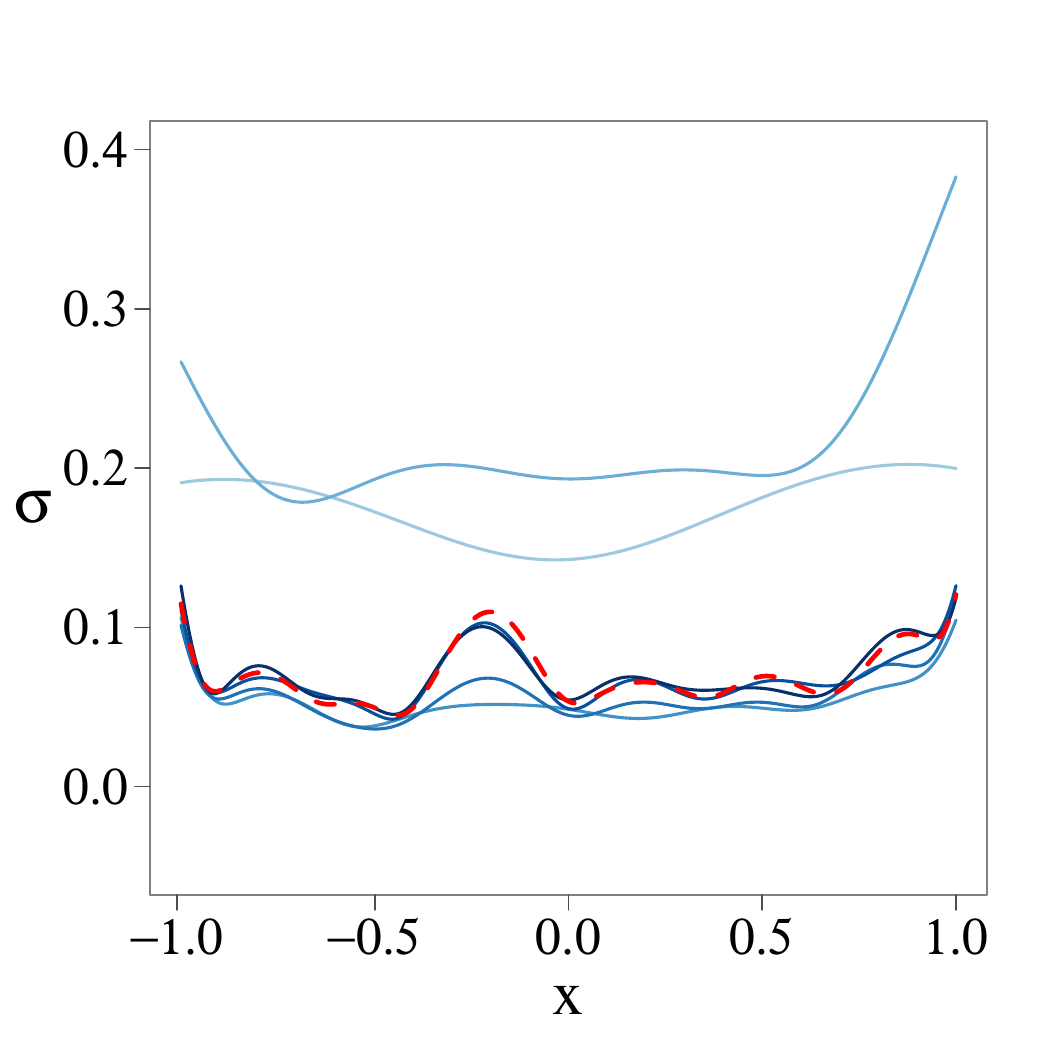}}
\subfigure{\includegraphics[scale=0.29, trim = 1mm 5mm 3mm 0mm, clip]{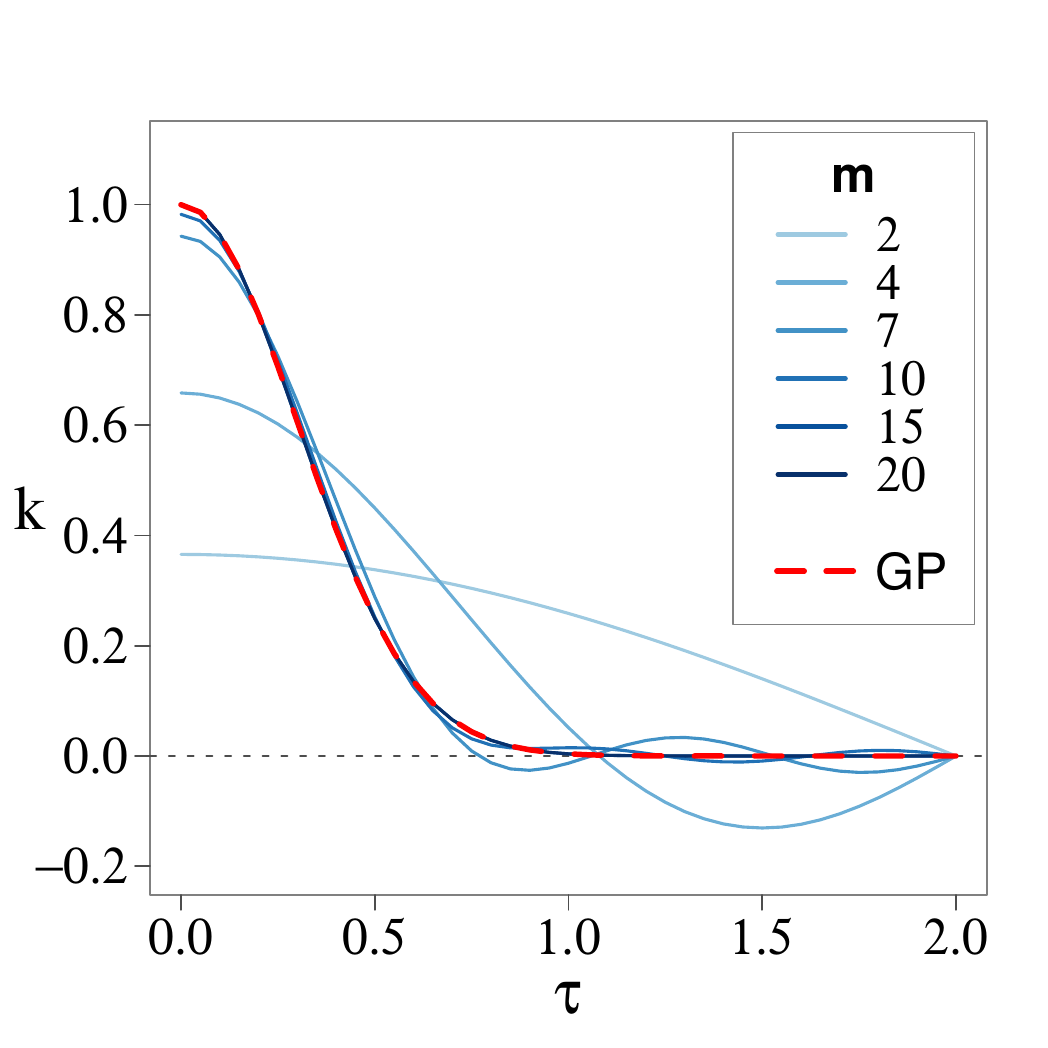}}
\vspace{-1mm}
\caption{Mean posterior predictive functions (left), posterior standard deviations (center), and covariance functions (right) of both the exact GP model (dashed red line) and the HSGP model for different number of basis functions $m$, with the boundary factor fixed to a large enough value.}
  \label{fig1_Post_J}
\end{figure*}

\begin{figure*}[h]
\centering
\subfigure{\includegraphics[scale=0.29, trim = 1mm 5mm 3mm 0mm, clip]{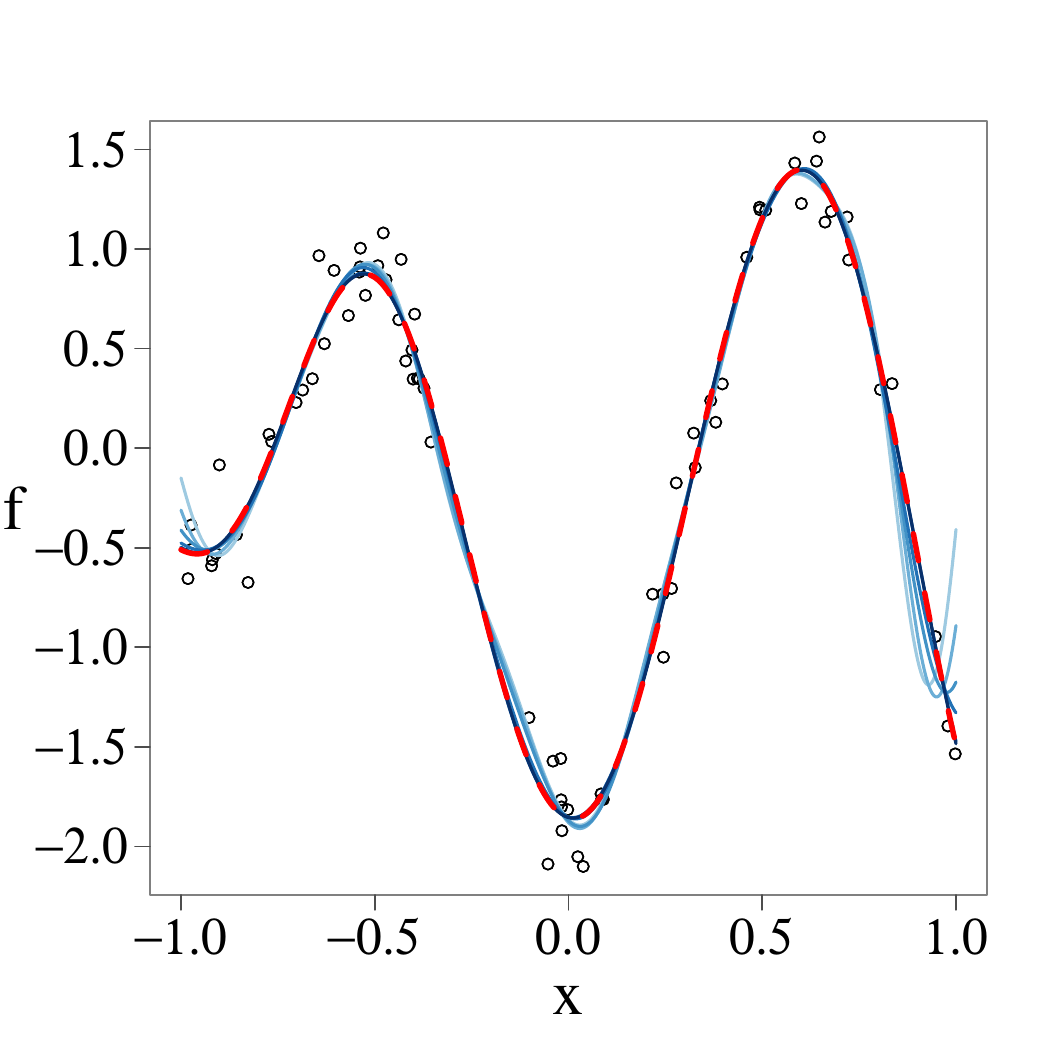}}
\subfigure{\includegraphics[scale=0.29, trim = 0mm 5mm 3mm 0mm, clip]{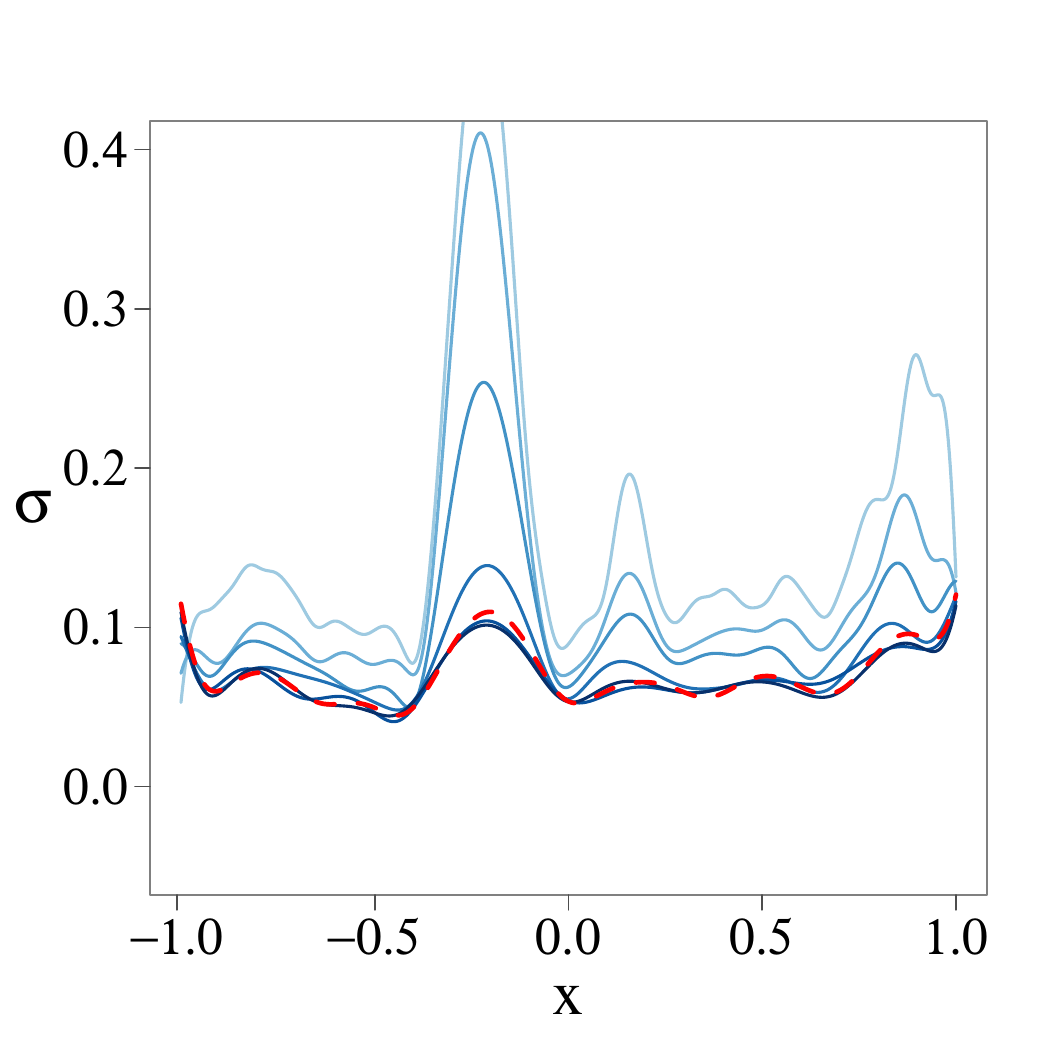}}
\subfigure{\includegraphics[scale=0.29, trim = 1mm 5mm 3mm 0mm, clip]{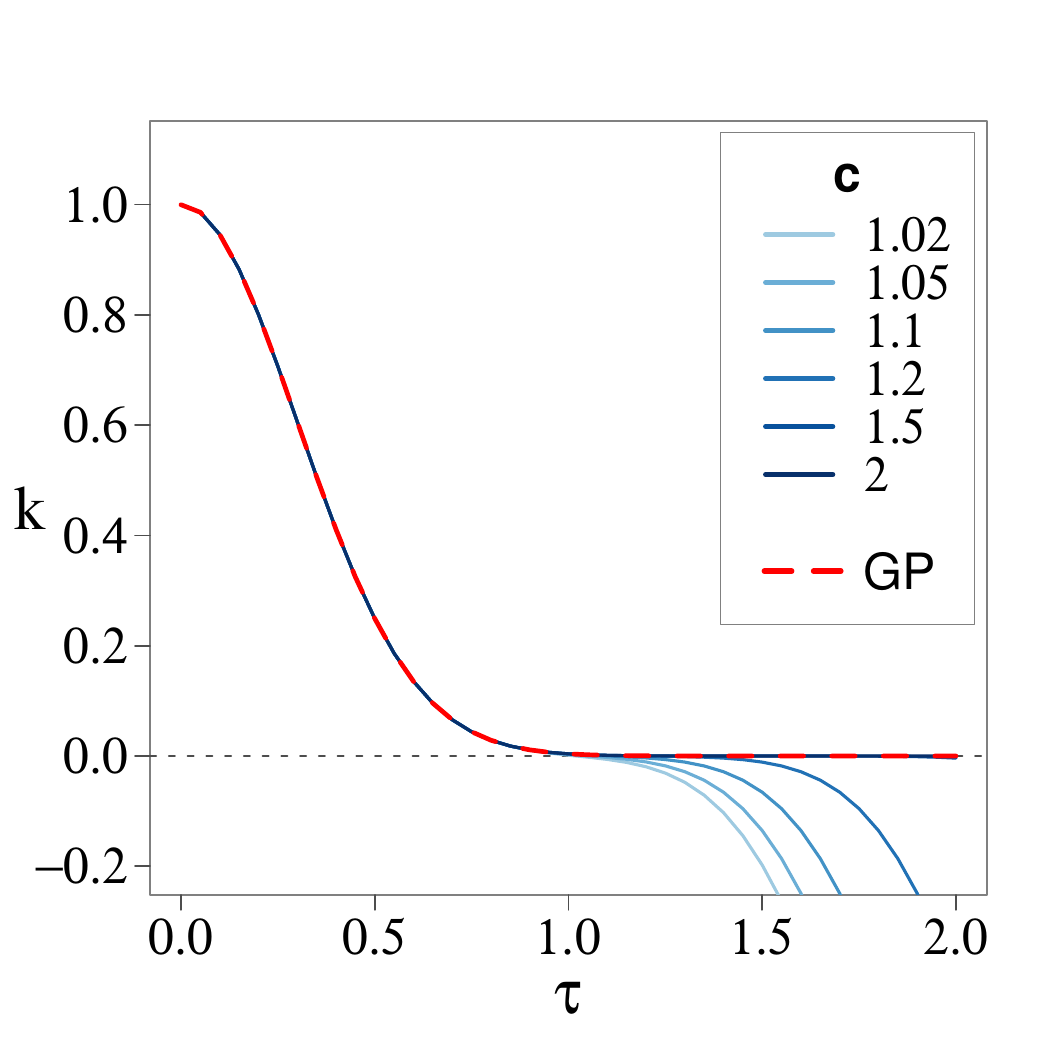}}
\vspace{-1mm}
\caption{Mean posterior predictive functions (left), posterior standard deviations (center), and covariance functions (right) of both the exact GP model (dashed red line) and the HSGP model for different values of the boundary factor $c$, with a large enough fixed number of basis functions.}
  \label{fig2_Post_L}
\end{figure*}

We start by illustrating how the number of basis functions $m$ and boundary factor $c$ influence the accuracy of the HSGP approximations individually. For this purpose, a set of noisy observations are drawn from an exact GP model with a squared exponential covariance function of length-scale $\ell=0.3$ and marginal variance $\alpha=1$, using input values from the zero-mean input domain with half-range $S=1$. Several HSGP models with varying $m$ and $c$ are fitted to this data. In this example, the length-scale and marginal variance parameters used in the HSGPs are fixed to the true values of the data-generating model. 
Figures~\ref{fig1_Post_J} and \ref{fig2_Post_L} illustrate the individual effects of $m$ and $c$, respectively, on the posterior predictive mean and standard deviation of the estimated function as well as on the covariance function itself. For a sufficiently large fixed value of $c$, Figure \ref{fig1_Post_J} shows clearly how $m$ affects the accuracy on the approximation for both the posterior mean or uncertainty. It is seen that if the number of basis functions $m$ is too small, the estimated function tend to be overly smooth because the necessary high frequency components are missing. In general, the higher the degree of wigglyness of the function to be estimated, the larger number of  basis functions will be required. If $m$ is fixed to a sufficiently large value, Figure~\ref{fig2_Post_L} shows that $c$ affects the approximation of the mean mainly near the boundaries, while the approximation of the standard deviation is affected across the whole domain. The approximation error tends to be bigger for the standard deviation than for the mean.

\begin{figure*}
\centering
\begin{tabular}{ c c c }
\arrayrulecolor{darkgray}\hline
c = 1.05 &
\includegraphics[scale=0.215, trim = 0mm 14mm 0mm 14mm, clip]{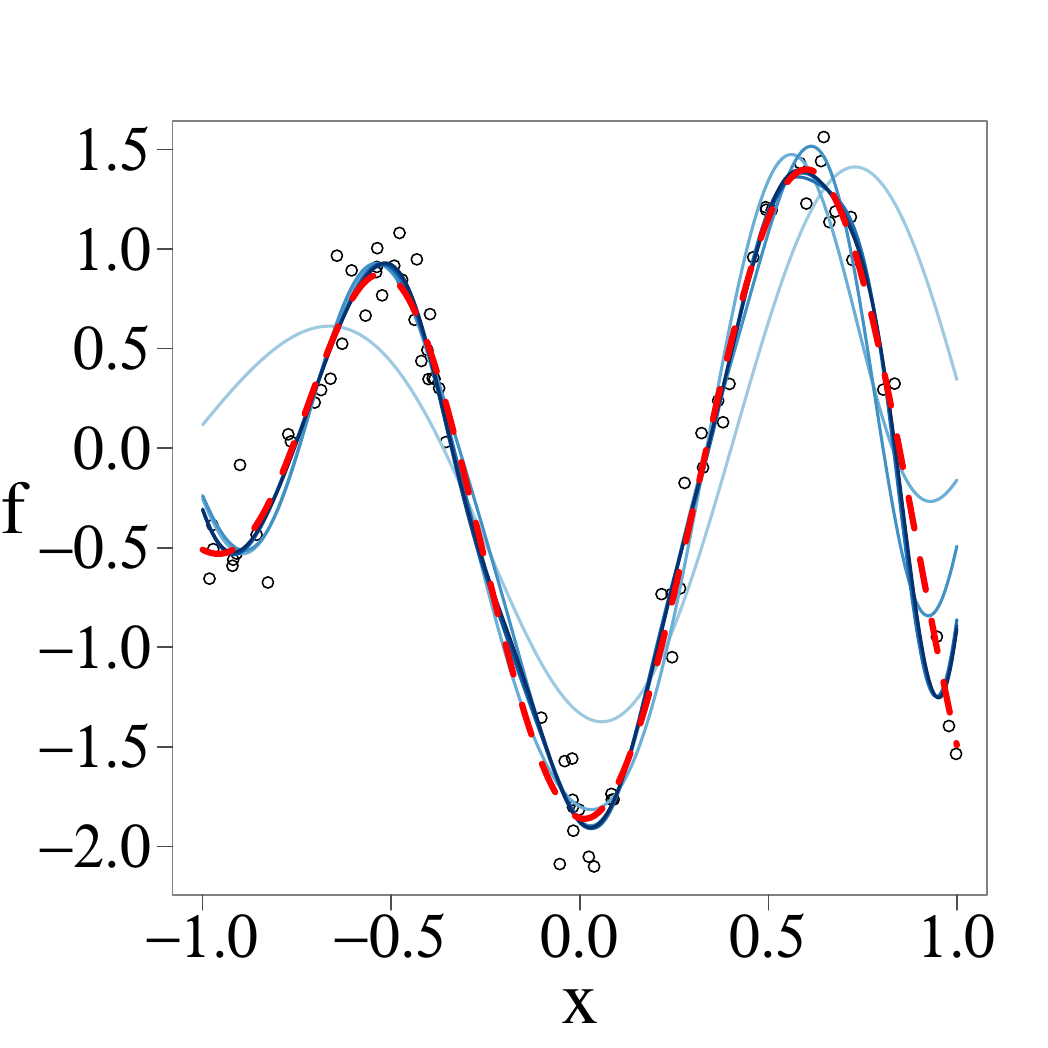}
\includegraphics[scale=0.215, trim = 0mm 14mm 0mm 14mm, clip]{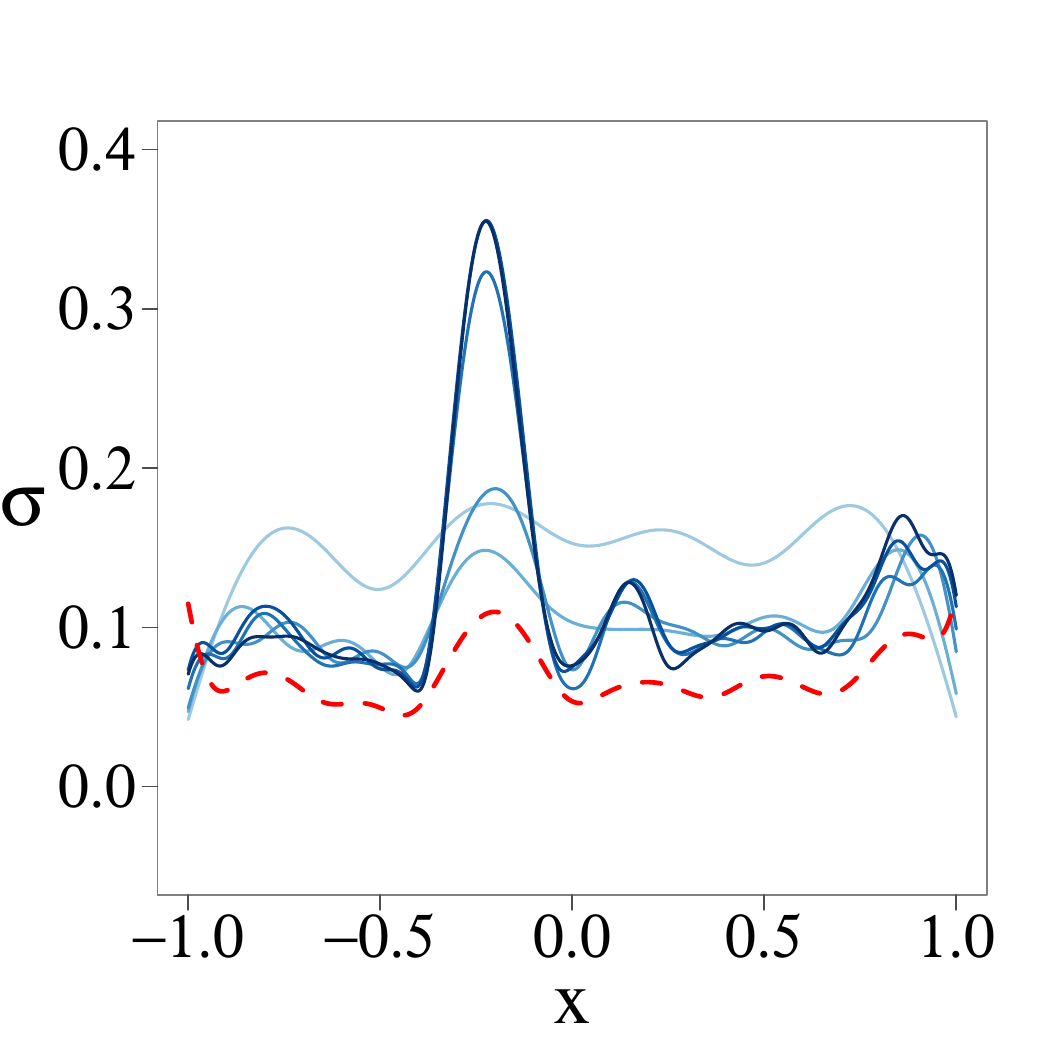} 
\includegraphics[scale=0.215, trim = 0mm 14mm 5mm 14mm, clip]{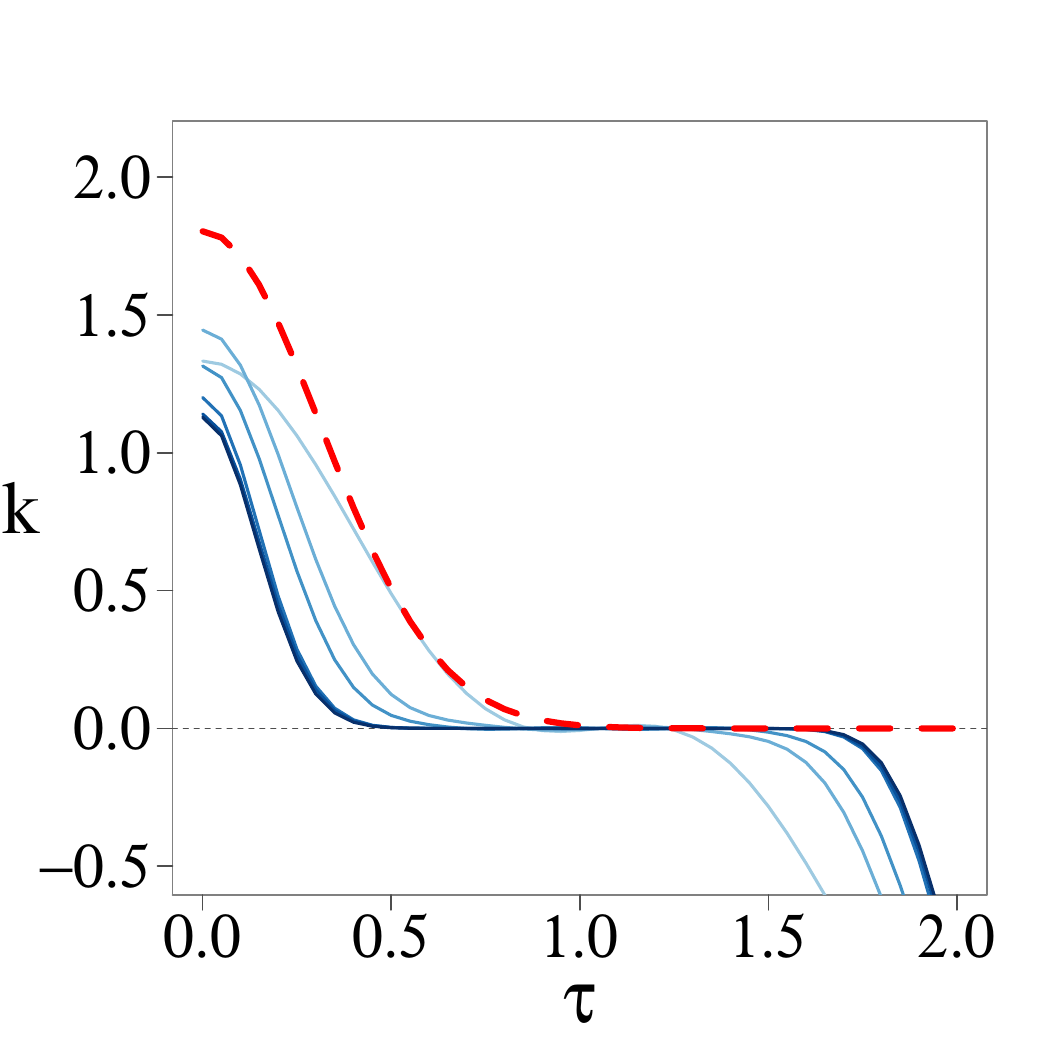} & 
\multirow{43}{*}{ \includegraphics[scale=0.35, trim = 28mm 30mm 100mm 30mm, clip]{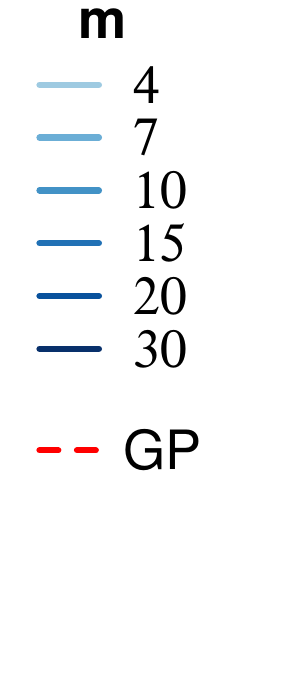}}\\ 
\arrayrulecolor{lightgray}\cline{1-2}
c = 1.1 &
\includegraphics[scale=0.215, trim = 0mm 14mm 0mm 14mm, clip]{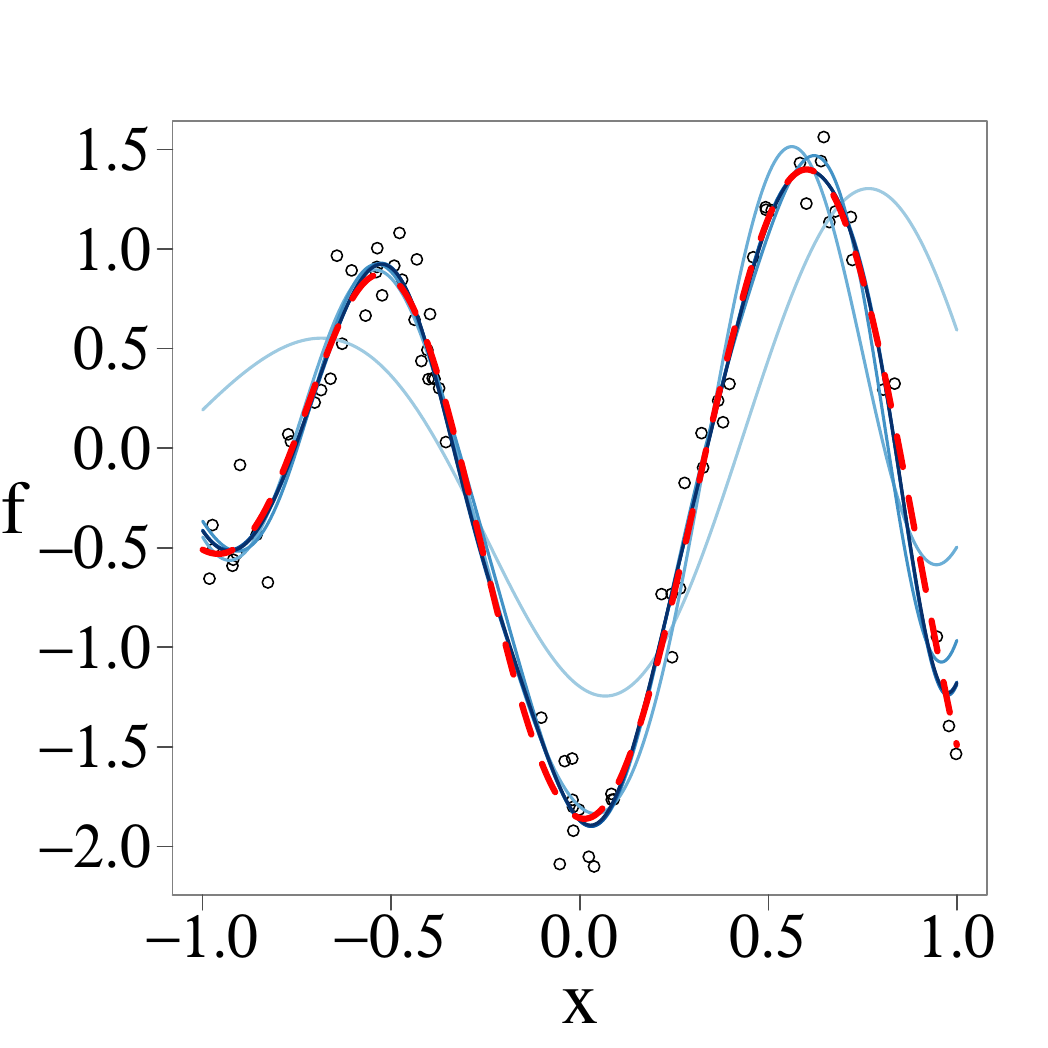} 
\includegraphics[scale=0.215, trim = 0mm 14mm 0mm 14mm, clip]{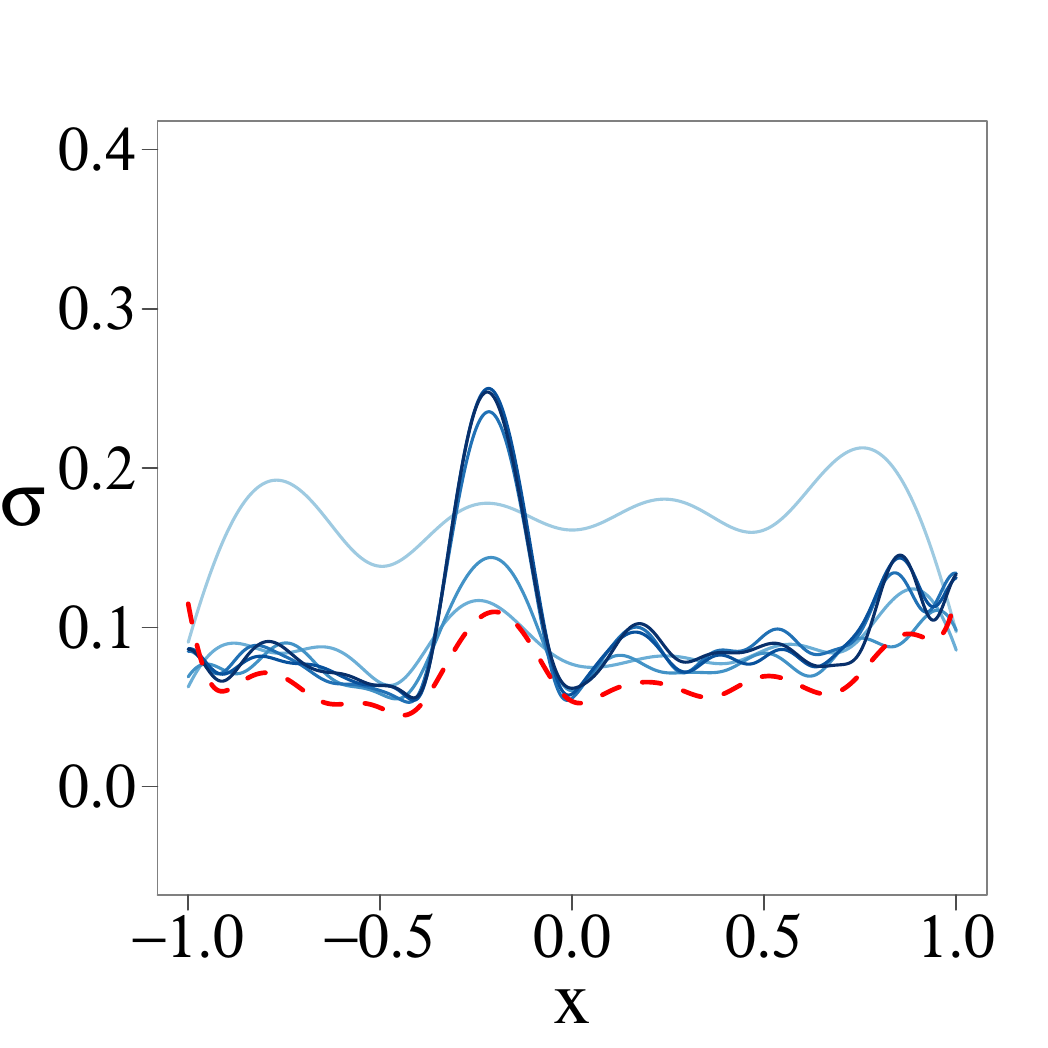} 
\includegraphics[scale=0.215, trim = 0mm 14mm 5mm 14mm, clip]{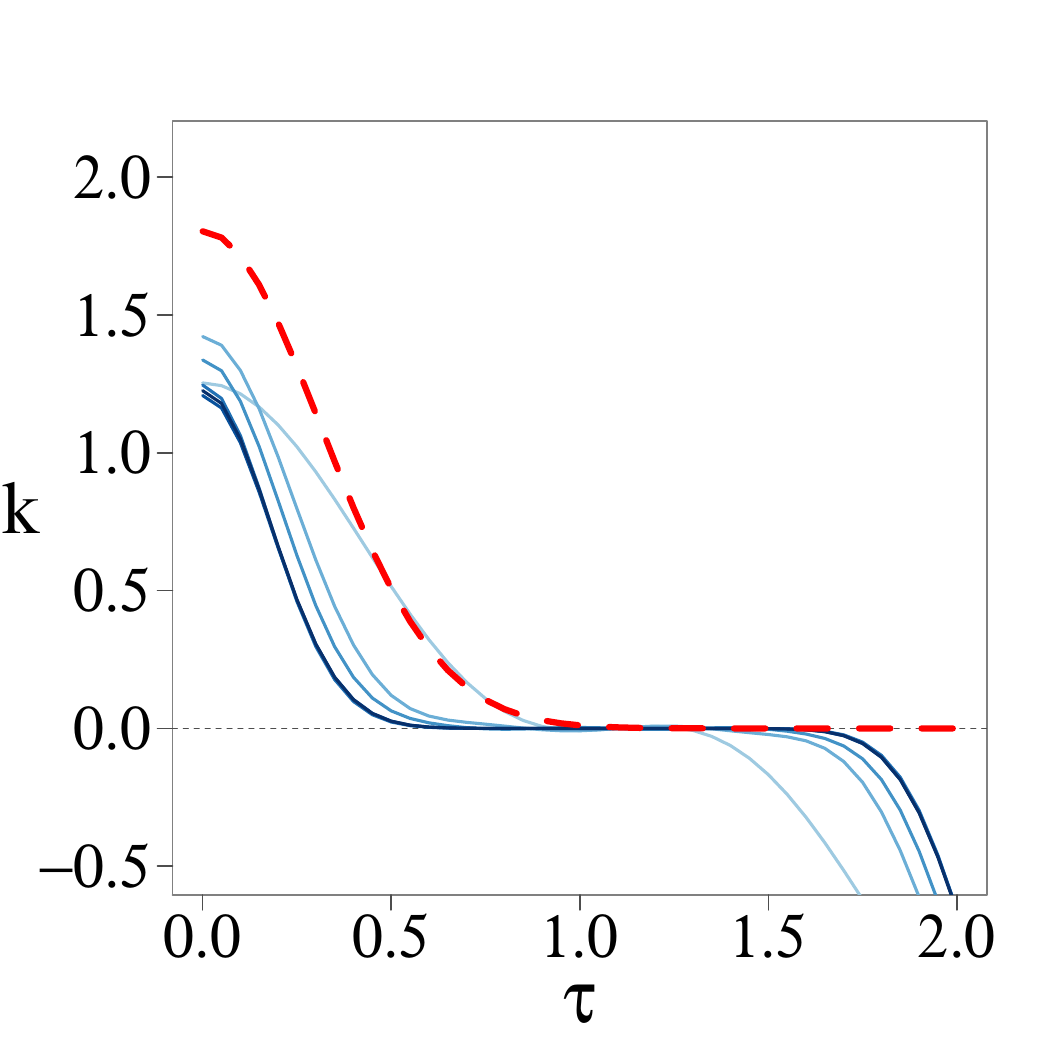} &\\
\cline{1-2}
c = 1.2 &
\includegraphics[scale=0.215, trim = 0mm 14mm 0mm 14mm, clip]{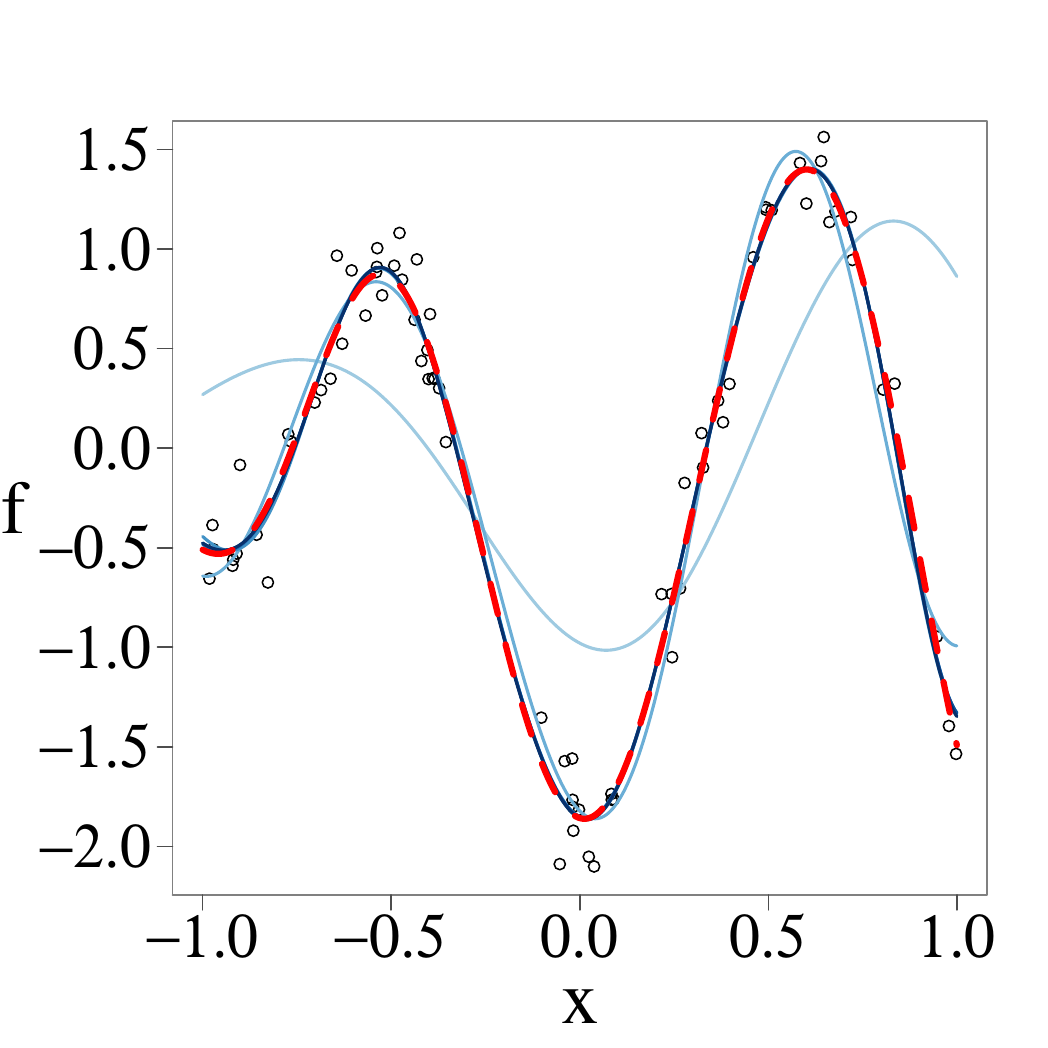} 
\includegraphics[scale=0.215, trim = 0mm 14mm 0mm 14mm, clip]{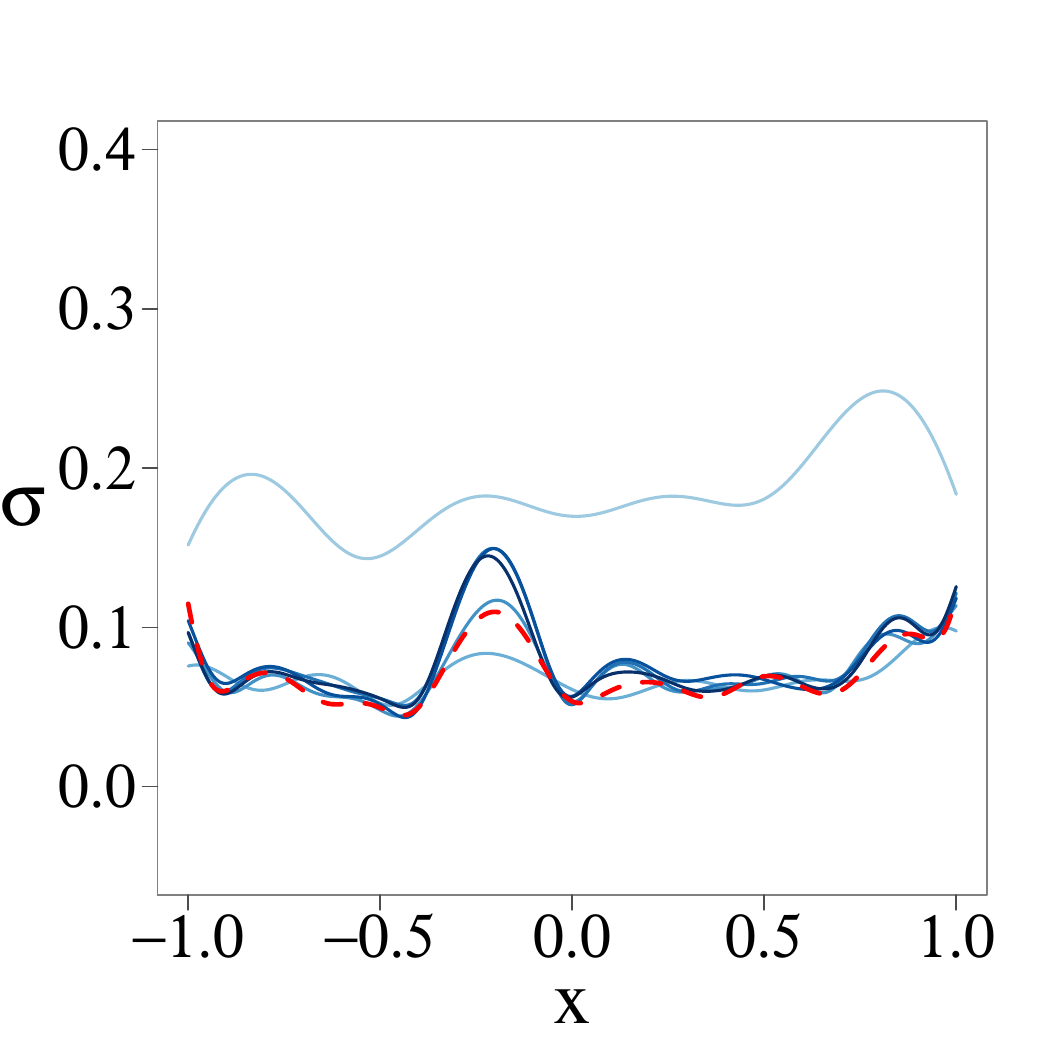} 
\includegraphics[scale=0.215, trim = 0mm 14mm 5mm 14mm, clip]{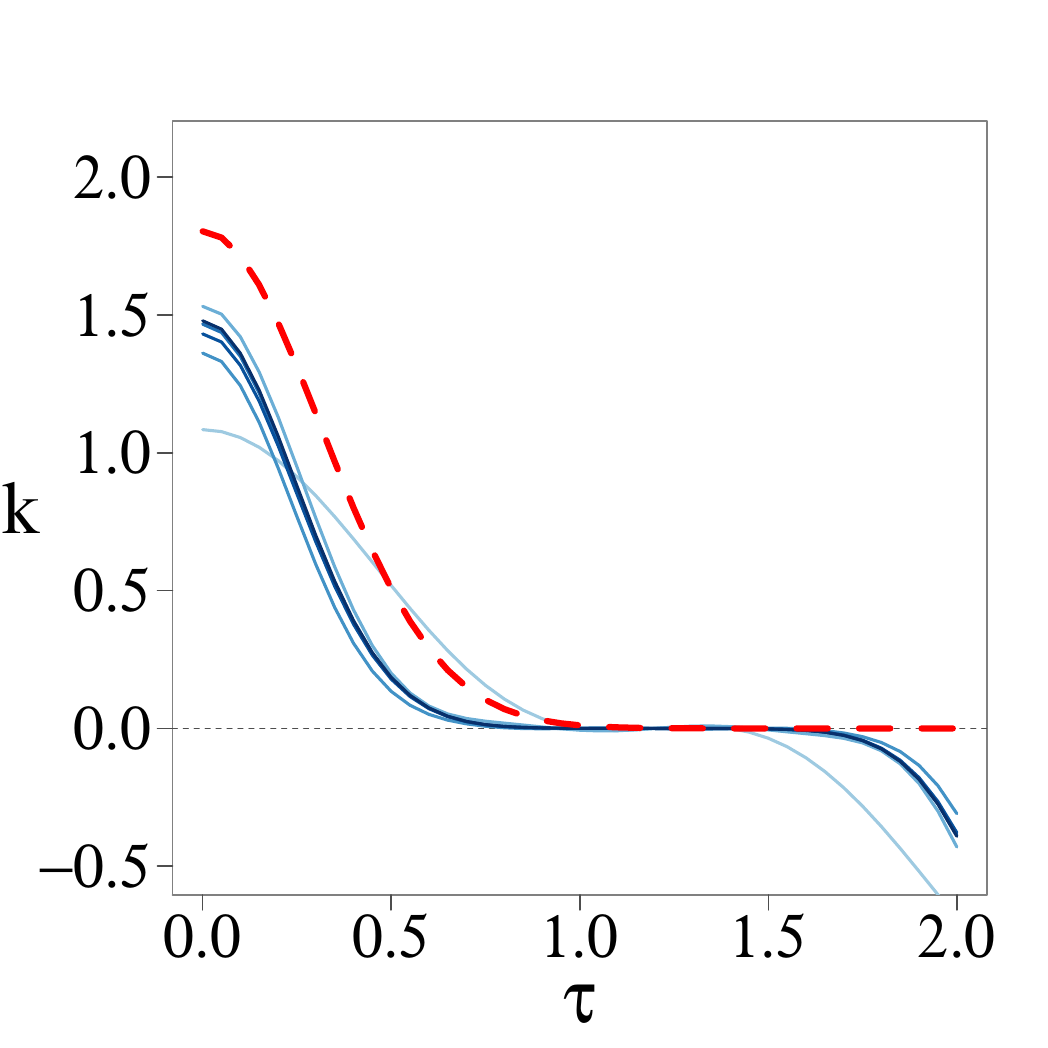} &\\
\cline{1-2}
c = 1.5 &
\includegraphics[scale=0.215, trim = 0mm 14mm 0mm 14mm, clip]{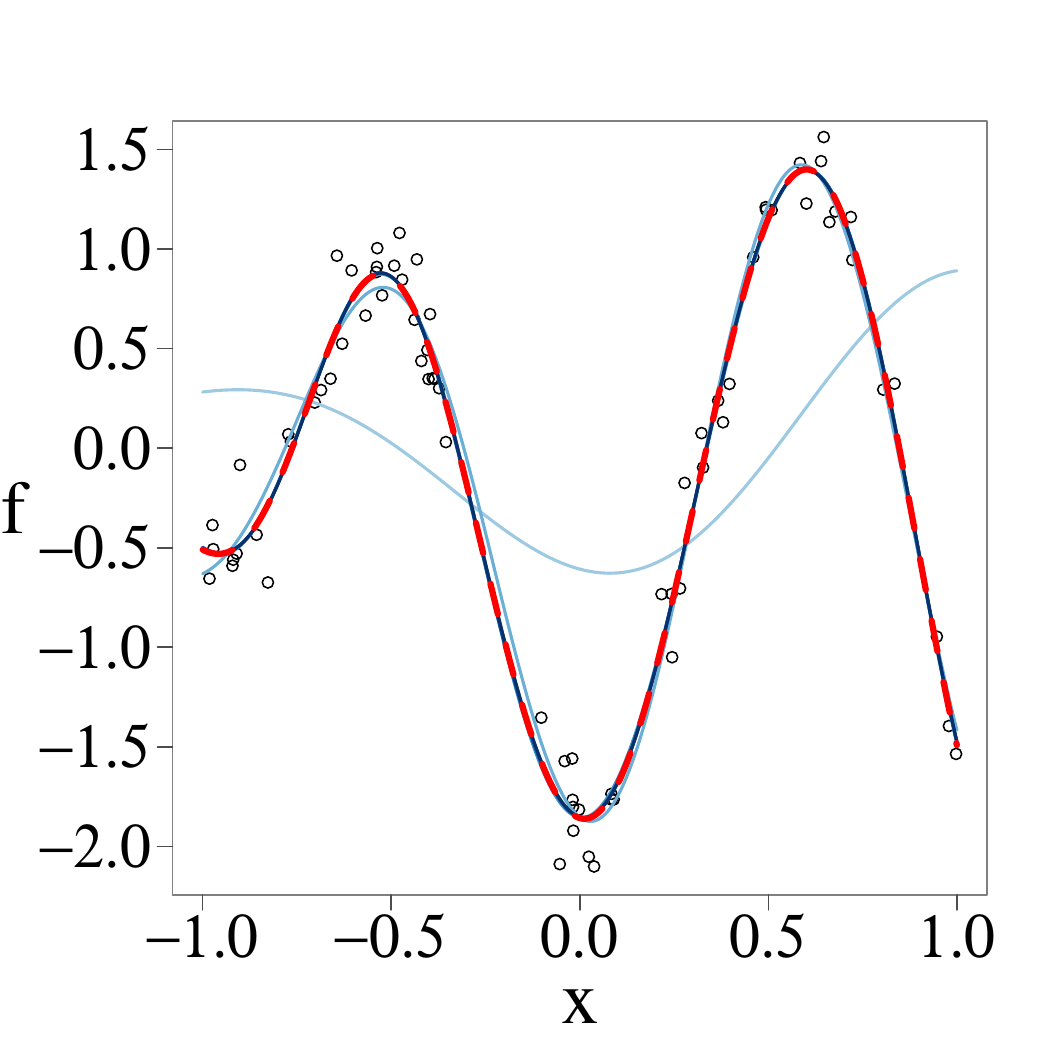} 
\includegraphics[scale=0.215, trim = 0mm 14mm 0mm 14mm, clip]{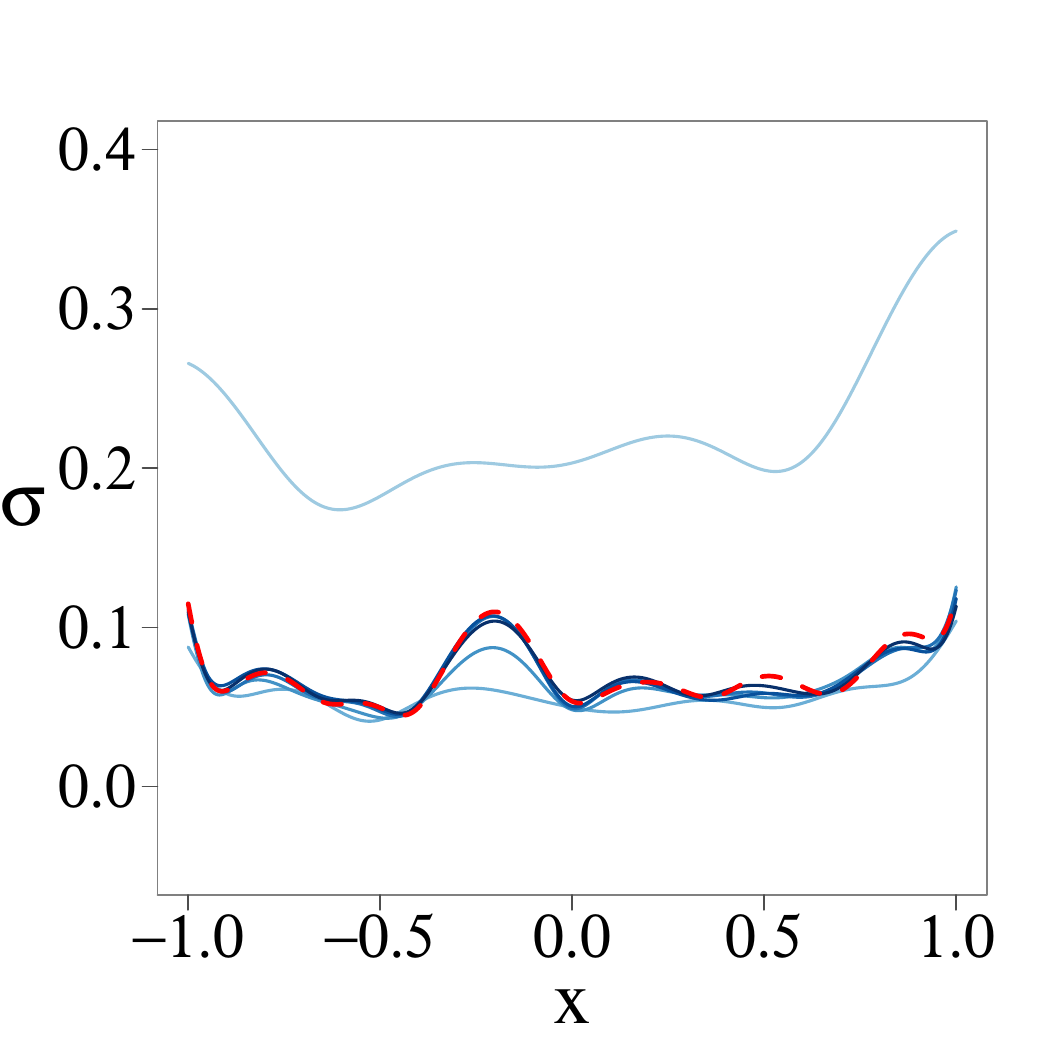} 
\includegraphics[scale=0.215, trim = 0mm 14mm 5mm 14mm, clip]{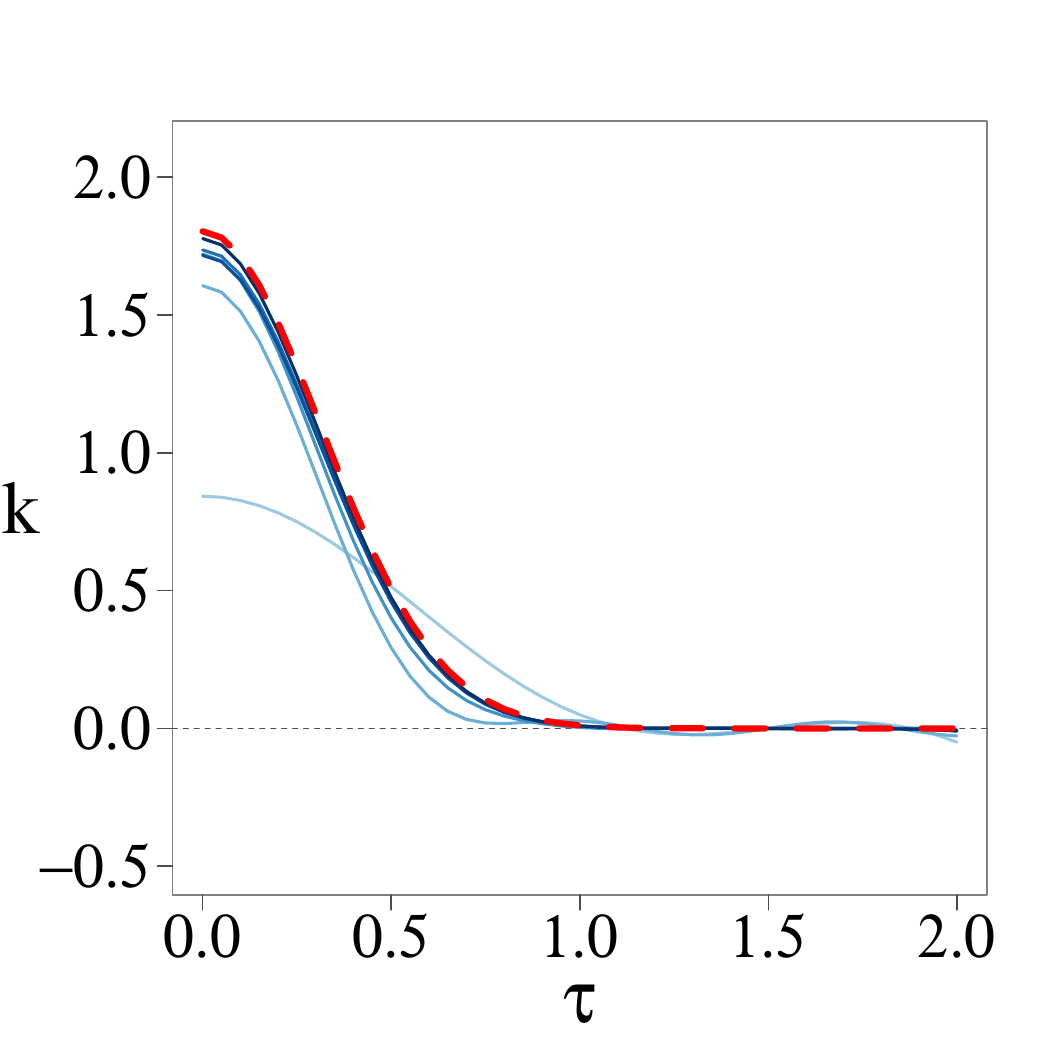} & \\
\cline{1-2}
c = 2 &
\includegraphics[scale=0.215, trim = 0mm 14mm 0mm 14mm, clip]{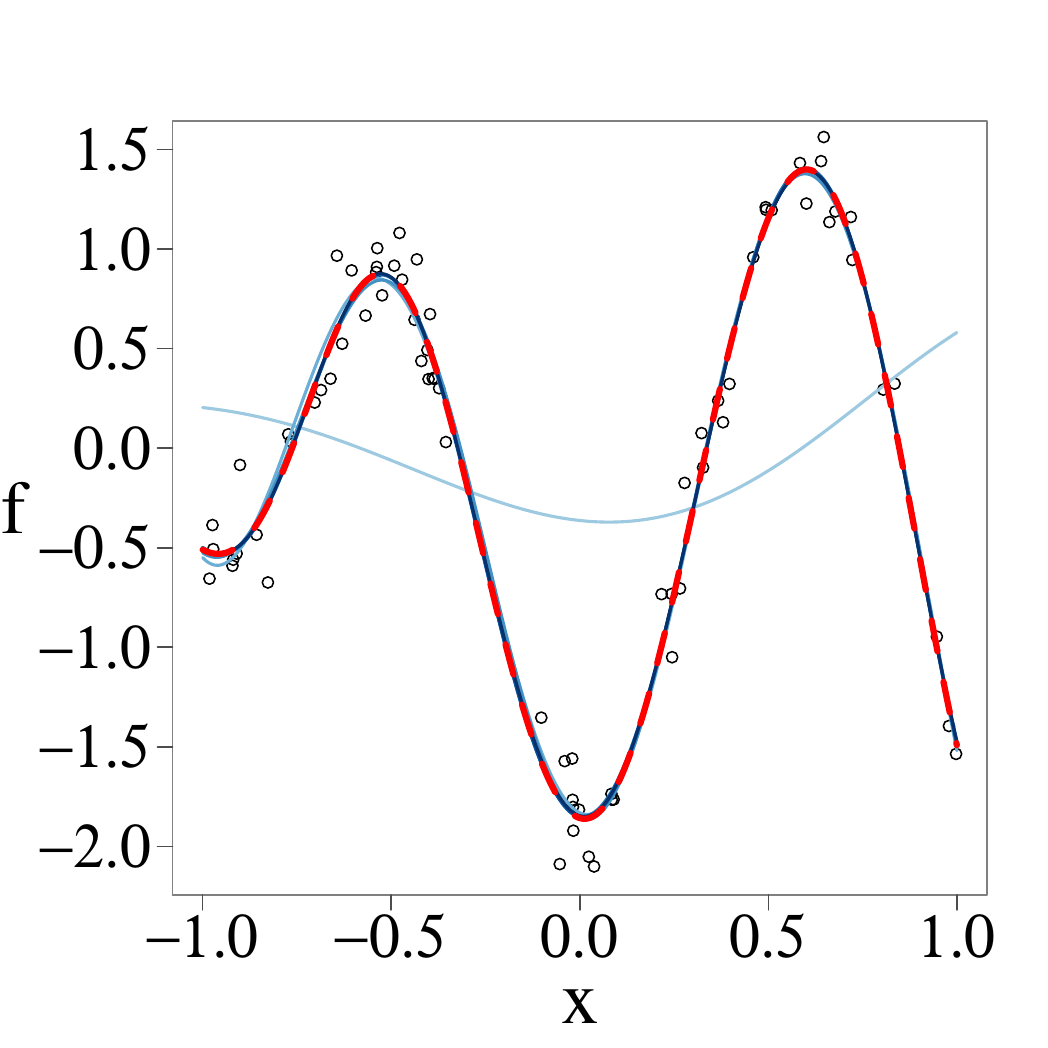} 
\includegraphics[scale=0.215, trim = 0mm 14mm 0mm 14mm, clip]{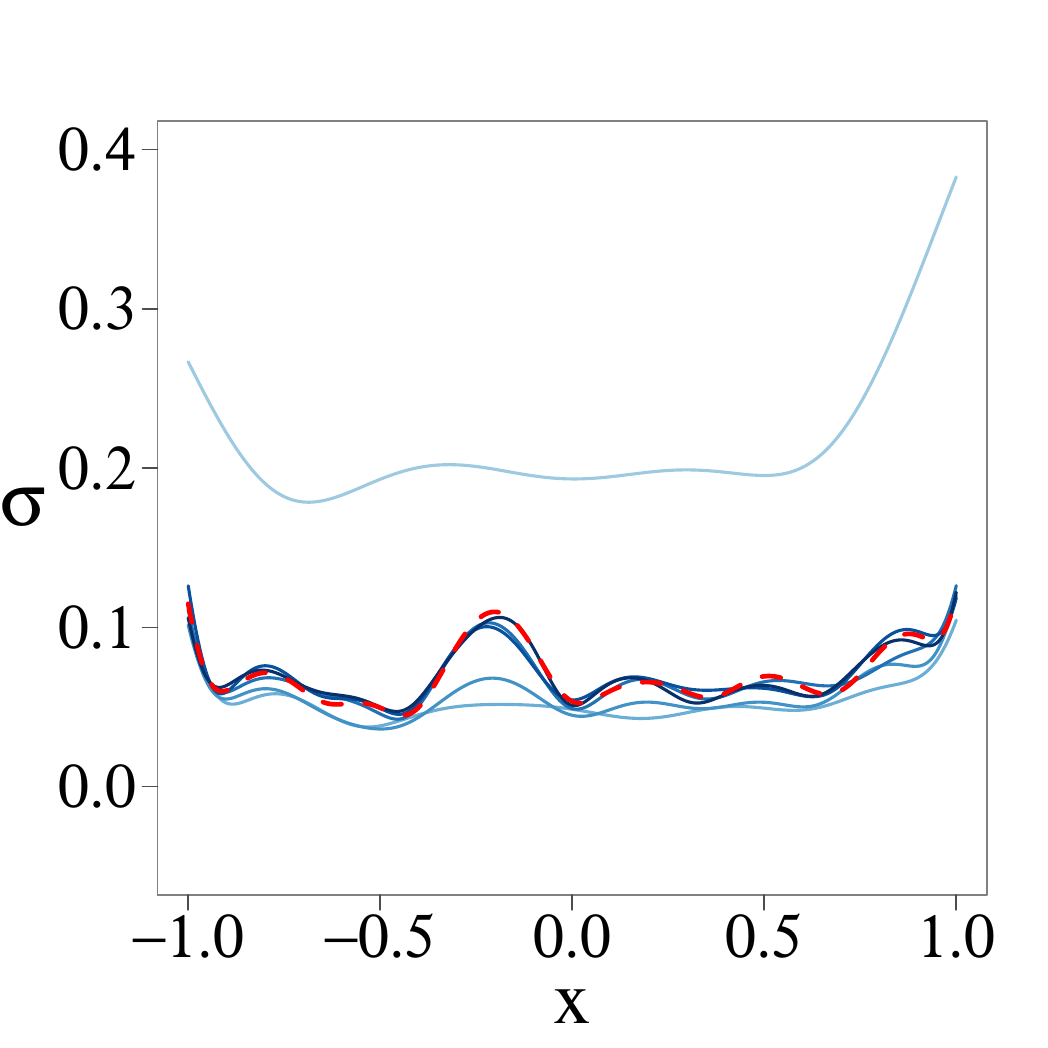} 
\includegraphics[scale=0.215, trim = 0mm 14mm 5mm 14mm, clip]{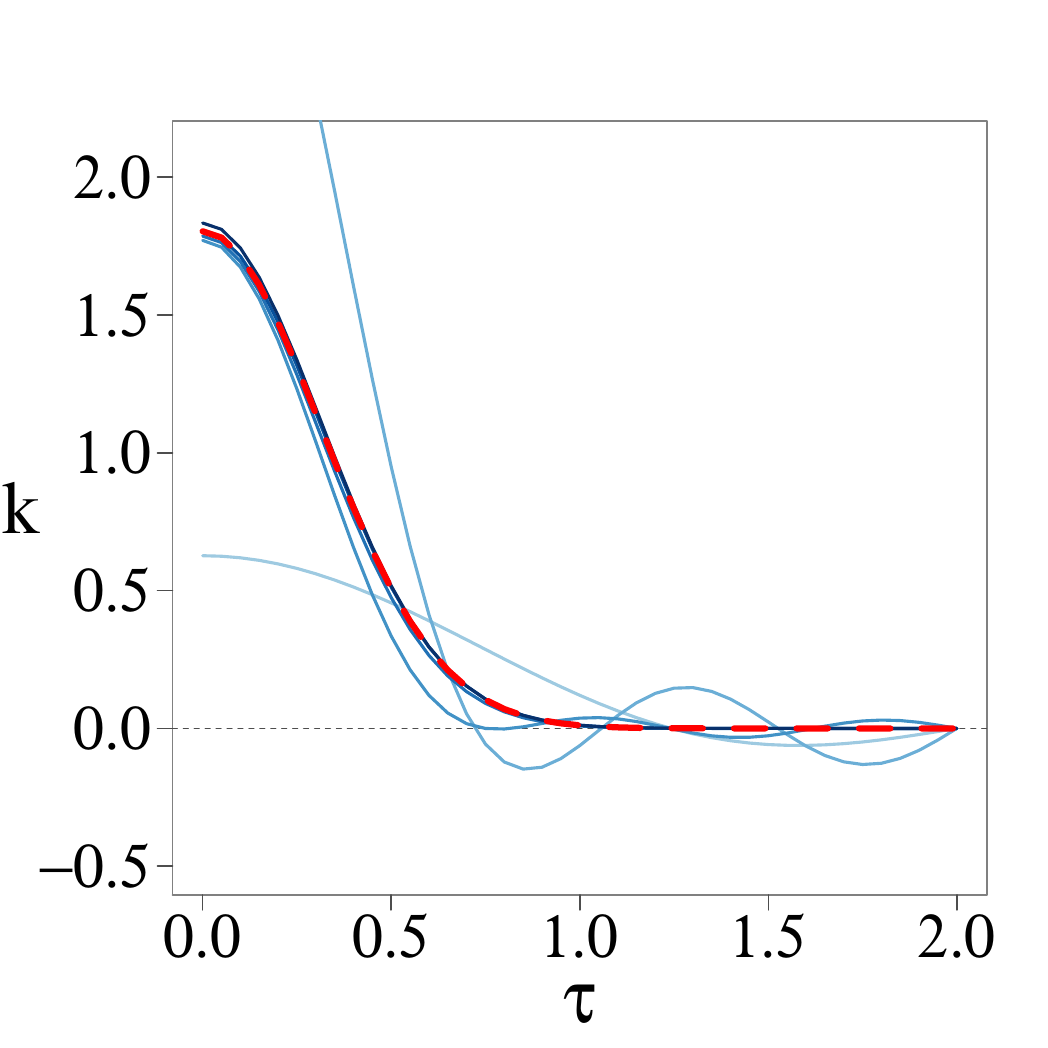} & \\
\cline{1-2}
c = 2.5 &
\includegraphics[scale=0.215, trim = 0mm 4mm 0mm 14mm, clip]{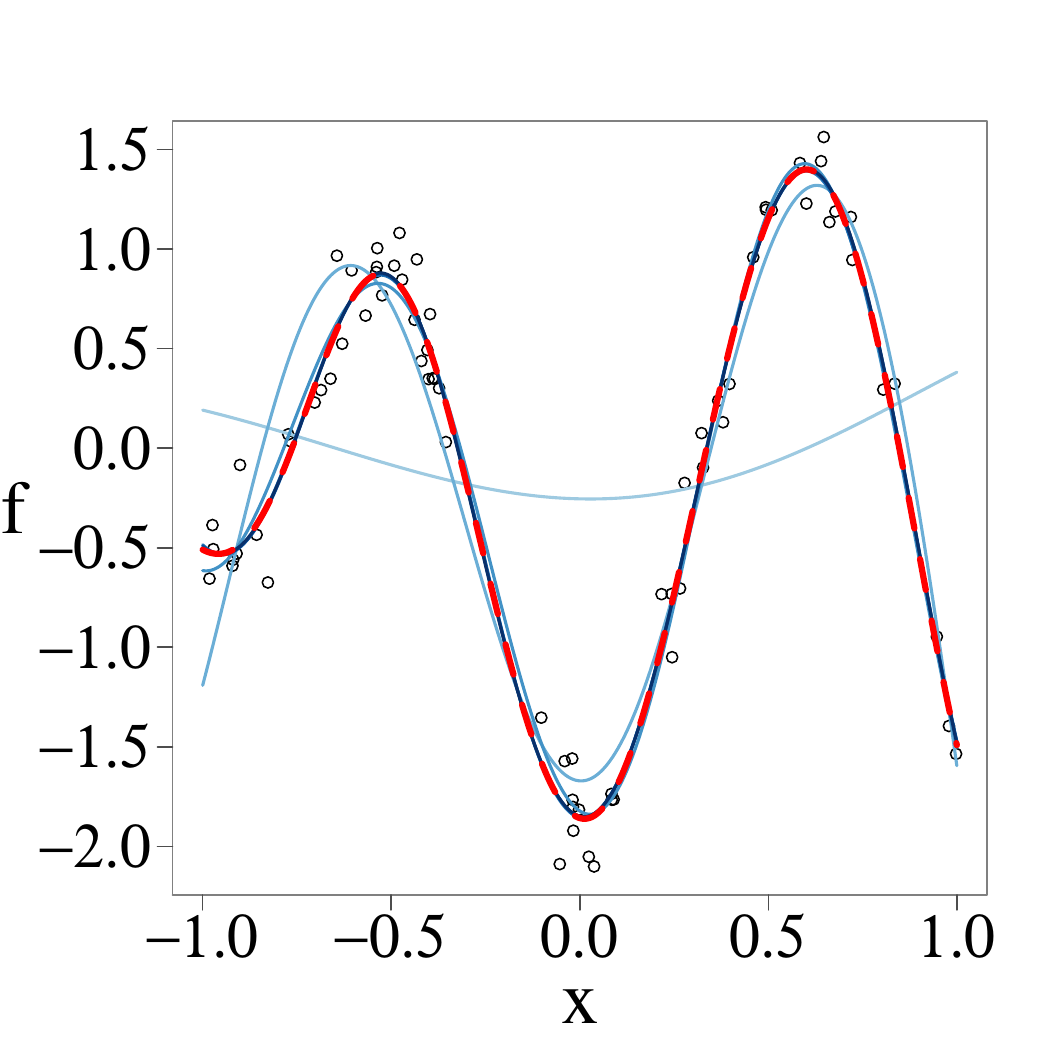}
\includegraphics[scale=0.215, trim = 0mm 4mm 0mm 14mm, clip]{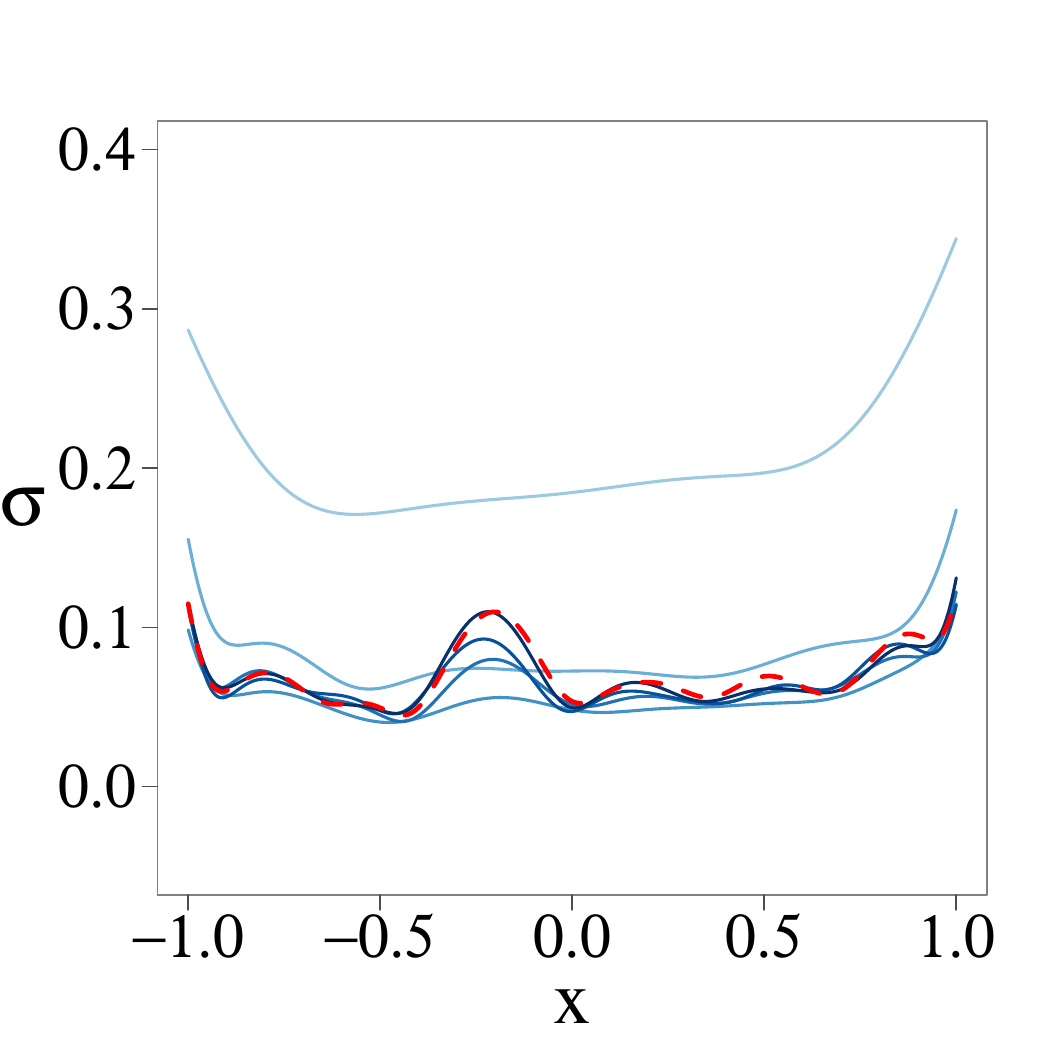}
\includegraphics[scale=0.215, trim = 0mm 4mm 5mm 14mm, clip]{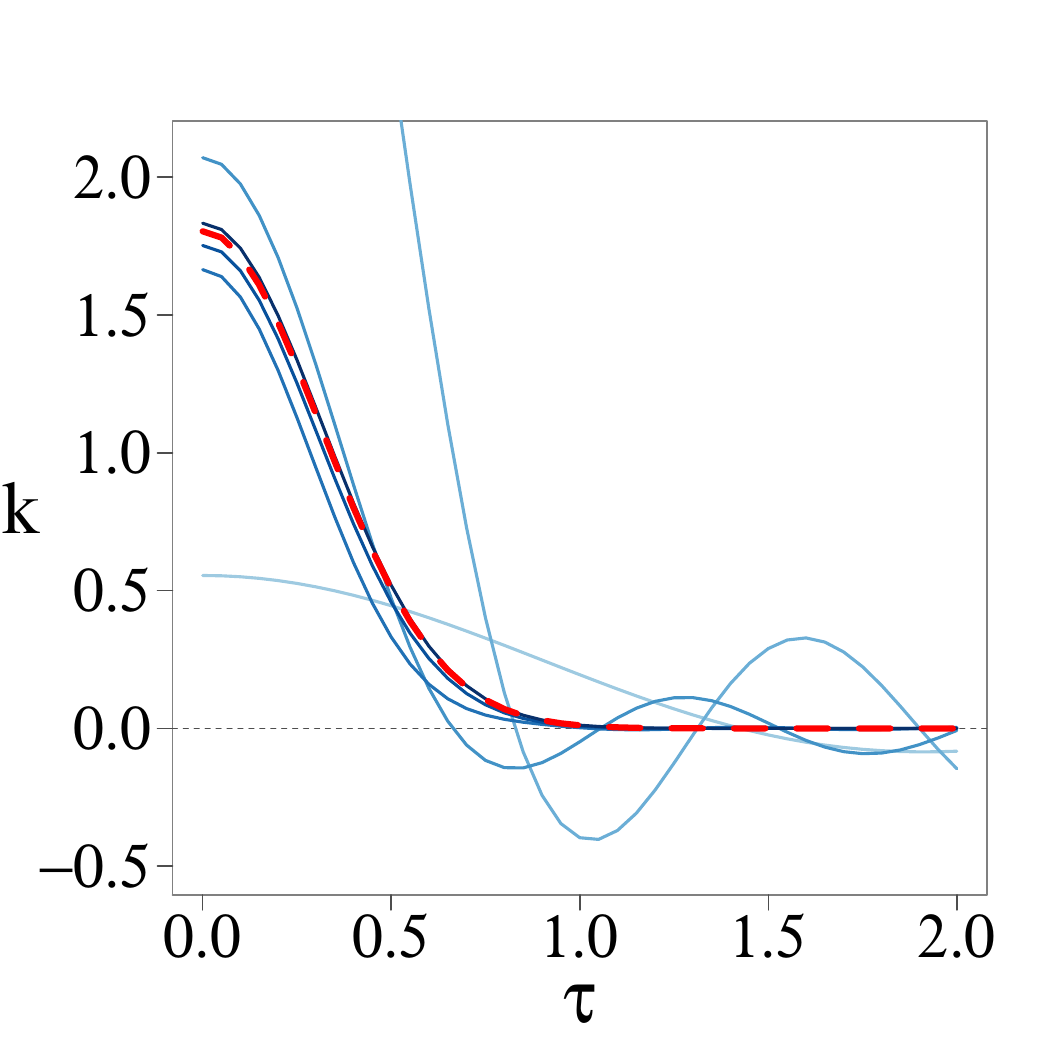} &\\
\arrayrulecolor{darkgray}\hline
\end{tabular}
\caption{Posterior mean predictive functions (left), posterior standard deviations (center) and  covariance functions (right) of both the exact GP model and the HSGP model for different number of basis functions $m$ and for different values of the boundary factor $c$.}
  \label{fig3_Post_part1}
\end{figure*}

\begin{figure*}[h]
\centering
\subfigure{\includegraphics[scale=0.38, trim = 0mm 8mm 0mm 0mm, clip]{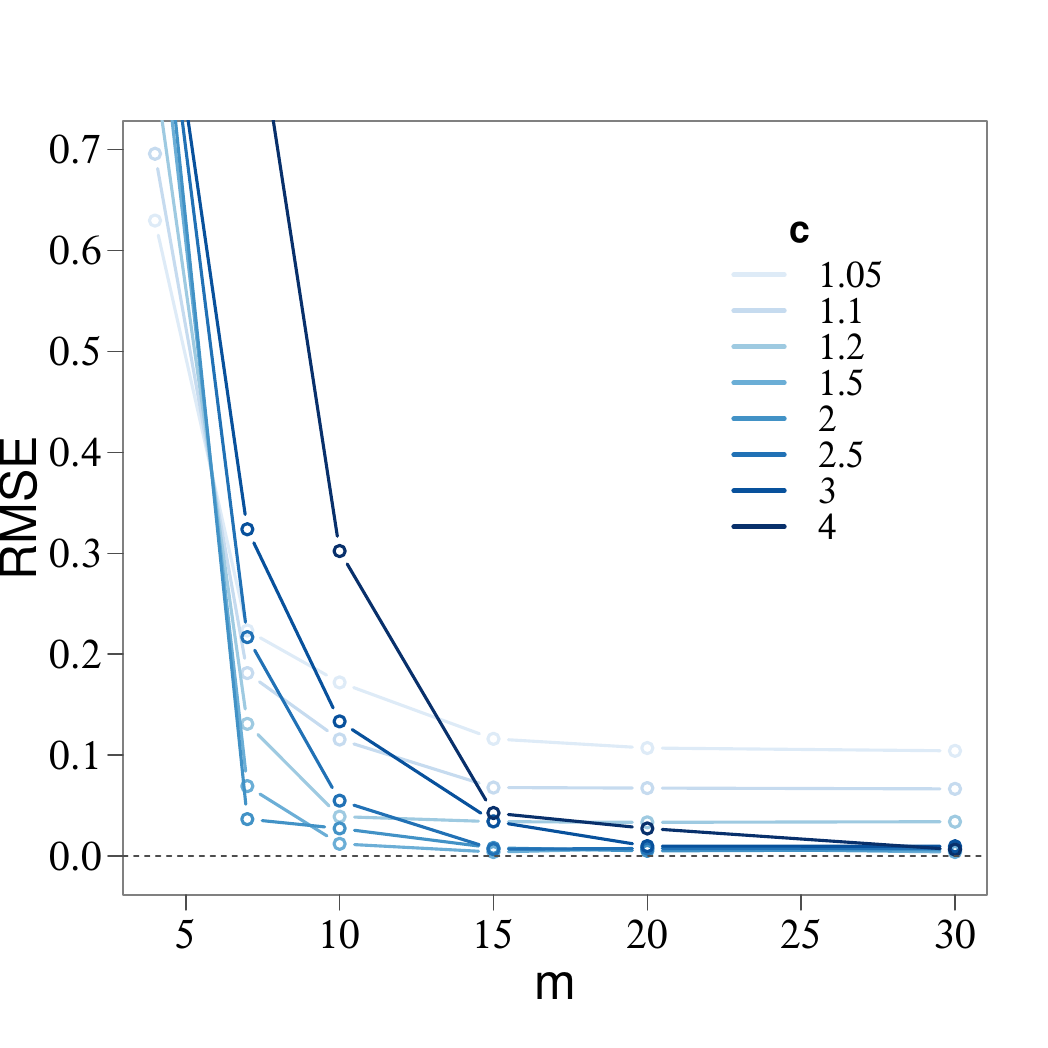}}
\subfigure{\includegraphics[scale=0.38, trim = 0mm 8mm 0mm 0mm, clip]{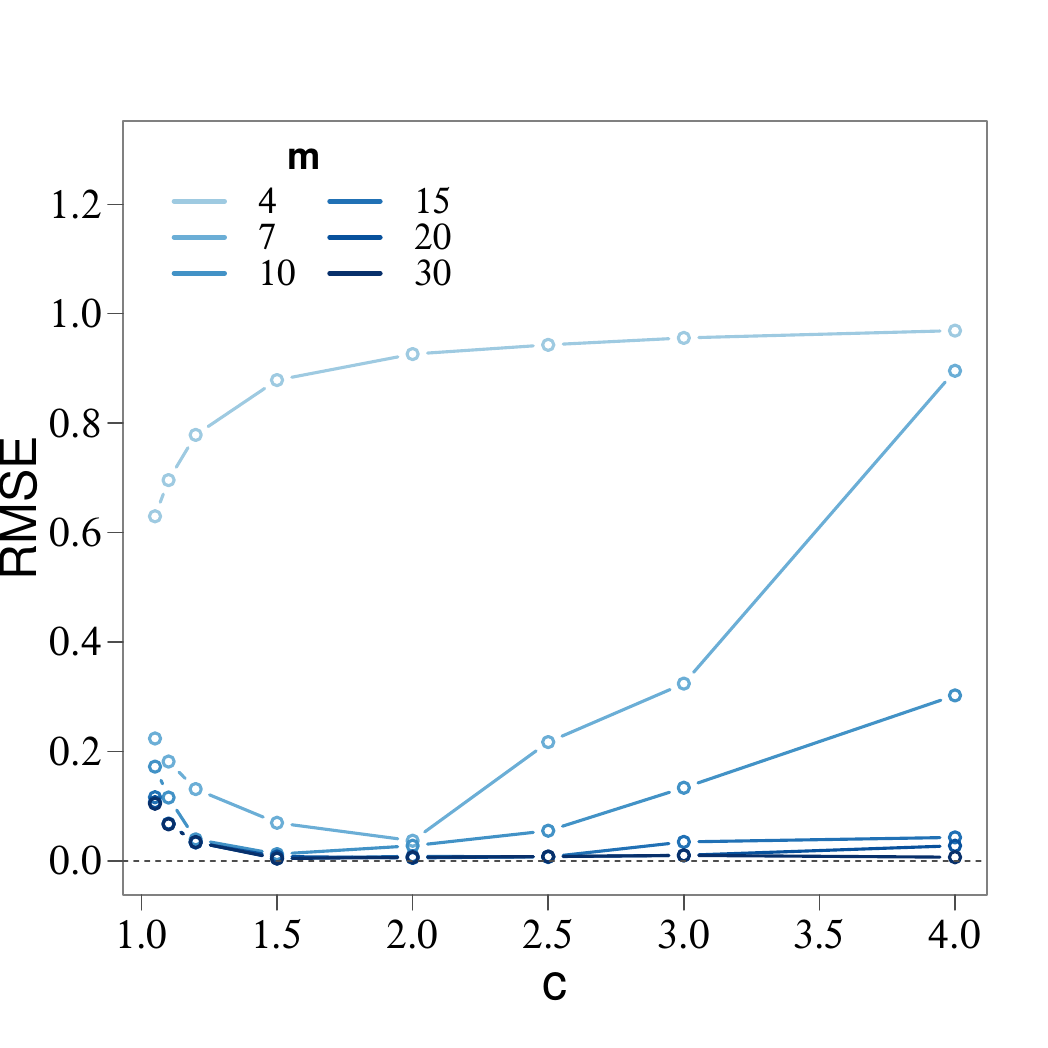}}
\caption{Root mean square error (RMSE) of the proposed HSGP models computed against the exact GP model. RMSE versus the number of basis functions $m$ and for different values of the boundary factor $c$ (left). RMSE versus the boundary factor $c$ and for different values of the number of basis functions $m$ (right). }
  \label{fig4_MSE_vs_J}
\end{figure*}

\begin{figure*}[h]
\centering
\subfigure{\includegraphics[scale=0.38, trim = 0mm 8mm 0mm 0mm, clip]{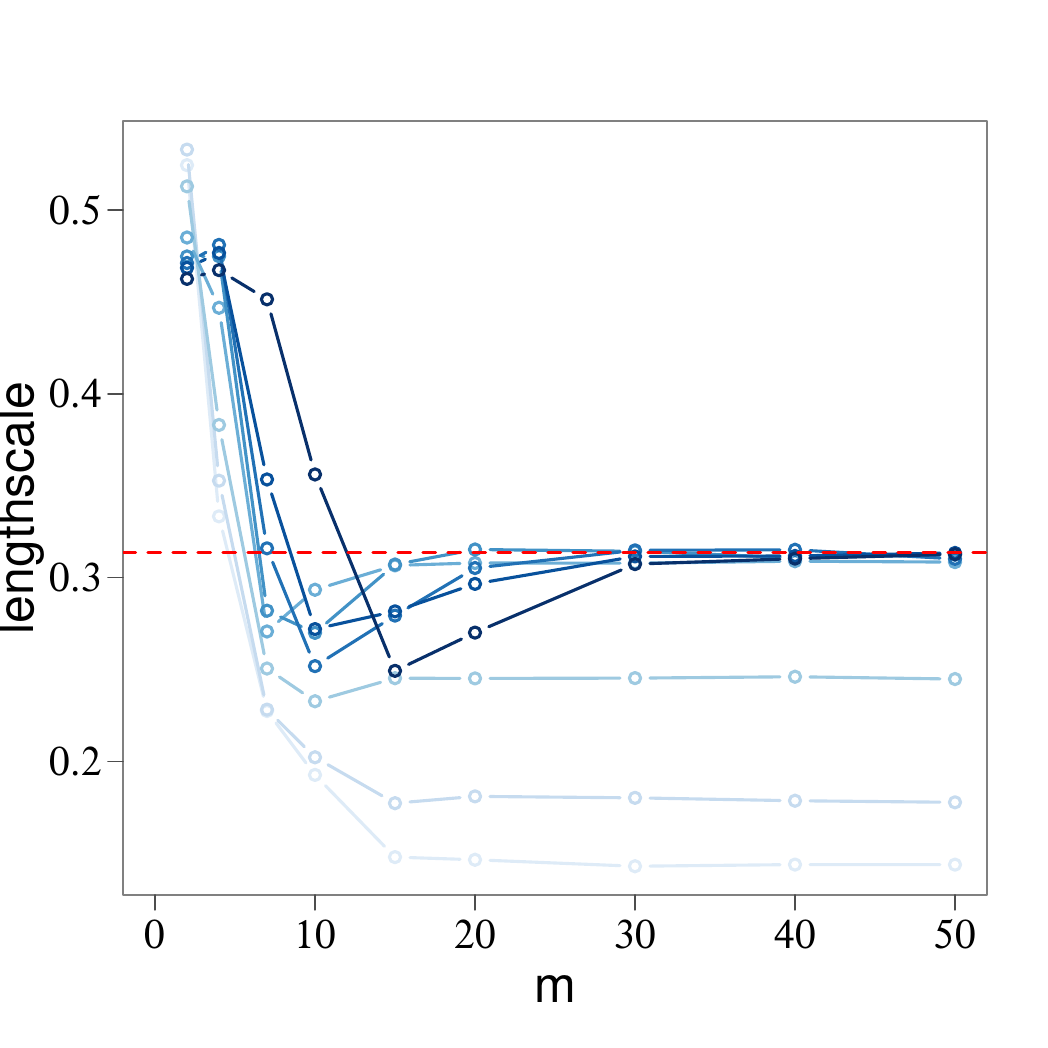}}
\subfigure{\includegraphics[scale=0.38, trim = 0mm 8mm 0mm 0mm, clip]{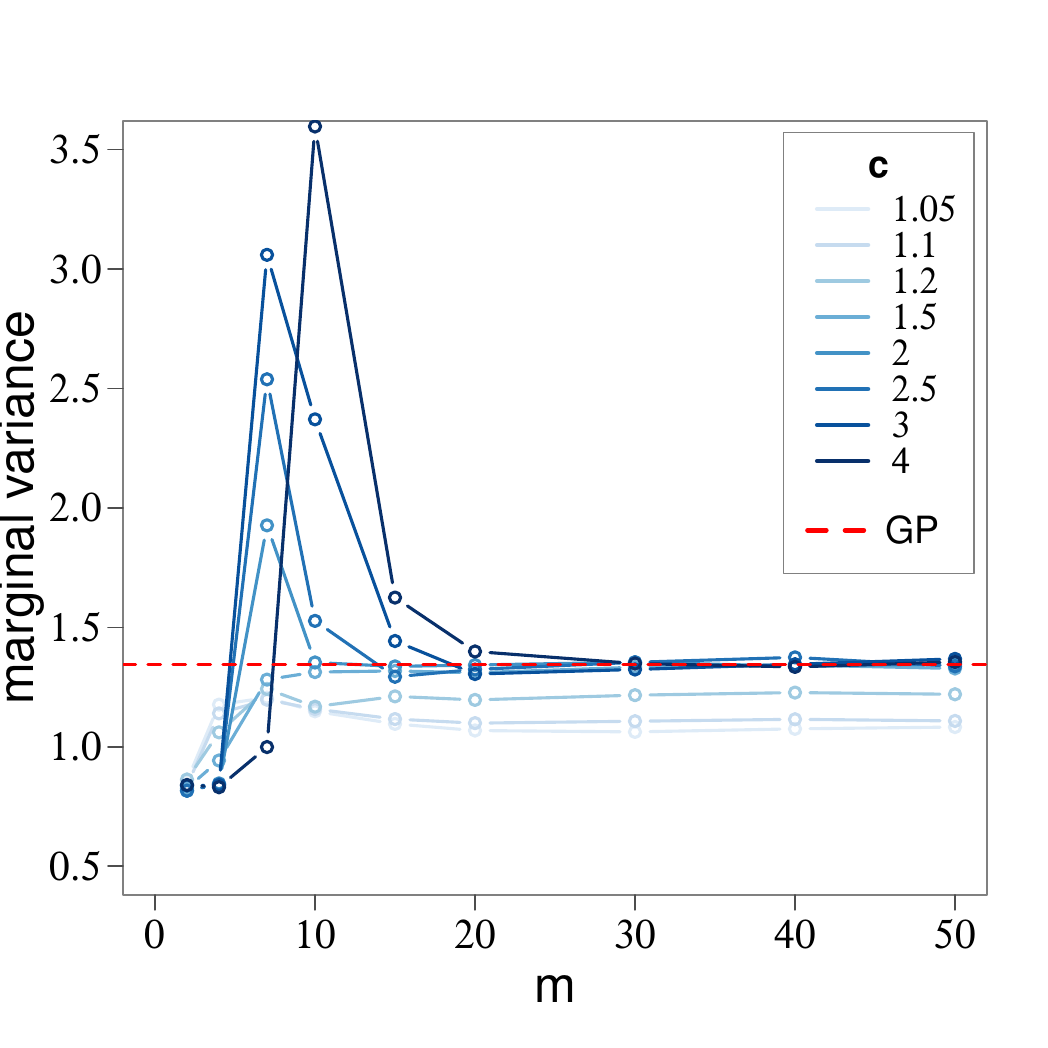}}
\caption{Estimated length-scale (left) and marginal variance (right) parameters of both exact GP and HSGP models, plotted versus the number of basis functions $m$ and for different values of the boundary factor $c$.}
  \label{fig5_lscale_vs_J}
\end{figure*}

Next, we analyze how the interaction between $m$ and $c$ affects the quality of the approximation. The length-scale and marginal variance of the covariance function will no longer be fixed but instead we compute the joint posterior distribution of the function values and the hyperparameters using the dynamic HMC algorithm implemented in Stan \citep{StanTeam:2021} for both the exact GP and the HSGP models. Figure~\ref{fig3_Post_part1} shows the posterior predictive mean and standard deviation of the function as well as the covariance function obtained after fitting the model for varying $m$ and~$c$. Figure~\ref{fig4_MSE_vs_J} shows the root mean square error (RMSE) of the HSGP models computed against the exact GP model. Figure~\ref{fig5_lscale_vs_J} shows the estimated length-scale and marginal variance for the exact GP model and the HSGP models. Looking at the RMSEs in Figure~\ref{fig4_MSE_vs_J}, we can conclude that the optimal choice in terms of precision and computation time for this example would be $m = 15$ basis functions and a boundary factor between $c = 1.5$ and $c = 2.5$. Further, the less conservative choice of $m = 10$ and $c = 1.5$ could also produce a sufficiently accurately approximation depending on the application. We may also come to the same conclusion by looking at the posterior predictions and covariance function plots in Figure~\ref{fig3_Post_part1}. From these results, some general conclusions may be drawn:
\begin{itemize}
\item As $c$ increases, $m$ has to increase as well (and vice versa). This is consistent with the expression for the eigenvalues in eq.~\eqref{eq_eigenvalue}, where $L$ appears in the denominator.
\item There exists a minimum $c$ below which an accurate approximation will never be achieved regardless of the number of basis functions $m$.
\end{itemize}

\subsection{Near linear proportionality between $m$, $c$ and $\ell$} \label{subsec_theoretic_linearity}

A priori, the terms in the series expansion \eqref{eq_approxf_multi} with very small spectral density are unlikely to contribute to the approximation. Given the boundary factor $c$ and the length-scale $\ell$, we can compute the cumulative sum of the spectral densities and find out how many basis functions are, a priori, explaining almost 100\% of the variation. Thus, given $c$ and $\ell$, we can estimate a good choice for the number of basis functions $m$ for any covariance function.

When considering squared exponential and Mat{\'e}rn covariance functions, we can show with simple algebra that when $c$ is larger than the minimal value recommendation, the number of $m$ first terms needed to explain almost 100\% of the variation has a near linear relationship with $\ell$ and $c$. With decreasing $\ell$, the $m$ should grow near linearly with $1/\ell$, and with increasing $c$, the $m$ should grow near linearly with $c$.
This is natural as with decreasing $\ell$, more higher frequency basis functions are needed. With increasing $c$, as a smaller range of the basis functions are used in the approximation, the expected number of zero up-crossings goes down linearly with $c$, and thus more higher frequency basis functions are needed to compensate this.
When $c$ is below our recommendations given $\ell$, the effect of $\ell$ and $c$ to the recommended $m$ is more non-linear, but as long as we stay in the recommended range the linearity assumption is useful thumb rule for how to change $m$, if $\ell$ or $c$ are changed.

\subsection{Empirical discovering of the functional form of the relationships between $m$, $c$ and $\ell$} \label{subsec_empiric_relations}

\begin{figure*}
\centering
\subfigure{\includegraphics[width=\textwidth, trim = 0mm 0mm 0mm 0mm, clip]{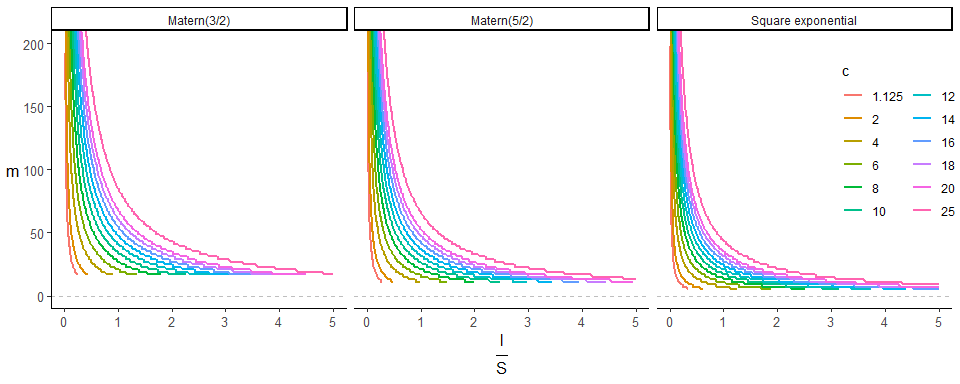}}
\caption{Relation among the minimum number of basis functions $m$, the boundary factor $c$ \hspace{0.1mm} ($c = \frac{L}{S}$) \hspace{0.1mm} and the length-scale normalized by the half-range of the data ($\frac{\ell}{S}$), for the squared exponential, Mat\'ern ($\nu$=3/2) and Mat\'ern ($\nu$=5/2) covariance functions.}
  \label{fig6_relationships}
\end{figure*}

Empirical simulations are carried out to analyze the relationships between $m$, $c$ and $\frac{\ell}{S}$ (lengthscale $\ell$ relative to the half-range $S$ of the input domain). Figure \ref{fig6_relationships} depicts how these three factors interact and affect the accuracy of the HSGP approximation for a GP with squared exponential, Mat\'ern ($\nu$=3/2), and Mat\'ern ($\nu$=5/2) covariance functions and a single input dimension. More precisely, for a given GP model (with a particular covariance function) with length-scale $\frac{\ell}{S}$ (From now to later on we will refer to $\frac{\ell}{S}$ by just $\ell$) and given a boundary factor $c$, Figure \ref{fig6_relationships} shows the minimum number of basis functions $m$ required to obtain an accurate approximation in the sense of satisfying eq. (\ref{eq_diff_covs}). We considered an approximation to be a sufficiently accurate when the total variation difference between the approximate and exact covariance functions, $\varepsilon$ in eq. (\ref{eq_diff_covs}), is below 1$\%$ of the total area under the curve of the exact covariance function $k$ such that
\begin{equation} \label{eq:rel_total_distance}
 \int | k(\tau) - \tilde{k}_m(\tau)|  \,\mathrm{d}\tau  < 0.01 \int k(\tau) \,\mathrm{d}\tau,
\end{equation}
where $\tilde{k}_m$ is the approximate covariance function with $m$ basis functions. Alternatively, these figures can be understood as providing the minimum $c$ that we should use for given $\ell$ and $m$. Of course, we may also read it as providing the minimum $\ell$ that can be approximated with high accuracy given $m$ and $c$. We obtain the following main conclusions:
\begin{itemize}
\item As $\ell$ increases, $m$ required for an accurate approximation decrease.
\item The lower $c$, the smaller $m$ can and $\ell$ must be to achieve an accurate approximation.
\item For a given $\ell$ there exist a minimum $c$ under which an accurate approximation is never going to be achieved regardless of $m$. This fact can be seen in Figure~\ref{fig6_relationships} as the contour lines which represent $c$ have an end in function of $\ell$ (Valid $c$ are restricted in function of $\ell$). As $\ell$ increases, the minimum valid $c$ also increases.
\end{itemize}

\subsubsection{Numerical equations}
\label{sec_num_equations}

As explained in Section~\ref{subsec_theoretic_linearity}, when $c$ is large enough, there is a near linear proportionality between $m$, $\ell$ and $c$. To obtain practical numerical functions that can be used to guide the selection of these parameters, we have empirically derived the practically useful constant terms. We require a lower bound for $c$ of $c \geq 1.2$ such that the equations below are precise enough for practical application.

\begin{itemize}
\item{Squared exponential}:
\begin{align}
&m= 1.75 \, \frac{c}{\nicefrac{\ell}{S}} \;\; \Leftrightarrow \;\; \nicefrac{\ell}{S}= 1.75 \, \frac{c}{m},	\label{eq_m_l_QE} \\
&\text{with} \nonumber \\
&c \geq \, 3.2 \, \nicefrac{\ell}{S} \;\; \& \;\, c \geq \, 1.2	\label{eq_c_vs_l_QE}
\end{align}

\item{Mat\'ern ($\nu$=5/2)}:
\begin{align}
&m= 2.65 \, \frac{c}{\nicefrac{\ell}{S}} \;\; \Leftrightarrow \;\; \nicefrac{\ell}{S}= 2.65 \, \frac{c}{m}, \label{eq_m_l_mat52} \\
&\text{with} \nonumber \\
&c \, \geq \, 4.1 \, \nicefrac{\ell}{S} \;\; \& \;\, c \geq \, 1.2  \label{eq_c_vs_l_mat52}
\end{align}

\item{Mat\'ern ($\nu$=3/2)}:
\begin{align}
&m= 3.42 \, \frac{c}{\nicefrac{\ell}{S}} \;\; \Leftrightarrow \;\; \nicefrac{\ell}{S}= 3.42 \, \frac{c}{m}, \label{eq_m_l_mat32}\\
&\text{with} \nonumber \\
&c \, \geq \, 4.5 \, \nicefrac{\ell}{S} \;\; \& \;\, c \geq \, 1.2 \label{eq_c_vs_l_mat32}
\end{align}

\end{itemize}
These constants vary monotonically with respect to $\nu$ (squared exponential corresponding to Mat\'ern with $\nu \to \infty$). Using the formula for Mat\'ern ($\nu$=3/2) provides the largest $m$ and $c$, and thus this formula alone could be used as a conservative choice for all Mat\'ern covariance functions with $\nu \geq 3/2$ and likely as a good initial guess for many other covariance functions. If the aim is to find minimal $m$ to speedup the computation, a further refined formula can be obtained for new covariance functions.

Figure~\ref{fig6_relationships} and previous equations~(\ref{eq_m_l_QE})-(\ref{eq_c_vs_l_mat52}) were obtained for a GP with a unidimensional covariance function, which result in a surfaces depending on three variables, $m$, $c$ and $\ell$. Equivalent results for a GP model with a two-dimensional covariance function would result in a surface depending on four variables, $m$, $c$, $\ell_1$ and $\ell_2$. More precisely, in the multi-dimensional case, whether the approximation is close enough might depend only on the ratio between wigglyness in every dimensions. For instance, in the two-dimensional case, it would depend on the ratio between $\ell_1$ and $\ell_2$. Future research will focus on building useful graphs or analytical models that provide these relations in multi-dimensional cases. However, as an approximation, we can use the unidimensional GP conclusions in Figure~\ref{fig6_relationships} or equations~(\ref{eq_m_l_QE})-(\ref{eq_c_vs_l_mat52}) to check the accuracy by analyzing individually the different dimensions of a multidimensional GP model.

\subsection{Relationships between $m$ and $\ell$ for a periodic squared exponential covariance function} \label{subsec_empiric_relations_periodic}

As commented in Section~\ref{sec_method_periodic}, in Appendix~\ref{sec_periodic} we present an approximate linear representation of a periodic squared exponential covariance function. In Appendix~\ref{sec_periodic}, we also analyze the accuracy of this linear representation and derive the minimum number of terms $m$ in the approximation required to achieve a close approximation to the exact periodic squared exponential kernel as a function of the length-scale $\ell$ of the kernel. Since this is a series expansion of sinusoidal functions, the approximation does not depend on any boundary condition. This relationship between $m$ and $\ell$ for a periodic squared exponential covariance function is gathered in Figure~\ref{figB1_m_lscale_periodic} and the numerical equation was estimated in eq.~\eqref{eq_J_l_periodic} which is depicted next:
\begin{align*}
&m \geq \frac{3.72}{\ell} \;\; \Leftrightarrow \;\; \ell \geq \frac{3.72}{m} .
\end{align*}

\subsection{Diagnostics of the approximation} \label{subsec_diagnostics}

Equations \eqref{eq_m_l_QE}, \eqref{eq_m_l_mat52}, \eqref{eq_m_l_mat32} and \eqref{eq_J_l_periodic} (depending on which kernel is used) provide the minimum length-scale that can be accurately inferred given $m$ and $c$. This information serves as a powerful diagnostic tool in determining if the obtained accuracy is acceptable. As the length-scale $\ell$ controls the wigglyness of the function, it strongly influences the difficulty of estimating the latent function from the data. Basically, if the length-scale estimate is accurate, we can expect the HSGP approximation to be accurate as well.

Having obtained an estimate $\hat{\ell}$ for a HSGP model with prespecified $m$ and $c$, we can check whether $\hat{\ell}$ exceeds the smallest length-scale, that can be accurately inferred, provided as a function of $m$ and $c$ by equations \eqref{eq_m_l_QE}, \eqref{eq_m_l_mat52}, \eqref{eq_m_l_mat32} and \eqref{eq_J_l_periodic} (depending on which kernel is used). If $\hat{\ell}$ exceeds this value, the approximation is assumed to be good. If $\hat{\ell}$ does not exceed this value, the approximation may be inaccurate, and $m$ and/or $c$ need to be increased. In Figures~\ref{fig4_MSE_vs_J} and \ref{fig5_lscale_vs_J}, $m = 10$ and $c = 1.5$ were sufficient for an accurate modeling of function with $\ell = 0.3$, which matches the diagnostic based on equations \eqref{eq_m_l_QE} and \eqref{eq_c_vs_l_QE}.

Equations in Section~\ref{sec_num_equations} to update $m$ and $c$ imply:
\begin{itemize}
	\item $c$ must be big enough for a given $\ell$, and
	\item $m$ must be big enough for given $\ell$ and $c$. 
\end{itemize}
If larger than minimal $c$ and $m$ (for a given $\ell$) are used in the initial HSGP model, it is likely that the results are already sufficiently accurate. As $\ell$ is initially unknown, we recommend using this diagnostic in an iterative procedure by starting with $c$ and $m$ based on some initial guess about $\ell$, and if the estimated $\hat{\ell}$ is below the diagnostic threshold, select new $c$ and $m$ using $\hat{\ell}$. This can be repeated until
\begin{itemize}
\item the estimated $\hat{\ell}$ is larger than the diagnostic threshold given $c$ and $m$, and
\item the predictive accuracy measures, for example, root mean square error (RMSE), coefficient of determination ($R^2$), or expected log predicitve density (ELPD) do not improve.
\end{itemize}

As commented above, the estimated $\hat{\ell}$ being larger than diagnostic threshold does not guarantee that the approximate is sufficiently accurate, and thus we recommend to look at the predicitve accuracy measures, too.

Apart from providing a powerful diagnostic tool in determining if the approxiamtion is sufficiently accurate, the equations in the previous Section~\ref{sec_num_equations} also provide the optimal values for $m$ (the minimum $m$ required for an accurate approximation) and $c$ (the minimum $c$ that allows for the minimum $m$) that van be used to minimize the computational cost in repeated computations (e.g., in cross-validation and simulation based calibration).
This is even more useful in multi-dimensional cases ($2 \leq D \leq 4$), where knowing the smallest useful value of $m$ for each dimension has even bigger effect on the total computational cost.

\subsubsection{A step-by-step user-guide to apply the diagnostics} \label{sec_user_guide}

Based on the above proposed diagnostics, we obtain a simple, iterative step-by-step procedure that users can follow in order to obtain an accurate HSGP approximation. The procedure is split into two phases, Phase~A and B, which have to be completed consecutively.

\paragraph*{Phase A:}

\begin{enumerate}

\item[A1.] Make an initial guess on the length-scale $\ell_1$. If there is no useful information available, we recommend to start with a large length-scale, about $\ell_1 \in [0.5, 1]$. Set the iteration index to $i=1$.

\item[A2.] Obtain the minimum valid boundary factor $c_i$ determined by $\ell = \ell_i$ for the given kernel as per Section~\ref{sec_num_equations}.

\item[A3.] Obtain the mimimum valid number of basis functions $m_i$ determined by $\ell = \ell_i$ and $c = c_i$ for the given kernel as per Section~\ref{sec_num_equations}. 

\item[A4.] Fit an HSGP model using $m_i$ and $c_i$ and ensure convergence of the MCMC chains.

\item[A5.] Perform the length-scale diagnostic by checking if $\hat{\ell}_i - 0.01 \geq \ell_i$. 

If the diagnostic is FALSE, the HSGP approximation is not yet sufficiently accurate. Set $\ell_{i+1} = \hat{\ell}_{i}$, increase the iteration index $i = i + 1$, and go back to A2.

If the diagnostic is TRUE, the HSGP approximation is close to be sufficiently accurate. Continue with Phase~B.

\end{enumerate}

\paragraph*{Phase B:}

\begin{enumerate}

\item[B1.] For the current HSGP model, compute measures of predictive accuracy, for example, RMSE, $R^2$, or ELPD.

\item[B2.] Set $m_{i+1} = m_{i} + 5$ and increase the iteration index $i = i + 1$.

\item[B3.] Obtain the minimum valid boundary factor $c_i$ determined by $\ell = \hat{\ell}_{i-1}$ for the given kernel as per Section \ref{sec_num_equations}.

\item[B4.] Obtain the minimum valid length-scale $\ell_i$ that can be accurately estimated by $m = m_i$ and $c = c_i$ for the given kernel as per Section~\ref{sec_num_equations}.

\item[B5.] Fit an HSGP model using $m_i$ and $c_i$ and ensure convergence of the MCMC chains.

\item[B6.] Perform the length-scale diagnostic by checking if $\hat{\ell}_i - 0.01 \geq \ell_i$. 

Check the stability of both $\hat{\ell}_i$ and the measures of predictive accuracy relative to the previous iteration.

If all the stability checks succeed, the HSGP approximation of the latest model should be sufficiently accurate and the procedure ends here.

Otherwise, go back to B1.
\end{enumerate}

\subsection{Performance analysis of the diagnostics}

\begin{figure*}
\centering
\subfigure{\includegraphics[width=\textwidth, trim = 0mm 0mm 0mm 0mm, clip]{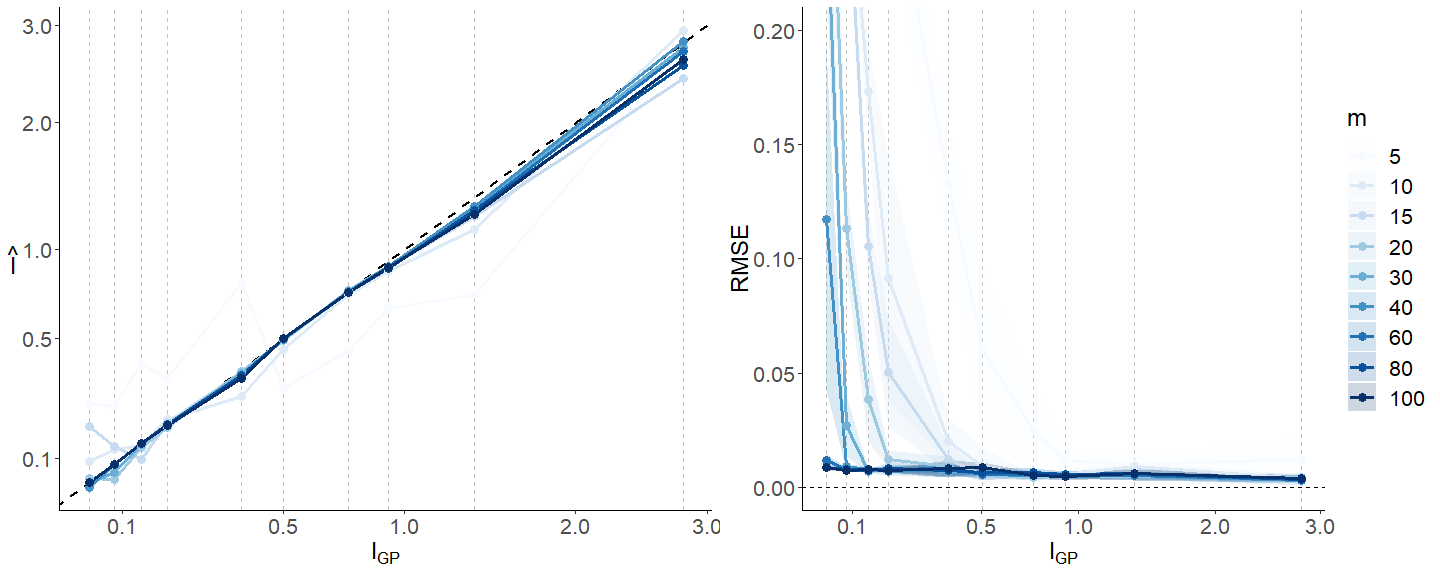}}
\caption{Exact GP length-scale ($\ell_{GP}$) and approximate HSGP length-scale ($\hat{\ell}$) for various datasets with varying smoothness (left) and their corresponding root mean squared errors (RMSE) computed with the exact GP model as the reference.}
  \label{fig7_lscale_comparison}
\end{figure*}

In this section, we first illustrate that accurate estimates of the length-scale implies accurate approximations via HSGPs. Figure~\ref{fig7_lscale_comparison} left shows a comparison of the length-scale estimates
obtained from the exact GP and HSGP models with a squared exponential kernel, from various noisy datasets drawn from underlying functions with varying smoothness. Different values for the number of basis functions $m$ are used when estimating the HSGP models, and the boundary factor $c$ is set to a valid and optimum value in every case by using eq.~(\ref{eq_c_vs_l_QE}). Figure~\ref{fig7_lscale_comparison} (right) shows the RMSE of the HSGP models with the exact GP model as the reference. It can be seen that accurate estimates of the length-scale imply small RMSEs.

\begin{figure}
\centering
\subfigure{\includegraphics[width=0.5\textwidth, trim = 0mm 0mm 0mm 0mm, clip]{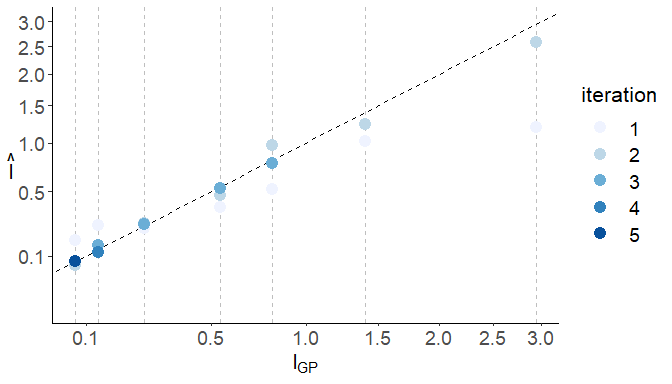}}
\caption{Comparison of the true and estimated length-scales in every iteration for the different data sets varying the length-scale.}
  \label{fig8_diagnostic_lscale_comparison}
\end{figure}

Table~\ref{tab1_diagnostic} shows the iterative steps of applying the diagnostic procedure explained in Section~\ref{sec_user_guide} over some of the data sets also used in the analysis in Figure~\ref{fig7_lscale_comparison} left. It is clearly visible that by following our recommendations, an optimum solution with minimum computational requirements is achieved in these cases. Figure~\ref{fig8_diagnostic_lscale_comparison} graphically compares the exact GP length-scale and the estimated HSGP length-scale in every iteration and data set. Between two and four iterations, depending on wigglyness of the function to be learned and the distance between the initial guess of the length scale and the true length scale, are sufficient to reach the optimal values of $m$ and $c$. 

As concrete examples, the iterative steps applied to perform diagnostic on two of the data sets in Table~\ref{tab1_diagnostic} are described in Appendix~\ref{sec_ex_iterative_steps}.

\begin{table*}
\centering
\setlength{\tabcolsep}{4pt}
\begin{tabular}{ c c c c c c c c c c | c c}
\arrayrulecolor{gray}\hline \\[-3mm]
$\ell_{\text{GP}}$ & iter. & $\ell$ & $c$ & $m$ & $\hat{\ell}$ & $\hat{\ell} + 0.01 \geq \ell$ & RMSE & $R^2$ & ELPD & RMSE* & \\ 
\arrayrulecolor{lightgray}\hline \\[-1mm]
\multirow{5}{*}{ 0.08} 
 & 1 & 0.50 & 1.60 & 6 & 0.17 & FALSE & 0.45 & 0.61 & -0.40 & 0.36 & \multirow{5}{*}{ \includegraphics[scale=0.20, trim = 0mm 10mm 0mm 5mm, clip]{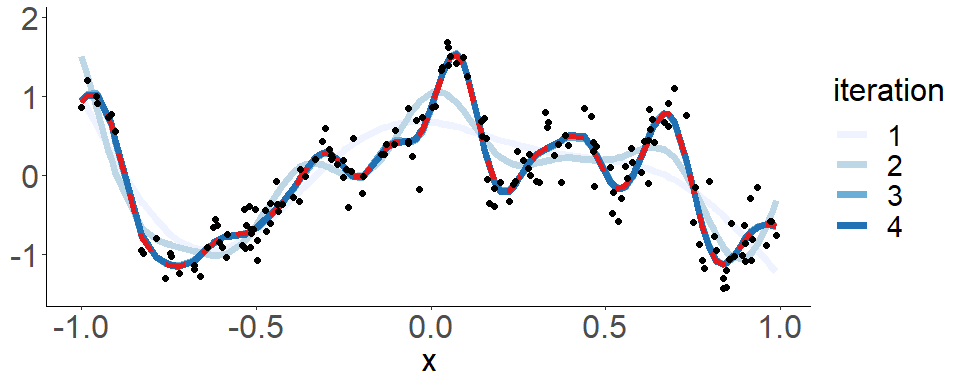}}\\
 & 2 & 0.17 & 1.20 & 13 & 0.07 & FALSE & 0.36 & 0.74 & -0.18 & 0.24 & \\
 & 3 & 0.07 & 1.20 & 31 & 0.08 & TRUE & 0.24 & 0.87 & 0.20 & 0.01 & \\
 & 4 & 0.06 & 1.20 & 36 & 0.08 & TRUE & 0.24 & 0.87 & 0.21 & 0.01 & \\
 \\
 \arrayrulecolor{lightgray}\hline \\[-1mm]
\multirow{3}{*}{ 0.13} 
 & 1 & 0.50 & 1.60 & 6 & 0.25 & FALSE & 0.46 & 0.63 & -0.49 & 0.33 & \multirow{5}{*}{ \includegraphics[scale=0.20, trim = 0mm 10mm 0mm 5mm, clip]{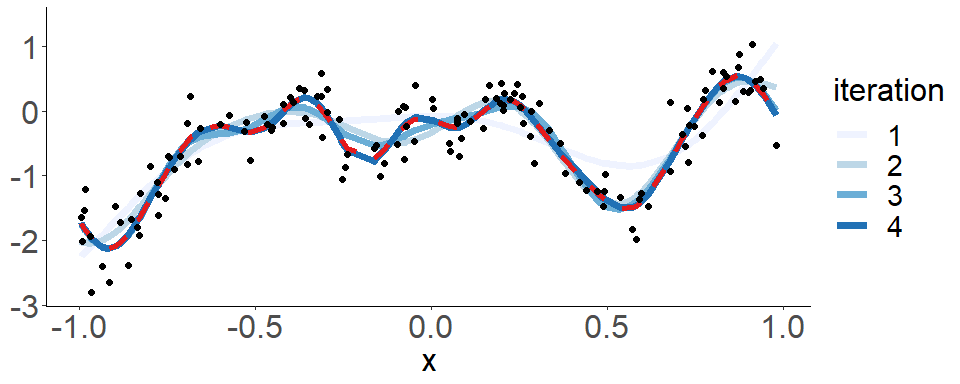}}\\
 & 2 & 0.25 & 1.20 & 9 & 0.15 & FALSE & 0.35 & 0.78 & -0.14 & 0.15 & \\
 & 3 & 0.15 & 1.20 & 15 & 0.15 & TRUE & 0.32 & 0.81 & -0.07 & 0.08 & \\
 & 4 & 0.11 & 1.20 & 20 & 0.12 & TRUE & 0.30 & 0.83 & -0.01 & 0.01 & \\
 \\[-1mm]
\arrayrulecolor{lightgray}\hline \\[-1mm]
\multirow{3}{*}{ 0.25} 
 & 1 & 0.50 & 1.60 & 6 & 0.24 & FALSE & 0.37 & 0.40 & -0.15 & 0.21 & \multirow{5}{*}{ \includegraphics[scale=0.20, trim = 0mm 10mm 0mm 5mm, clip]{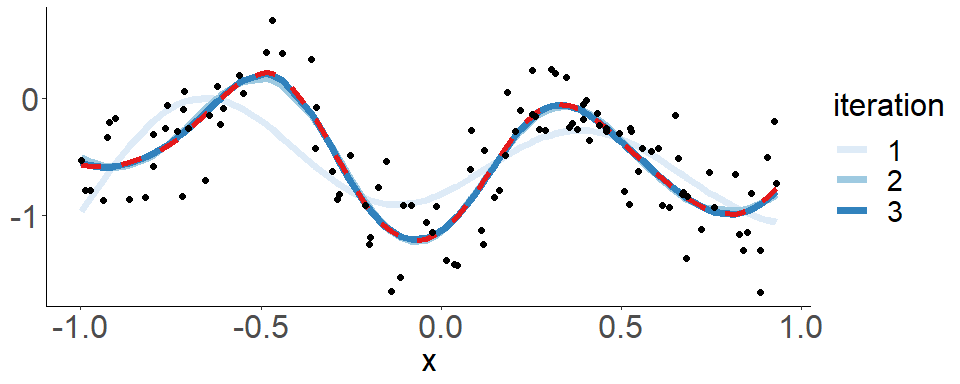}}\\
 & 2 & 0.24 & 1.20 & 9 & 0.26 & TRUE & 0.30 & 0.60 & -0.09 & 0.02 & \\
 & 3 & 0.25 & 1.20 & 14 & 0.26 & TRUE & 0.29 & 0.62 & -0.05 & 0.01 & \\
 \\
 \\[-1mm]
\arrayrulecolor{lightgray}\hline \\[-1mm]
\multirow{3}{*}{ 0.54} 
 & 1 & 0.50 & 1.60 & 6 & 0.38 & FALSE & 0.27 & 0.86 & 0.03 & 0.01 & \multirow{5}{*}{ \includegraphics[scale=0.20, trim = 0mm 10mm 0mm 3mm, clip]{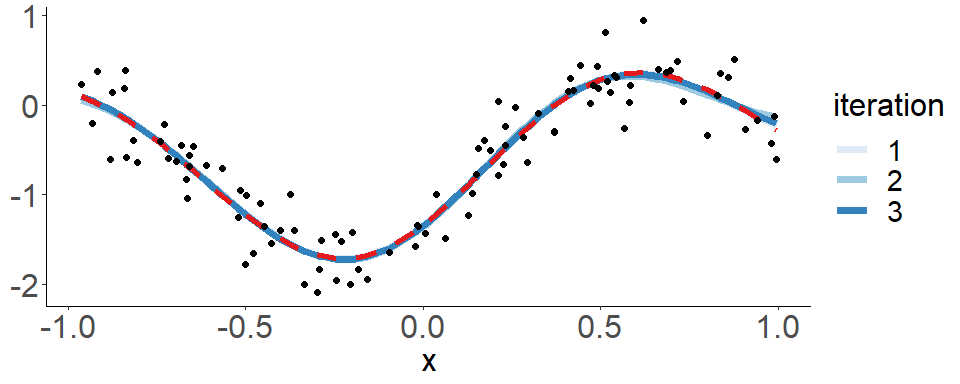}}\\
 & 2 & 0.38 & 1.20 & 6 & 0.47 & TRUE & 0.28 & 0.86 & 0.03 & 0.05 & \\
 & 3 & 0.24 & 1.50 & 11 & 0.53 & TRUE & 0.27 & 0.86 & 0.04 & 0.01 & \\
 \\
 \\[-1mm]
\arrayrulecolor{lightgray}\hline \\[-1mm]
\multirow{3}{*}{ 0.80} 
 & 1 & 1.00 & 3.20 & 6 & 0.52 & FALSE & 0.29 & 0.92 & 0.13 & 0.03 & \multirow{5}{*}{ \includegraphics[scale=0.20, trim = 0mm 10mm 0mm 3mm, clip]{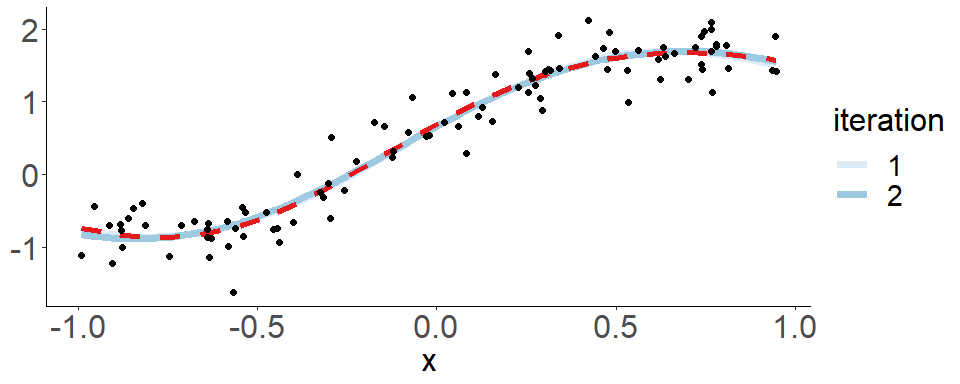}}\\
 & 2 & 0.52 & 1.65 & 6 & 0.98 & TRUE & 0.29 & 0.92 & 0.11 & 0.03 & \\
 & 3 & 0.50 & 3.12 & 11 & 0.77 & TRUE & 0.28 & 0.92 & 0.11 & 0.01 & \\
 \\
 \\[-1mm]
\arrayrulecolor{lightgray}\hline \\[-1mm]
\multirow{2}{*}{ 1.40} 
 & 1 & 1.00 & 3.20 & 6 & 1.02 & TRUE & 0.28 & 0.56 & 0.14 & 0.01 & \multirow{5}{*}{ \includegraphics[scale=0.20, trim = 0mm 10mm 0mm 3mm, clip]{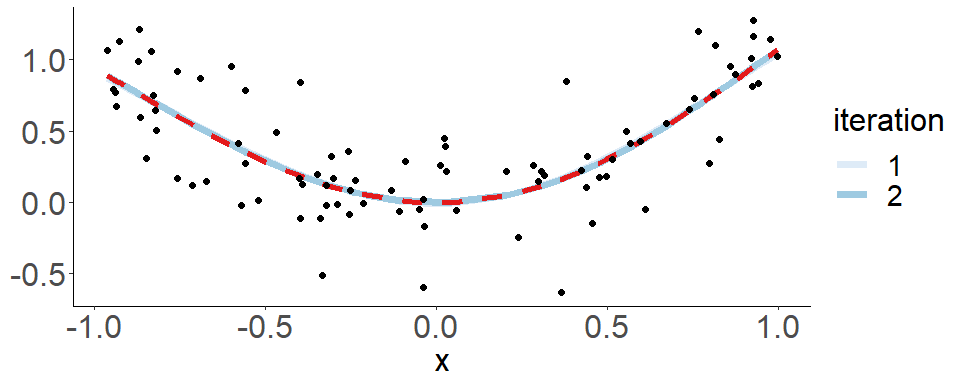}}\\
 & 2 & 0.52 & 3.27 & 11 & 1.23 & TRUE & 0.28 & 0.57 & 0.14 & 0.00 & \\
 \\
 \\
 \\[-1mm]
 \arrayrulecolor{lightgray}\hline \\[-1mm]
\multirow{2}{*}{ 2.94} 
 & 1 & 0.50 & 1.60 & 6 & 1.19 & TRUE & 0.34 & 0.45 & -0.10 & 0.06 & \multirow{5}{*}{ \includegraphics[scale=0.20, trim = 0mm 10mm 0mm 3mm, clip]{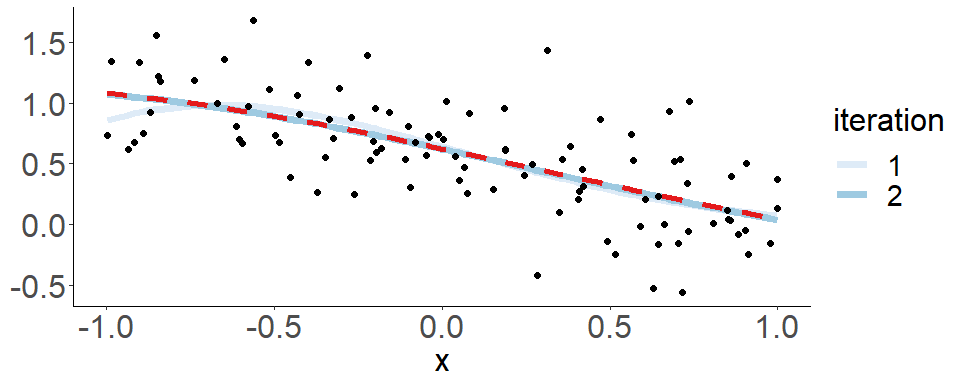}}\\
 & 2 & 0.61 & 3.81 & 11 & 2.58 & TRUE & 0.34 & 0.45 & -0.07 & 0.01 & \\
 \\
 \\
 \\[-1mm]
\arrayrulecolor{gray}\hline
\end{tabular}
\caption{Diagnostic on various data sets with varying smoothness. $\ell_{GP}$ refers to the exact GP length-scale estimate. $\ell$ refers to the minimum valid length-scale that can be accurately estimated determined by $m$ and $c$ as per Section \ref{sec_num_equations}, except for the first iteration which is the initial guess of the length-scale. $\hat{\ell}$ refers to the HSGP length-scale estimate. RMSE* refers to the root mean squared error computed with the exact GP model as the reference.}
  \label{tab1_diagnostic}
\end{table*}

\subsection{Other covariance functions}

Above, we thoroughly studied the relationship between the number of basis functions in the approximation and the approximation accuracy across different configurations. We specifically focused on the Matérn and squared exponential families of covariance functions, yet there exists other families of stationary covariance functions. The basis function approximation can easily be implemented for any stationary covariance function, where the spectral density function is available. The assess the accuracy of a basis function approximation for a kernel, where the diagnostic plots like Figure \ref{fig6_relationships} or equations (\ref{eq_m_l_QE})-(\ref{eq_c_vs_l_mat52}) are not available, we suggest to use the relative total variational distance between the true covariance function and the approximate covariance function as given in eq. \eqref{eq:rel_total_distance}. Ensuring that the relative distance is bounded by a small constant for the relevant lengthscale implies a high quality approximation. Another possibility to asses the accuracy of the approximation is to look at the cumulative sum of the spectral densities terms used in the series expansion and find out how much of the total variation they are actually explaining. 

To select $c$ for many other covariance functions, users can be guided by equations~\eqref{eq_c_vs_l_QE}, \eqref{eq_c_vs_l_mat52} and \eqref{eq_c_vs_l_mat32}, as pointed in Section~\ref{sec_num_equations}.

\subsection{The computational cost in the multi-dimensional setting}

The HSGP model is computationally superior in $1$D and $2$D even for highly wiggly functions, except when the number of data points is so small ($n \lesssim 300$, i.e., $n$ smaller than some value around 300) that exact GPs are already reasonably fast themselves. However, the computation time of the HSGP model increases rapidly with the number of input dimensions ($D$) since the number of multivariate basis functions $m^*=m_1\times \cdots \times m_D$ in the approximation increases exponentially with $D$ (see eq.~(\ref{eq_m_multi})). Yet, the HSGP method can still be computationally faster than the exact GP for larger datasets due the latter's cubic scaling in $n$. 

In our experiments of multivariate problems (see Section~\ref{sec_computation_nD}), the computation time for the HSGP model was faster than for the exact GP for most of the non-linear $2D$ functions or moderate-to-large sized $3$D datasets ($n \gtrsim 1000$, i.e., $n$ greater than some value around 1000), even for highly wiggly 3D functions (e.g., $\ell_1, \ell_2, \ell_3 \approx 0.1$).

For small sized datasets ($n\lesssim 1000$), HSGPs are likely to be slower than exact GPs already for highly to moderated wiggly $3$D functions (e.g., $\ell_1 \lesssim 0.1$, and $\ell_2, \ell_3 \gtrsim 0.3$) and for overall smooth $4$D functions (e.g., $\ell_1 \lesssim 0.1$, and $\ell_2, \ell_3, \ell_4 \gtrsim 0.4$).

As it has been shown in case study III (Section~\ref{sec_caseIII}), the proposed diagnostic tool can be very useful for multivariate problems as it allows one to reduce $m^*$ to the minimum sufficient value, reducing computational time drastically, and still getting an accurate approximation. For example, assuming a squared exponential covariance function, choosing the optimal value for $c$ allows one to use few basis functions in every single dimension (i.e., $m \lesssim 10$ for $\ell \gtrsim 0.3$; $20 \gtrsim m \gtrsim 10$ for $0.3 \gtrsim \ell \gtrsim 0.1$; and $m \gtrsim 20$ for $\ell \lesssim 0.1$), which, from results presented in Figure~\ref{fig15_time_nD_log}, implies that the HSGP model can be, in general terms, useful for highly wiggly $3$D functions and smooth $4$D functions.

Whether HSGP or exact GP is faster will also depend on the specific implementation details, which can have big effects on the scaling constants. Thus, more detailed recommendations would depend on the specific software implementation.

In our experiments, we observed roughly similar or even worse scaling behavior for the thin plate splines in comparison to HSGPs, with serious challenges with computation time when doing inference with spline models for $D=3$ and more than 10 basis functions per dimension, or even for $D=2$ and more than 40 basis functions per dimension (see Figures~B.3 right and C.3 in the online supplement).

\section{Case studies}\label{sec_cases}

In this section, we will present several simulated and real case studies in which we apply the developed HSGP models and the recommended steps to fit them. More case studies are presented in the online supplemental materials.

\subsection{Simulated data for a univariate function}\label{sec_univariate_simu}

In this experiment, we analyze a synthetic dataset with $n = 250$ observations, where the true data generating process is a Gaussian process with additive noise. The data points are simulated from the model $y_i = f(x_i) + \epsilon_i$, where $f$ is a sample from a Gaussian process $f(x) \sim \GP(0, k(x, x', \theta))$ with a Mat{\'e}rn ($\nu$=3/2) covariance function $k$ with marginal variance $\alpha=1$ and length-scale $\ell=0.2$ at inputs values $\bm{x}=(x_1,x_2,\dots,x_n)$ with $x_i \in [-1,1]$. $\epsilon_i$ is additive Gaussian noise with standard deviation $\sigma=0.2$. 

In the HSGP model, the latent function values $f(x)$ are approximated as in eq.~(\ref{eq_approxf}), with the Mat{\'e}rn ($\nu$=3/2) spectral density $S$ as in eq.~(\ref{eq_specdens_32}), and eigenvalues $\lambda_j$ and eigenfunctions $\phi_j$ as in equations (\ref{eq_eigenvalue}) and (\ref{eq_eigenfunction}), respectively.  

The joint posterior parameter distributions are estimated by sampling using the dynamic HMC algorithm implemented in Stan \citep{StanTeam:2021}. $\mathrm{Normal}(0,1)$, \linebreak $\mathrm{Normal}(0,3)$ and $\mathrm{Gamma}(1.2,0.2)$ prior distributions has been used for the observation noise $\sigma$, covariance function marginal variance $\alpha$, and length-scale $\ell$, respectively. We use the same prior distributions to fit the exact GP model.

The HSGP model is fitted following the recommended iterative steps as in Section~\ref{sec_user_guide}. A initial value for the minimum lengthscale $\ell$ to use at the first iteration is guessed to be 0.5. While diagnostic $\hat{\ell} + 0.01 \geq \ell$ is \textit{false}, $c$ and $m$ are updated by equations~\eqref{eq_c_vs_l_mat32} and \eqref{eq_m_l_mat32}, respectively, and the minimum $\ell$ is updated with the estimated $\hat{\ell}$. After the diagnostic generated the first \textit{true}, $m$ is updated by increasing the $m$ of the previous iteration by 5 additional basis functions, $c$ is updated by eq.~\eqref{eq_c_vs_l_mat32} as a function of the estimated $\hat{\ell}$ at previous iteration, and the minimum $\ell$ is set by eq.~\eqref{eq_m_l_mat32} as function of $c$ and $m$. Table~\ref{tab_caseI} contains the values for the parameters $\ell$, $c$, $m$, the estimated $\hat{\ell}$, the diagnostic $\hat{\ell} + 0.01 \geq \ell$ and the RMSE compute with both the data and the GP as the reference for every iterative steps of the fitting process.

\begin{table}
\centering
\setlength{\tabcolsep}{4pt}
\begin{tabular}{cccccccc}
\arrayrulecolor{gray}\hline \\[-3mm]
iter. & $\ell$ & $c$ & $m$ & $\hat{\ell}$ & {\scriptsize $\hat{\ell} + 0.01 \geq \ell$} & {\scriptsize RMSE} & {\scriptsize RMSE*} \\ 
\arrayrulecolor{lightgray}\hline \\[-1mm]
 1 & 0.50 & 2.25 & 16 & 0.12 & FALSE & 0.26 & 0.17 \\
 2 & 0.12 & 1.20 & 35 & 0.32 & TRUE & 0.18 & 0.01 \\
 3 & 0.02 & 1.20 & 40 & 0.28 & TRUE & 0.18 & 0.01 \\[1mm]
\arrayrulecolor{gray}\hline
\end{tabular}
\caption{Diagnostic on various data sets with varying smoothness. $\ell$ refers to the minimum valid length-scale that can be accurately estimated determined by $m$ and $c$, except for the first iteration which is the initial guess of the length-scale. RMSE* refers to the root mean squared error computed with the exact GP model as the reference.}
  \label{tab_caseI}
\end{table}

\begin{figure}
\centering
\includegraphics[width=0.5\textwidth]{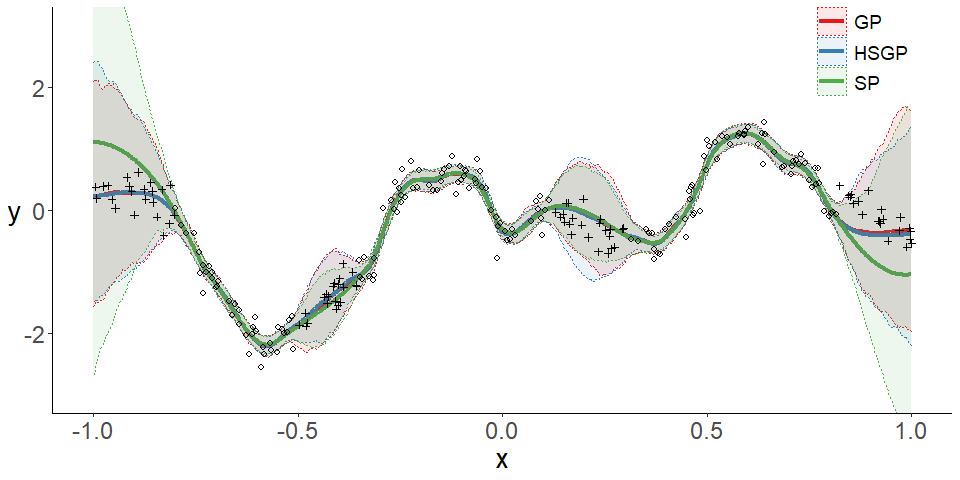}
\caption{Posterior predictive means of the proposed HSGP model, the exact GP model, and the spline model (SP). 95\% credible intervals are plotted as dashed lines.}
  \label{fig9_Posteriors_exI}
\end{figure}

Figure~\ref{fig9_Posteriors_exI} shows the posteriors predictive distributions of the exact GP and the HSGP models, the later using the parameter values as the third iterative step, $c=1.2$ ($L=c\cdot 1= 1.2$; see eq.~\eqref{eq_boundary}) and $m=40$ basis functions. In order to do model comparison a spline-based model with 60 knots is also fitted and plotted jointly with the GP and HSGP models. The spline-based model is fitted using the thin plate regression spline approach by \citet{wood2003thin} and implemented in the R-package \textit{mgcv} \citep{wood2011mgcv}. A Bayesian approach is used to fit this spline model using the R-package \textit{brms} \citep{burkner2017brms}. The sample observations are plotted as circles and the out-of-sample or test data are plotted as crosses. The posteriors of the three models, exact GP, HSGP and spline, are similar in the interpolation regions of the input space. However, when extrapolating the spline model solution clearly differs from the exact GP and HSGP models as well as the actual observations. 

\begin{figure}[!h]
\centering
\subfigure{\includegraphics[width=0.5\textwidth, trim = 0mm 0mm 0mm 0mm, clip]{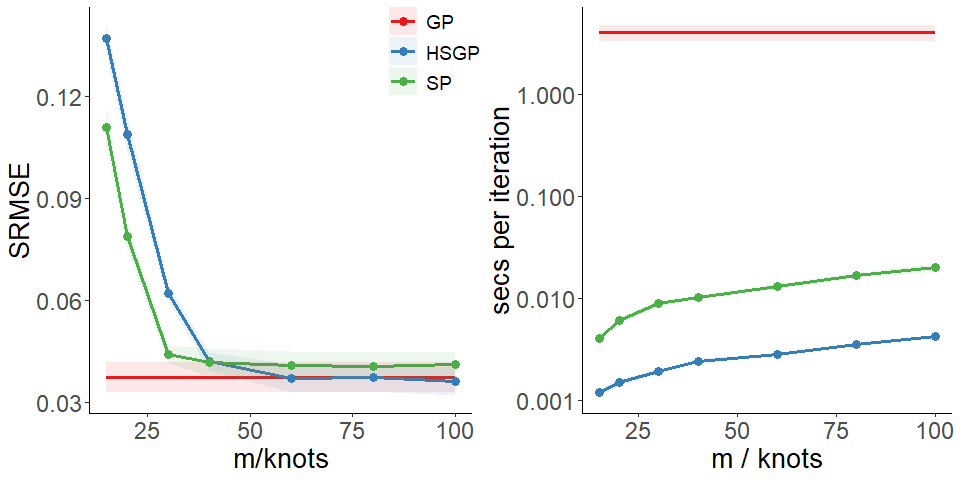}}
\caption{Computational time (y-axis), in seconds per iteration (iteration of the HMC sampling method), as a function of the number of basis functions $m$, for the HSGP model, and knots, for the spline model. The y-axis is on a logarithmic scale. The standard deviation of the computational time is plotted as dashed lines.}
  \label{fig10_time_exI}
\end{figure}

Figure~\ref{fig10_time_exI} shows both the standardized root mean squared error (SRMSE) of the models for the sample data and the computational times in seconds per iteration (iteration of the HMC sampling method), as a function of the number of basis functions $m$, for the HSGP model, and knots, for the spline model. Both the HSGP and splines models show roughly similar performance at sample data. The HSGP method yields a good approximation of the exact GP model for both interpolation and extrapolation. However, the spline model does not extrapolate data properly (Figure~\ref{fig9_Posteriors_exI}). The HSGP model is on average roughly 400 times faster than the exact GP and 10 times faster than the spline model for this particular model and data. Also, it is seen that the computation time increases slowly as a function of the number of basis functions.

The Stan model code for the exact GP, the HSGP and the spline models of this case study and the R code where it is implemented can be found online at {\small \url{ https://github.com/gabriuma/basis_functions_approach_to_GP/tree/master/Paper/Case-study_1D-Simulated-data}}\,.

\subsection{Birthday data}\label{sec_birthday}
This example is an analysis of patterns in birthday frequencies in a dataset containing records of all births in the United States on each day during the period 1969–1988. The model decomposes the number of births along all the period in longer-term trend effects, patterns during the year, day-of-week effects, and special days effects. The special days effects cover patterns such as possible fewer births on Halloween, Christmas or new year, and excess of births on Valentine’s Day or the days after Christmas (due, presumably, to choices involved in scheduled deliveries, along with decisions of whether to induce a birth for health reasons). \citet{gelman2013bayesian} presented an analysis using exact GP and maximum a posteriori inference. As the total number of days within the period is $T=7305$ ($t=1,2,\dots,T$), a full Bayesian inference with MCMC for a exact GP model is memory and time consuming. We will use the HSGP method as well as the low-rank GP model with a periodic covariance function described in Appendix~\ref{sec_periodic} which is based on expanding the periodic covariance function into a series of stochastic resonators \citep{solin2014explicit}.

Let $y_t$ denote the number of births on the $t$'th day. The observation model is a normal distribution with mean function $\mu(t)$ and noise variance $\sigma^2$,
\begin{equation*}
y_{t} \sim \Normal(\mu(t),\sigma^2).
\end{equation*}
The mean function $\mu(t)$ will be defined as an additive model in the form
\begin{equation} \label{eq_mean_brithday}
\mu(t) = f_1(t) + f_2(t) + f_3(t) + f_4(t).
\end{equation}

The component $f_1(t)$ represents the long-term trends modeled by a GP with squared exponential covariance function,
\begin{equation*}
f_1(t) \sim \GP(0,k_1), \hspace{5mm} k_1(t,t') = \alpha_1 \exp\!\!\left(\!-\frac{1}{2} \frac{(t-t')^2}{\ell_1^2}\right), 
\end{equation*}
which means the function values $\bm{f}_1=\{f_1(t)\}_{t=1}^T$ are multivariate Gaussian distributed with covariance matrix $\bm{K}_1$, where $K_{1_{t,s}}=k_1(t,s)$, with $t,s=1,\dots,T$. $\alpha_1$ and $\ell_1$ represent the marginal variance and length-scale, respectively, of this GP prior component. 
The component $f_2(t)$ represents the yearly smooth seasonal pattern, using a periodic squared exponential covariance function (with period 365.25 to match the average length of the year) in a GP model,
\begin{align*}
&f_2(t) \sim \GP(0,k_2), \\
&k_2(t,t') = \alpha_2 \exp\!\!\left(\!-\frac{2\,\sin^{\!2}\!(\pi(t-t')/365.25}{\ell_2^2}\right).
\end{align*}

The component $f_3(t)$ represents the weekly smooth pattern using a periodic squared exponential covariance function (with period 7 of length of the week) in a GP model,
\begin{align*}
&f_3(t) \sim \GP(0,k_3), \\
&k_3(t,t') = \alpha_3 \exp\!\!\left(\!-\frac{2\,\sin^{\!2}\!(\pi(t-t')/7}{\ell_3^2}\right). 
\end{align*}

The component $f_4(t)$ represents the special days effects, modeled as a Student's $t$ prior model with 1 degree of freedom and variance $\tau^2$:
\begin{equation*}
f_4(t) \sim t(1,0,\tau^2).
\end{equation*}
The component $f_1$ will be approximated using the HSGP model and the function values $f_1$ are approximated as in eq.~(\ref{eq_approxf}), with the squared exponential spectral density $S$ as in eq.~(\ref{eq_specdens_inf}), and eigenvalues $\lambda_j$ and eigenfunctions $\phi_j$ as in equations (\ref{eq_eigenvalue}) and (\ref{eq_eigenfunction}). The year effects $f_2$ and week effects $f_3$ use a periodic covariance function  and thus do no fit under the main framework of the HSGP approximation covered in this paper. However, they do have a representation based on expanding periodic covariance functions into a series of stochastic resonators (Appendix~\ref{sec_periodic}). Thus, the functions $f_2$ and $f_3$ are approximated as in eq.~(\ref{eq_f_period}), with variance coefficients $\tilde{q}_j^2$ as in eq.~(\ref{eq_q_2}).

The joint posterior parameter distributions are estimated by sampling using the dynamic HMC algorithm implemented in Stan \citep{StanTeam:2021}. $\mathrm{Normal}(0,1)$, \linebreak $\mathrm{Normal}(0,10)$ and $\mathrm{Normal}(0,2)$ prior distributions has been used for the observation noise $\sigma$, covariance function marginal variances $\bm{\alpha}=\{\alpha_1, \alpha_2, \alpha_3\}$, and length-scales $\bm{\ell}=\{\ell_1, \ell_2, \ell_3\}$, respectively. A $\mathrm{Normal}(0,0.1)$ prior distribution has been used for the standard deviation $\tau$ of the Student's $t$ distribution with 1 degree of freedom used to model $f_4$ (i.e., the special days effects).

The HSGP model is fitted following the recommended iterative steps as in Section~\ref{sec_user_guide}, where in each iteration the diagnosis is applied on $f_1$, $f_2$ and $f_3$, where for each one these functions the parameters $c$, $m$, minimum $\ell$, estimated $\hat{\ell}$ and diagnostic $\hat{\ell} + 0.01 \geq \ell$ are updated. For functions $f_2$ and $f_3$ there are not boundary factor $c$ as they use periodic covariance functions, and $m$ and minimum $\ell$ are updated by eq.~\eqref{eq_J_l_periodic}. A initial value for the minimum lengthscale $\ell_1$ of $f_1$ is guessed to correspond to around 3 years (i.e., $\ell_1=3$ years $= 3\cdot 365=1095$ days) of the input dimension. Initial values for the minimum lengthscales $\ell_2$ and $\ell_3$ of $f_2$ and $f_3$, respectively, are guessed to correspond to half of the period (i.e., $l_2=l_3=0.5$). After the diagnostic generated the first \textit{true}, $m$ is updated by increasing the $m$ of the previous iteration by 5 additional basis functions and $c$ and the minimum $\ell$ updated accordingly as explained in Section~\ref{sec_user_guide}. The full diagnosis process is applied until two \textit{trues} are achieved for each function. Table~\ref{tab_caseII} contains the values for the parameters $\ell$, $c$, $m$, the estimated $\hat{\ell}$, the diagnostic $\hat{\ell} + 0.01 \geq \ell$ for each function, and the RMSE compute with the data as the reference for every iterative steps of the fitting process.

\begin{figure}[h]
\centering
\subfigure{\includegraphics[width=0.5\textwidth, trim = 0mm 0mm 0mm 0mm, clip]{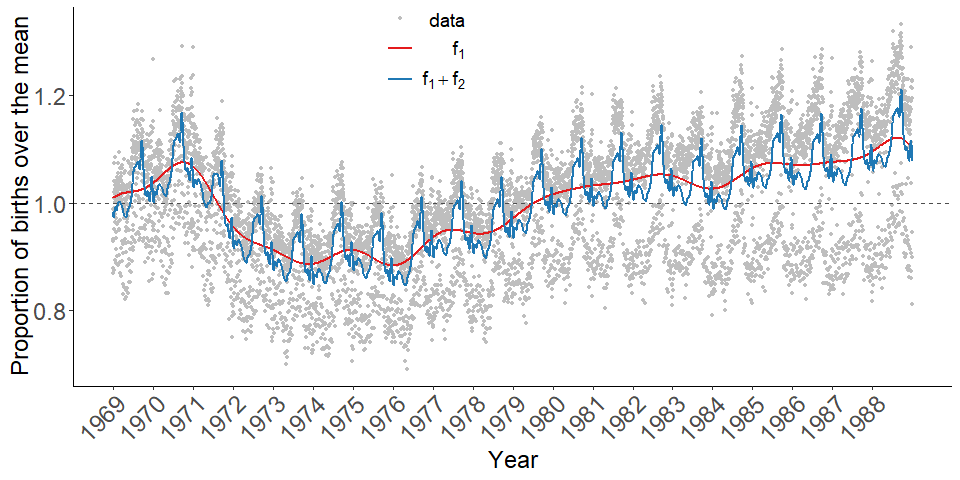}}
\caption{Posterior means of the long-term trend $f_1$ and year effects pattern ($f_2$) for the whole series. }
  \label{fig11_posteriors_birthday}
\end{figure}

\begin{figure}[h]
\centering
\subfigure{\includegraphics[width=0.5\textwidth, trim = 0mm 0mm 0mm 0mm, clip]{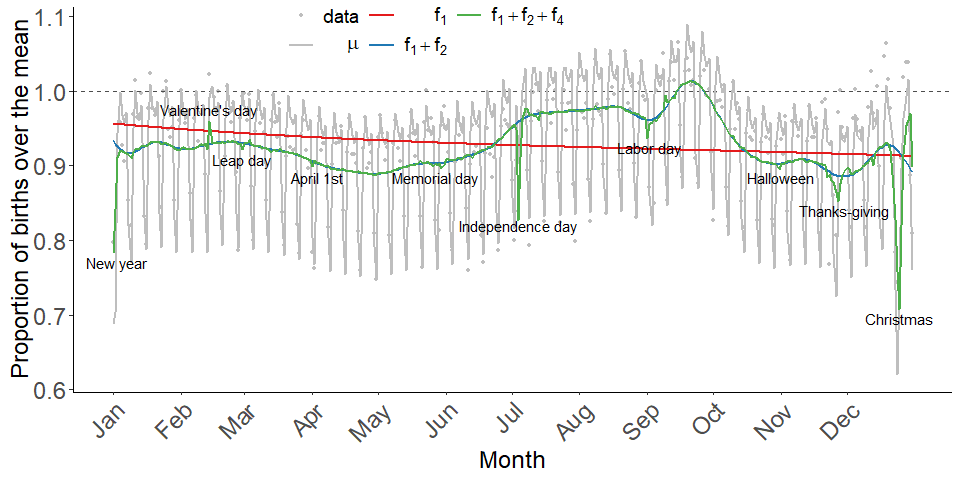}}
\caption{Posterior means of the function $\mu$ for the year 1972 of the series. The special days effects pattern $f_4$ in the year is also represented, as well as the long-term trend $f_1$ and year effects pattern $f_2$. }
  \label{fig12_posteriors_oneyear_birthday}
\end{figure}

\begin{figure}[h]
\centering
\subfigure{\includegraphics[width=0.5\textwidth, trim = 0mm 0mm 0mm 0mm, clip]{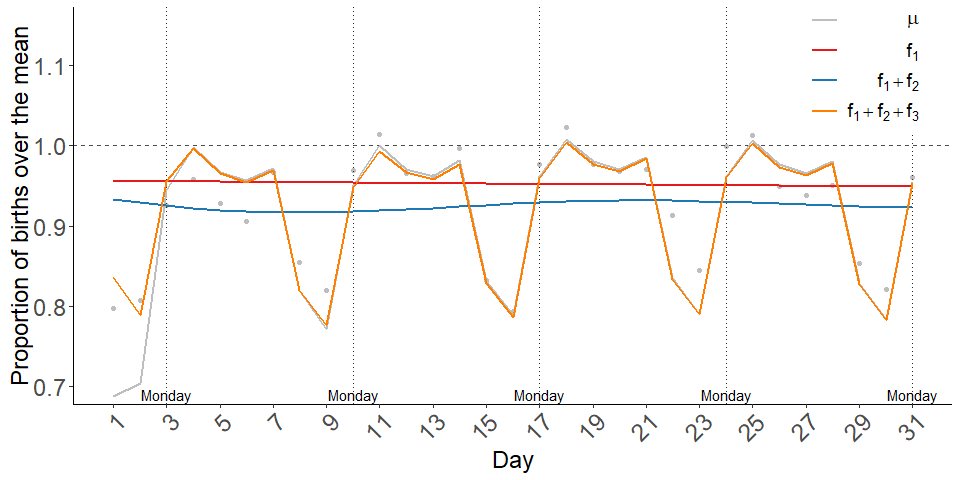}}
\caption{Posterior means of the function $\mu$ for the month of January of 1972. The week effects pattern $f_3$ in the month is also represented, as well as the long-term trend $f_1$, year effects pattern $f_2$ and special days effects pattern $f_4$. }
  \label{fig13_posteriors_onemonth_birthday}
\end{figure}

Figure~\ref{fig11_posteriors_birthday} shows the posterior means of the long-term trend $f_1$ and yearly pattern $f_2$ for the whole period, jointly with the observed data. Figure \ref{fig12_posteriors_oneyear_birthday} shows the model for one year (1972) only. In this figure, the special days effects $f_4$ in the year can be clearly represented. The posterior means of the the function $\mu$ and the components $f_1$ (long-term trend) and $f_2$ (year pattern) are also plotted in this Figure \ref{fig12_posteriors_oneyear_birthday}. Figure \ref{fig13_posteriors_onemonth_birthday} show the process in the month of January of 1972 only, where the week pattern $f_3$ can be clearly represented. The mean of the the function $\mu$ and components $f_1$ (long-term trend), $f_2$ (year pattern) and $f_4$ (special-days effects) are also plotted in this Figure \ref{fig13_posteriors_onemonth_birthday}. 

R code and the Stan model code for the HSGP model of this case study can be found online at {\small \url{https://github.com/gabriuma/basis_functions_approach_to_GP/tree/master/Paper/Case-study_Birthday-data}}\,.

\begin{table}[t]
\centering
\setlength{\tabcolsep}{4pt}
\begin{tabular}{ccccccccc}
\arrayrulecolor{gray}\hline \\[-3mm]
iter. & f & $\ell$ & $c$ & $m$ & $\hat{\ell}$ & {\scriptsize $\hat{\ell} + 0.01 \geq \ell$} & {\scriptsize RMSE} & {\scriptsize ELPD} \\ 
\arrayrulecolor{lightgray}\hline \\[-2mm]
   & $f_1$ & 0.52 & 1.20 & 7  & 0.53 & FALSE &      &  	   \\
 1 & $f_2$ & 0.50 & -    & 8  & 0.34 & FALSE & 0.32 & 0.05 \\
   & $f_3$ & 0.50 & -    & 8  & 1.62 & TRUE  &      &  	   \\
\arrayrulecolor{lightgray}\hline \\[-2mm]
   & $f_1$ & 0.50 & 1.20 & 8  & 0.54 & TRUE  &      &  	   \\
 2 & $f_2$ & 0.34 & -    & 11 & 0.29 & FALSE & 0.32 & 0.06 \\
   & $f_3$ & 0.29 & -    & 13 & 1.53 & TRUE  &      &  	   \\
\arrayrulecolor{lightgray}\hline \\[-2mm]
   & $f_1$ & 0.28 & 1.20 & 13 & 0.33 & TRUE  &      &  	   \\
 3 & $f_2$ & 0.29 & -    & 13 & 0.24 & FALSE & 0.30 & 0.12 \\
   & $f_3$ & 0.29 & -    & 13 & 1.64 & TRUE  &      &  	   \\
\arrayrulecolor{lightgray}\hline \\[-2mm]
   & $f_1$ & 0.28 & 1.20 & 18 & 0.23 & TRUE  &      &  	   \\
 4 & $f_2$ & 0.24 & -    & 16 & 0.23 & TRUE  & 0.30 & 0.15 \\
   & $f_3$ & 0.29 & -    & 13 & 1.68 & TRUE  &      &  	   \\
\arrayrulecolor{lightgray}\hline \\[-2mm]
   & $f_1$ & 0.16 & 1.20 & 23 & 0.20 & TRUE  &      &  	   \\
 5 & $f_2$ & 0.24 & -    & 16 & 0.23 & TRUE  & 0.29 & 0.18 \\
   & $f_3$ & 0.29 & -    & 13 & 1.63 & TRUE  &      &  	   \\[1mm]
\arrayrulecolor{gray}\hline
\end{tabular}
\caption{Iterative steps in HSGP model fitting of \textit{Birthday data} case study. The diagnosis procedure updating the parameters minimum $\ell$, $c$, $m$, estimated $\hat{\ell}$ and diagnostic $\hat{\ell} + 0.01 \geq \ell$ is applied on the three underlying functions $f_1$, $f_2$ and $f_3$. Notice that $f_2$ and $f_3$ do not have boundary factor $c$ as they use periodic covariance functions. RMSE and ELPD are evaluated for the model underlying function $\mu=f_1 + f_2 + f_3 + f_4$ and sample data. The full diagnosis process is applied until two \textit{trues} are achieved for each function.}
  \label{tab_caseII}
\end{table}

\subsection{Case study: Simulated data for $2$D and $3$D functions}\label{sec_caseIII}

In this case study, we apply the diagnostic tool to fit and diagnose two different data sets, one data set simulated from a bivariate (D=$2$) function and another data set simulated from a $3$-dimensinal (D=$3$) function. Furthermore, in Section~\ref{sec_computation_nD} we present results of the computational time required to fit the HSGP model in $2$D, $3$D and $4$D input spaces, with different sized data sets and as a function of the number of multivariate basis functions $m^*=m_1\times \cdots \times m_D$ used in the approximation.

$2$D and $3$D synthetic functions were drawn from $2$D and $3$D GP priors, with input values $\bm{x}_i \in [-1,1]^2$ and $\bm{x}_i \in [-1,1]^3$, respectively. Squared exponential covariance functions with marginal variance $\alpha=1$ and length-scales $\ell_1=0.10$, $\ell_2=0.3$, and $\ell_3=0.4$, where $\ell_i$ is the length scale for the $i$'th dimension, were used in the GP priors. $200$ and $1000$ data points were sampled from the $2$D and $3$D drawn functions, respectively, and independent Gaussian noise with standard deviation $\sigma=0.2$ was added to the data points to form the final noisy sets of observations.

In the HSGP models with $2$ and $3$ input dimensions, the underlying functions $f(\bm{x})$ are approximated as in eq. (13)
, with the $D$-dimensional squared exponential spectral densities $S$ as in eq. \eqref{eq_specdens_inf}, and the $D$-vectors of eigenvalues $\bm{\lambda}_j$ and the multivariate eigenfunctions $\phi_j$ as in equations \eqref{eq_eigenfunction_multi} and \eqref{eq_eigenvalue_multi}, respectively.

The joint posterior parameter distributions are estimated via the dynamic HMC sampling algorithm implemented in Stan \citep{StanTeam:2021}. $\mathrm{Normal}(0,1)$, $\mathrm{Normal}(0,3)$, and $\mathrm{InverseGamma}(2,0.5)$ priors were used for the observation noise $\sigma$, marginal variance $\alpha$, and length-scales $\ell$, respectively. We used the same priors to fit the exact GP model serving as benchmark.

The HSGP models are fitted following the recommended iterative procedure detailed in Section~\ref{sec_user_guide}, where the diagnostic is applied on every dimension, separately. For each dimension, the parameters $c$, $m$, minimum $\ell$, estimated $\hat{\ell}$, and the diagnostic $\hat{\ell} + 0.01 \geq \ell$ are updated using the equations in Section~\ref{sec_num_equations}. The values $0.5$, $1$, and $1$ are set as initial values for the minimum lengthscales $\ell_1$, $\ell_2$ and $\ell_3$, respectively. In order to be as efficient as possible, after the diagnostic generated the first \textit{true} for a certain dimension, its corresponding $m$ is updated by increasing the $m$ of the previous iteration by only 2 additional basis functions and, after the second \textit{true}, $m$ is no longer increased. The full diagnostic process is applied until two \textit{trues} are achieved for each dimension. 

Table~\ref{tab_caseIII_2D} and \ref{tab_caseIII_3D} contain the iterative steps to fit and diagnose the $2$D and $3$D data sets, respectively. 
The minimum requirements to fit the models were easily achieved by performing 4 iterations: from a $1^{\text{st}}$ iteration that uses few multivariate basis functions $m^*$ ($m^*=m_1(=6) \times m_2(=6)=36$ and $m^*=m_1(=6) \times m_2(=6) \times m_3(=6)=216$ for the $2$D and $3$D data sets, respectively) to the $4^{\text{th}}$ iteration that uses the minimum multivariate basis functions ($m^*=m_1(=22) \times m_2(=11)=242$ and $m^*=m_1(=21) \times m_2(=10) \times m_3(=12)=2520$ for the $2$D and $3$D data sets, respectively) required to accurately approximate these notably wiggly functions. Without using our recommended guidelines and diagnostics, it would likely have taken the user several more trials to get the optimal solution and to spend a significant amount of time, as using a larger number of basis functions in multivariate cases increases time of computation drastically.

\begin{figure*}[h]
\centering
\subfigure{\includegraphics[scale=0.315, trim = 1mm 18mm 39.2mm 20mm, clip]{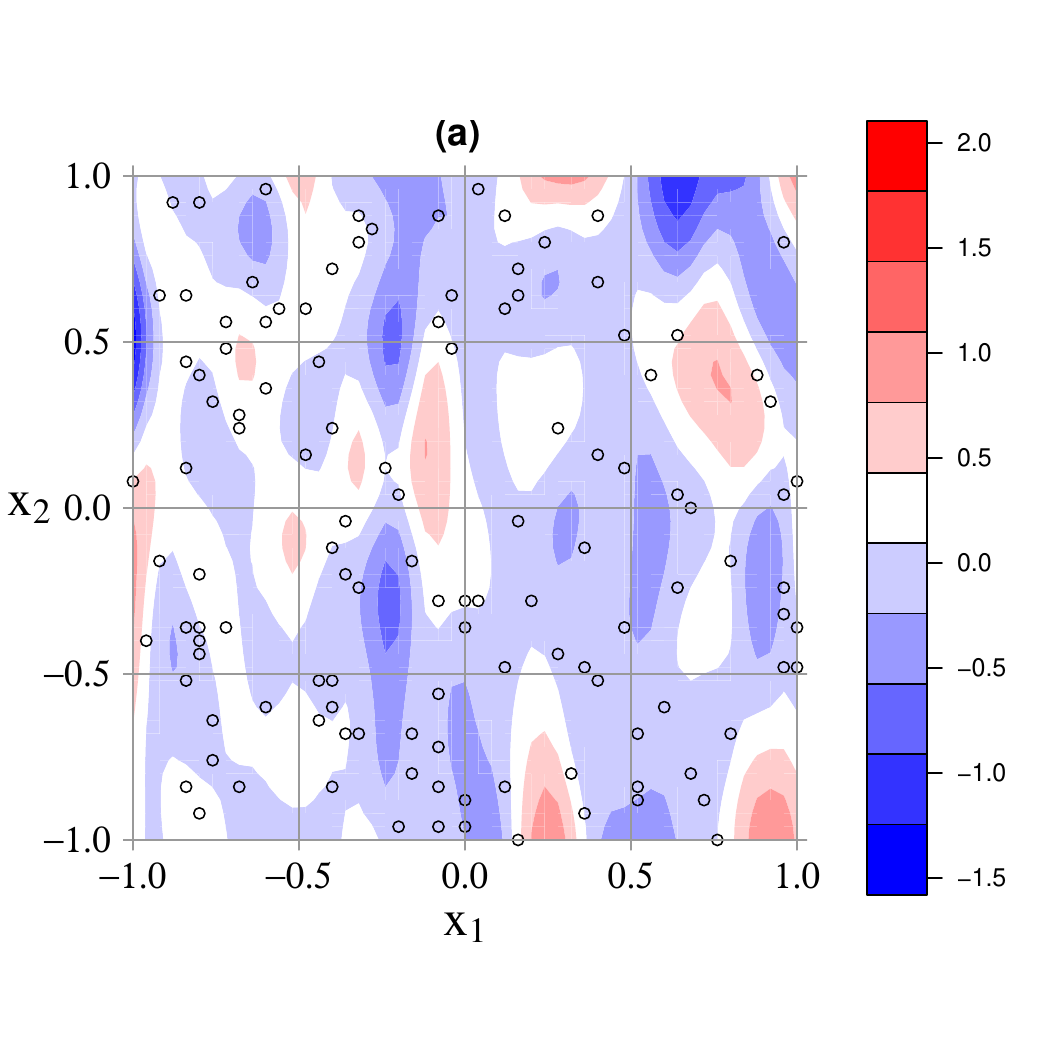}}
\subfigure{\includegraphics[scale=0.315, trim = 19mm 18mm 39.2mm 20mm, clip]{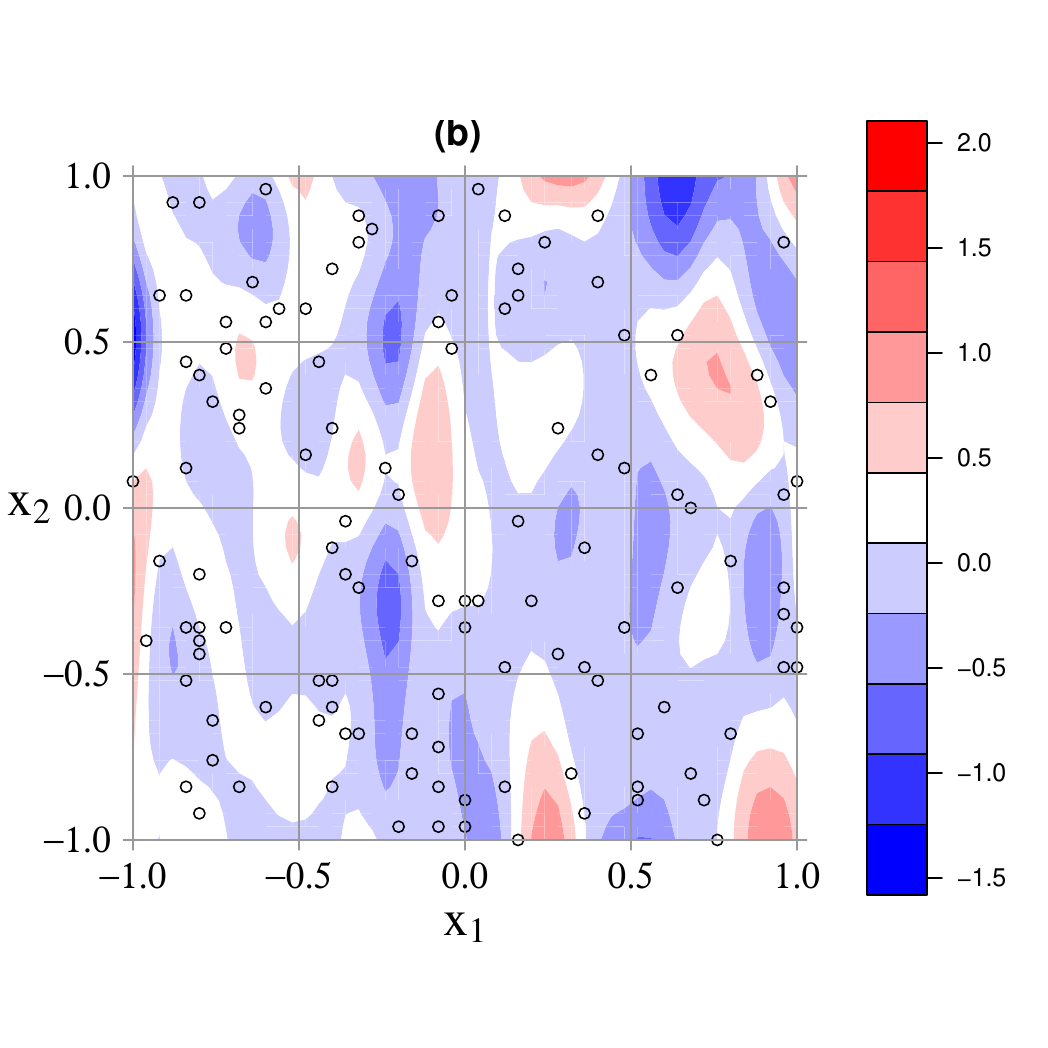}}
\subfigure{\includegraphics[scale=0.315, trim = 19mm 18mm 39.2mm 20mm, clip]{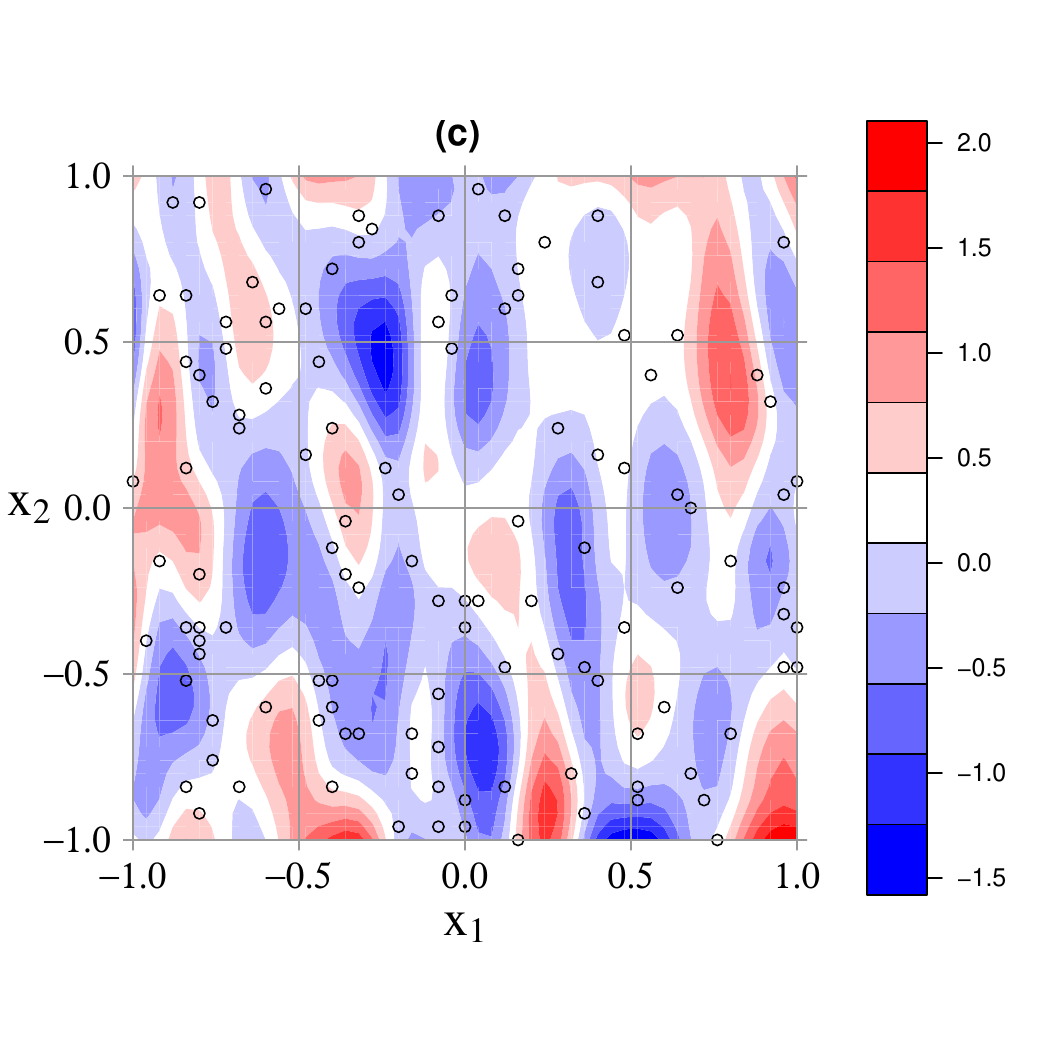}}
\subfigure{\includegraphics[scale=0.30, trim = 151mm 14mm 2mm 20mm, clip]{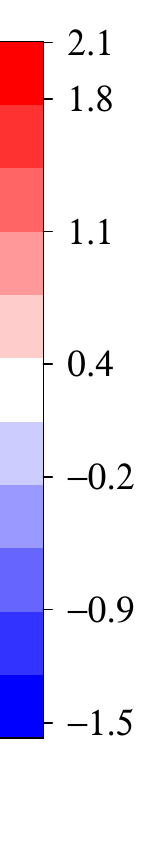}}
\hspace{1pt}\subfigure{\includegraphics[scale=0.27, trim =0mm 5mm 10mm 0mm, clip]{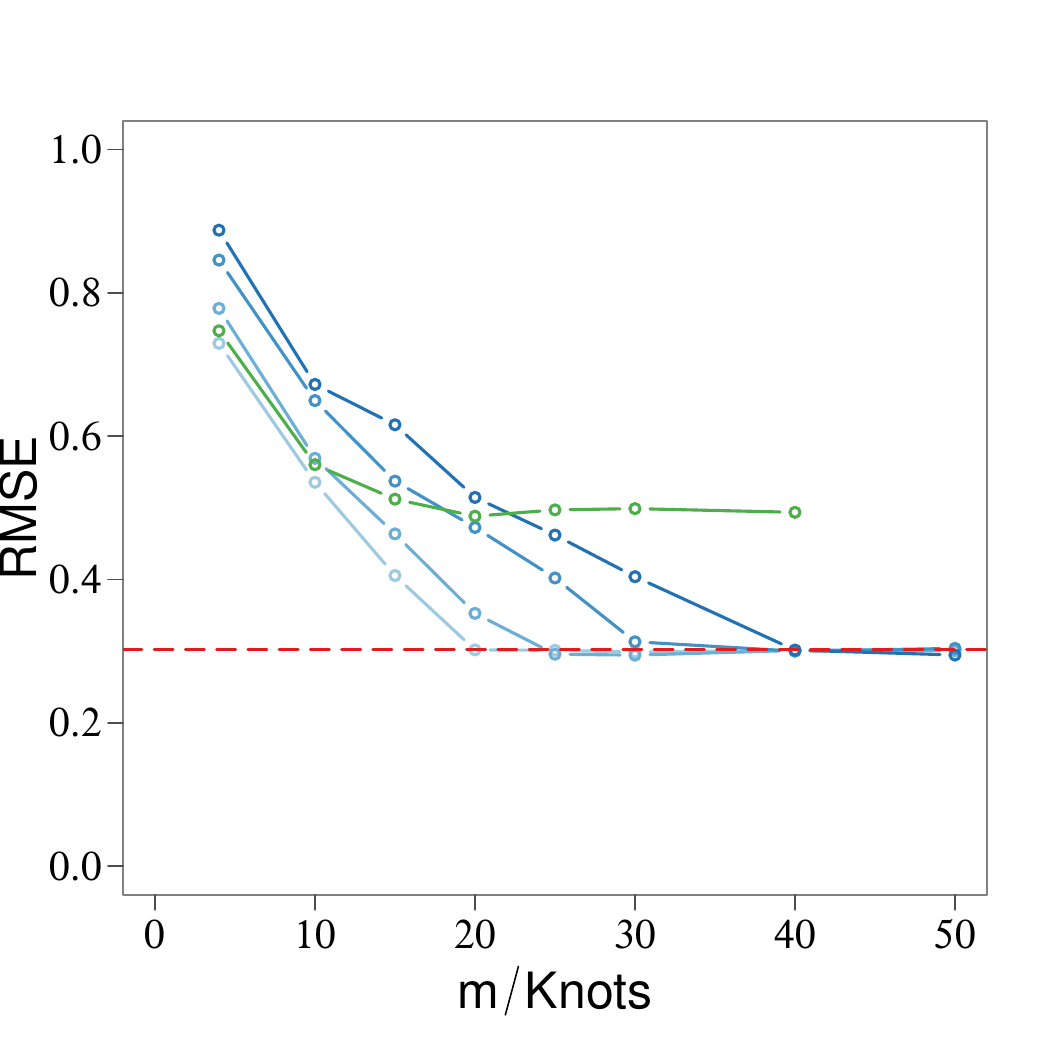}}
\hspace{-1.4cm} \subfigure{\raisebox{1.65cm}{\includegraphics[scale=0.30, trim = 28mm 70mm 112mm 20mm, clip]{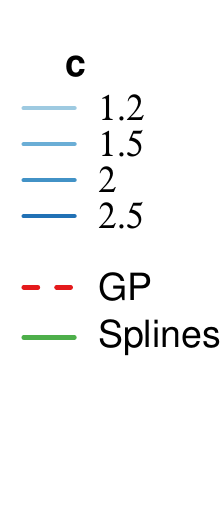}}}
\caption{Error between the $2$D data-generating function and posterior mean functions of the GP (a), HSGP (b) and spline-based (c) models. Sample points are plotted as circles. The right side plot shows the root mean square error (RMSE) of the different methods, and plotted as a function of the boundary factor $c$, number of univariate basis functions $m$ and knots (the same number of basis functions $m$ and knots per dimension are used, resulting in $m^D$ total number of multivariate basis functions).}
  \label{fig14_errors_exII}
\end{figure*}

\begin{table}[h]
\centering
\setlength{\tabcolsep}{3.5pt}
\begin{tabular}{cccccccccc}
\arrayrulecolor{gray}\hline \\[-3mm]
 {\small iter.} & {\small Dim.} & {\small $\ell$} & {\small $c$} & {\small $m$} & {\small $\hat{\ell}$} & {\scriptsize $\hat{\ell} + 0.01 \geq \ell$} & {\scriptsize RMSE} & {\scriptsize RMSE*} \\ 
\arrayrulecolor{lightgray}\hline \\[-2mm]
 \multirow{2}{*}{ 1 } & $1$ & 0.50 & 1.6 & 6 & 0.19 & FALSE & \multirow{2}{*}{ 0.75 } & \multirow{2}{*}{ 0.72 }\\ 
 & $2$ & 1.00 & 3.2 & 6 & 0.25 & FALSE &  &  \\
\arrayrulecolor{lightgray}\hline \\[-2mm]
 \multirow{2}{*}{ 2 } & $1$ & 0.19 & 1.2 & 12 & 0.11 & FALSE  & \multirow{2}{*}{ 0.18 } & \multirow{2}{*}{ 0.12 }\\
 & $2$ & 0.25 & 1.2 & 9 & 0.25 & TRUE &  &  \\
\arrayrulecolor{lightgray}\hline \\[-2mm]
 \multirow{2}{*}{ 3 } & $1$ & 0.11 & 1.2 & 20 & 0.11 & TRUE  & \multirow{2}{*}{ 0.11 } & \multirow{2}{*}{ 0.03 }\\
 & $2$ & 0.19 & 1.2 & 11 & 0.27 & TRUE &  &  \\
\arrayrulecolor{lightgray}\hline \\[-2mm]
 \multirow{2}{*}{ 4 } & $1$ & 0.10 & 1.2 & 22 & 0.11 & TRUE  & \multirow{2}{*}{ 0.10 } & \multirow{2}{*}{ 0.02 }\\
 & $2$ & 0.19 & 1.2 & 11 & 0.27 & TRUE  &  & \\[1mm] 
\arrayrulecolor{lightgray}\hline
 \multicolumn{8}{l}{\scriptsize Exact GP length-scales: $\ell_{1_{GP}}=0.10$, \, $\ell_{2_{GP}}=0.29$ }
\end{tabular}
\caption{Iterative steps in HSGP model fitting on the $2$D data. The diagnostic procedure updating the parameters minimum $\ell$, $c$, $m$, estimated $\hat{\ell}$, and diagnostic $\hat{\ell} + 0.01 \geq \ell$ is applied on the two input dimensions. The full diagnostic process is applied until two consecutive \textit{trues} in the diagnostic are achieved for each dimension. RMSE and RMSE* refer to the root mean squared error of the HSGP computed against the true function and the exact GP, respectively.}
  \label{tab_caseIII_2D}
\end{table}

\begin{table}[h]
\centering
\setlength{\tabcolsep}{3.5pt}
\begin{tabular}{cccccccccc}
\arrayrulecolor{gray}\hline \\[-3mm]
 {\scriptsize iter.} & {\scriptsize Dim.} & {\scriptsize $\ell$} & {\scriptsize $c$} & {\scriptsize $m$} & {\scriptsize $\hat{\ell}$} & {\scriptsize $\hat{\ell} + 0.01 \geq \ell$} & {\scriptsize RMSE} & {\scriptsize RMSE*} \\ 
\arrayrulecolor{lightgray}\hline \\[-2mm]
 \multirow{3}{*}{ 1 } & $1$ & 0.50 & 1.6 & 6 & 0.14 & FALSE & \multirow{3}{*}{ 0.63 } & \multirow{3}{*}{ 0.58 }\\ 
 & $2$ & 0.50 & 1.6 & 6 & 0.27 & FALSE &  &  \\
 & $3$ & 1.00 & 3.2 & 6 & 0.23 & FALSE &  &  \\
\arrayrulecolor{lightgray}\hline \\[-2mm]
 \multirow{3}{*}{ 2 } & $1$ & 0.13 & 1.2 & 16 & 0.11 & FALSE  & \multirow{3}{*}{ 0.13 } & \multirow{3}{*}{ 0.07 }\\
 & $2$ & 0.27 & 1.2 & 8 & 0.29 & TRUE &  &  \\
 & $3$ & 0.23 & 1.2 & 19 & 0.40 & TRUE &  &  \\
\arrayrulecolor{lightgray}\hline \\[-2mm]
 \multirow{3}{*}{ 3 } & $1$ & 0.11 & 1.2 & 19 & 0.12 & TRUE  & \multirow{3}{*}{ 0.11 } & \multirow{3}{*}{ 0.04 }\\
 & $2$ & 0.21 & 1.2 & 10 & 0.29 & TRUE &  &  \\
 & $3$ & 0.19 & 1.3 & 12 & 0.43 & TRUE &  &  \\
 \arrayrulecolor{lightgray}\hline \\[-2mm]
 \multirow{3}{*}{ 4 } & $1$ & 0.10 & 1.2 & 21 & 0.12 & TRUE  & \multirow{3}{*}{ 0.11 } & \multirow{3}{*}{ 0.03 }\\
 & $2$ & 0.21 & 1.2 & 10 & 0.29 & TRUE &  &  \\
 & $3$ & 0.19 & 1.3 & 12 & 0.43 & TRUE &  &  \\[1mm] 
\arrayrulecolor{lightgray}\hline
 \multicolumn{9}{l}{\scriptsize Exact GP length-scales: $\ell_{1_{GP}}=0.11$, \,$\ell_{2_{GP}}=0.32$, \, $\ell_{3_{GP}}=0.43$ }
\end{tabular}
\caption{Iterative steps in HSGP model fitting on the $3$D data. The diagnostic procedure updating the parameters minimum $\ell$, $c$, $m$, estimated $\hat{\ell}$ and diagnostic $\hat{\ell} + 0.01 \geq \ell$ is applied on the three input dimensions. The full diagnostic process is applied until two consecutive \textit{trues} in the diagnostic are achieved for each dimension. RMSE and RMSE* refer to the root mean squared error of the HSGP computed against the true function and the exact GP, respectively.}
  \label{tab_caseIII_3D}
\end{table}

Figure \ref{fig14_errors_exII} shows the difference  between the true underlying data-generating function and the posterior mean for GP, HSGP and spline-based model solutions for the $2$D data set. For the $3$D data set plots with the posterior functions are not shown because it is difficult to plot functions in a $3$D input space. The splines-based model was fitted using a cubic spline basis, penalized by the conventional integrated square second derivative cubic spline penalty \citep{wood2017generalized}, and implemented in the R-package \textit{mgcv} \citep{wood2011mgcv}. A Bayesian approach is used to fit this spline-based model using the R-package \textit{brms} \citep{burkner2017brms}. For the spline-based model, $40$ knots in each dimension were used. Figure \ref{fig14_errors_exII}-left shows the root mean squared error (RMSE), computed against the data-generating function, as a function of the boundary factor $c$ and number of univariate basis functions $m$, for the HSGP, and knots, for the $2$D spline-based model. It can be seen that the posterior mean for HSGP and exact GP accurately represents the true data-generating function. In contrast, the accuracy of the spline-based model is significantly worse. 

R code and the Stan model codes for the exact GP, the HSGP and the spline-based models of this case study can be found online at {\small \url{https://github.com/gabriuma/}} \linebreak {\small \url{basis_functions_approach_to_GP/tree/master/Paper/Case-study_2D&3D-Simulated-data}}\,.

\subsubsection{Computation requirements in 2D, 3D, and 4D input spaces}\label{sec_computation_nD}

\begin{figure*}
\centering
\includegraphics[width=\textwidth, trim = 0mm 0mm 0mm 0mm, clip]{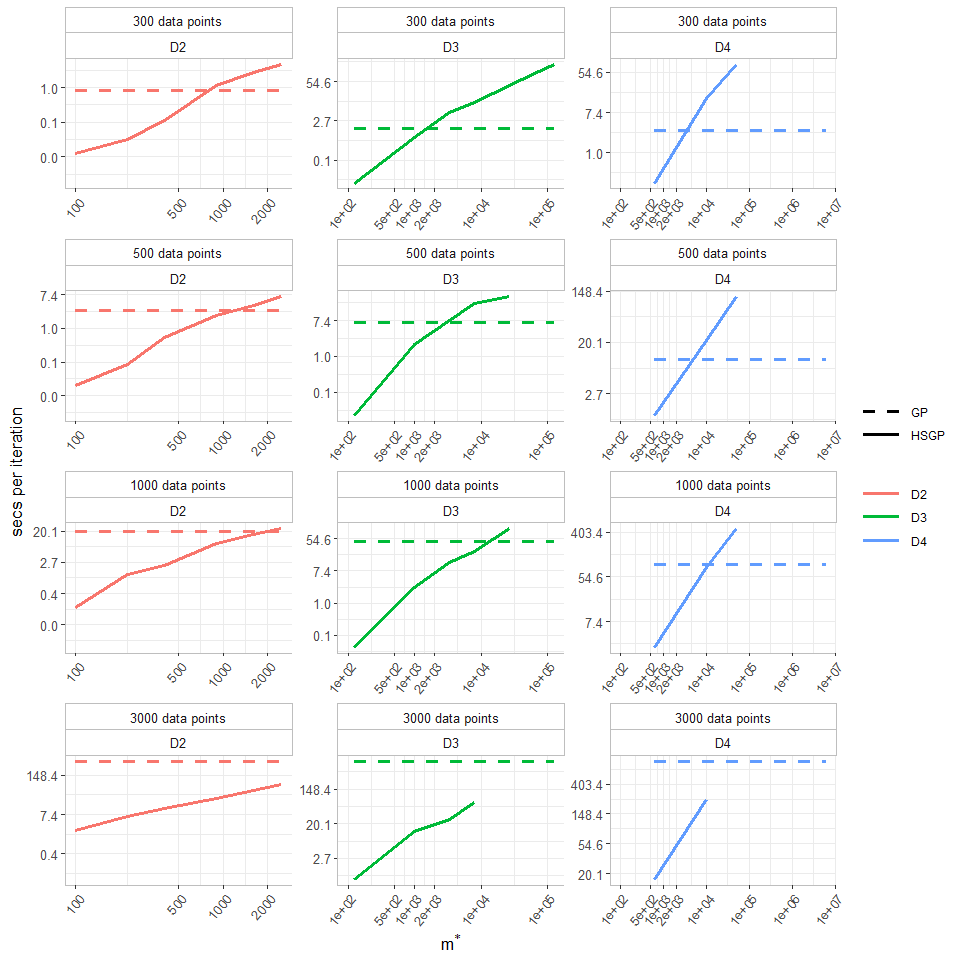}
\caption{Computational time in seconds per iteration (iteration of the HMC sampling method) for different datasets with different dimensionalties and data points, and plotted as a function of the number of multivariate basis functions $m^*=m_1\times \cdots \times m_D$.}
  \label{fig15_time_nD_log}
\end{figure*}

Figure~\ref{fig15_time_nD_log} shows the computation time for dynamic HMC (in seconds per iteration) in $2$D, $3$D and $4$D input spaces and different sized data sets, $n=300, n=500, n=1000$, and $n=3000$, as a function of the number of multivariate basis functions $m^*=m_1\times \cdots \times m_D$ used in the approximation.
Both the time of computation and $m^*$ are plotted in the logarithmic scale. 

Looking back at Figure~\ref{fig6_relationships}, or equations~\eqref{eq_m_l_QE}, \eqref{eq_m_l_mat32}, and \eqref{eq_m_l_mat52}, and assuming squared exponential covariance function, any univariate function (or single dimensions of a multivariate function) with true lengthscales bigger than $0.3$ can be accurately fitted using only $10--12$ basis functions. For lengthscales between $0.1$ and $0.3$, $10--22$ basis functions are sufficient.

For a very wiggly $2$D function (say, with $\ell_1=0.1$ and $\ell_2=0.1$), the approximate number of multivariate basis functions needed is $m^*=22\times 22= 484$, which results in significantly faster computation than the exact GP, even with small data sets (i.e., $n \lesssim 300$) (see Figure~\ref{fig15_time_nD_log}). For a very wiggly $3$D function (say, with $\ell_1=0.1$, $\ell_2=0.1$ and $\ell_3=0.1$), the approximate number of multivariate basis functions needed is $m^*=22\times 22\times 22= 10648$, which, unless for small data sets (i.e., $n \lesssim 1000$), the method is still significantly faster than the regular GP. A $3$D function where each dimension has a lengthscale around $0.1$ is not that common in statistical analysis, and thus in many cases the approximation will be significantly faster than excat GP. For $4$D data sets, the method can still be more efficient than the exact GP for moderate-to-large data sets (i.e., $n \lesssim 1000$). 

Finally, for $D>5$ the method starts to be impractical even for smooth univariate functions. However, in these cases, the method may still be used for lower dimensional componenets in an additive modeling scheme.

\subsection{Leukemia data}\label{sec_bf_caseVI}

The next example presents a survival analysis in acute myeloid leukemia (AML) in adults, with data recorded between 1982 and 1998 in the North West Leukemia Register in the United Kingdom. The data set consists of survival and censoring times $t_i$ and censoring indicator $z_i$ (0 for observed and 1 for censored) for $n=1043$ cases ($i=1,\dots,n$). About 16\% of cases were censored. Predictors are \textit{age} ($x_1$), \textit{sex} ($x_2$), \textit{white blood cell} (WBC) ($x_3$) count at diagnostic with 1 unit = $50\times109/L$, and the \textit{Townsend deprivation index} (TDI) ($x_4$) which is a measure of deprivation for district of residence. We denote $\bm{x}_i=(x_{i1},x_{i2},x_{i3},x_{i4}) \in {\rm I\!R}^{4}$ as the vector of predictor values for observation $i$.

As the WBC predictor values were strictly positive and highly skewed, a logarithm transformation is used. Continuous predictors were normalized to have zero mean and unit standard deviation.  
We assume a log-normal observation model for the observed survival time, $t_i$, with a function of the predictors, $f(\bm{x}_i):{\rm I\!R}^4 \to {\rm I\!R}$, as the location parameter, and $\sigma$ as the Gaussian noise: 
\begin{equation*}
  p(t_i \mid f_i)= \LogNormal(t_i \mid f(\bm{x}_i),\sigma^2).
\end{equation*}

We do not have a full observation model, as we do not have a model for the censoring process. We use the complementary cumulative log-normal probability distribution for the censored data conditionally on the censoring time $t_i$:
\begin{align*}
p(y_i > t_i \mid f) &= \int_{t_i}^{\infty} \LogNormal(y_i \mid f(\bm{x}_i),\sigma^2) \,\mathrm{d}y_i \\
&=  1 - \Phi\! \left( \frac{\log(y_i)-f(\bm{x}_i)}{\sigma} \right),
\end{align*}
where $y_i>t_i$ denotes the unobserved survival time.
The latent function $f(\cdot)$ is modeled as a Gaussian process, centered around a linear model of the predictors $\bm{x}$, and with a squared exponential covariance function $k$. Due to the predictor \textit{sex} ($x_2$) being a categorical variable (`1' for female and `2' for male), we apply indicator variable coding for the GP functions, in a similar way such coding is applied in linear models \citep{Gelman+Hill+Vehtari:2020:ROS}. The latent function $f(\bm{x})$, besides of being centered around a linear model, is composed of a general mean GP function, $h(\bm{x})$, defined for all observations, plus a second GP function, $g(\bm{x})$, that only applies to one of the predictor levels ('male' in this case) and is set to zero otherwise:
\begin{align*}
h(\bm{x}) &\sim \GP\left(\bm{0}, \, k(\bm{x},\bm{x}', \theta_0) \right),  \\[1pt] 
g(\bm{x}) &\sim \GP\left(\bm{0}, \, k(\bm{x},\bm{x}', \theta_1)\right),\\[1pt] 
f(\bm{x}) &= c + \bm{\beta} \bm{x} + h(\bm{x}) + \mathbb{I}\!\left[x_2=2\right]g(\bm{x}),
\end{align*}
where $\mathbb{I}\left[\cdot\right]$ is an indicator function. Above, $c$ and $\bm{\beta}$ are the intercept and vector of coefficients, respectively, of the linear model. $\theta_0$ contains the hyperparameters $\alpha_0$ and $\ell_0$ which are the marginal variance and length-scale of the general mean GP function, and $\theta_1$ contains the hyperparameters $\alpha_1$ and $\ell_1$ which are the marginal variance and length-scale, respectively, of a GP function specific to the male sex ($x_2=2$). Scalar length-scales, $\ell_0$ and $\ell_1$, are used in both multivariate covariance functions, assuming isotropic functions.
\begin{figure*}[h]
\centering
\includegraphics[scale=0.70, trim = 0mm 5mm 0mm 0mm, clip]{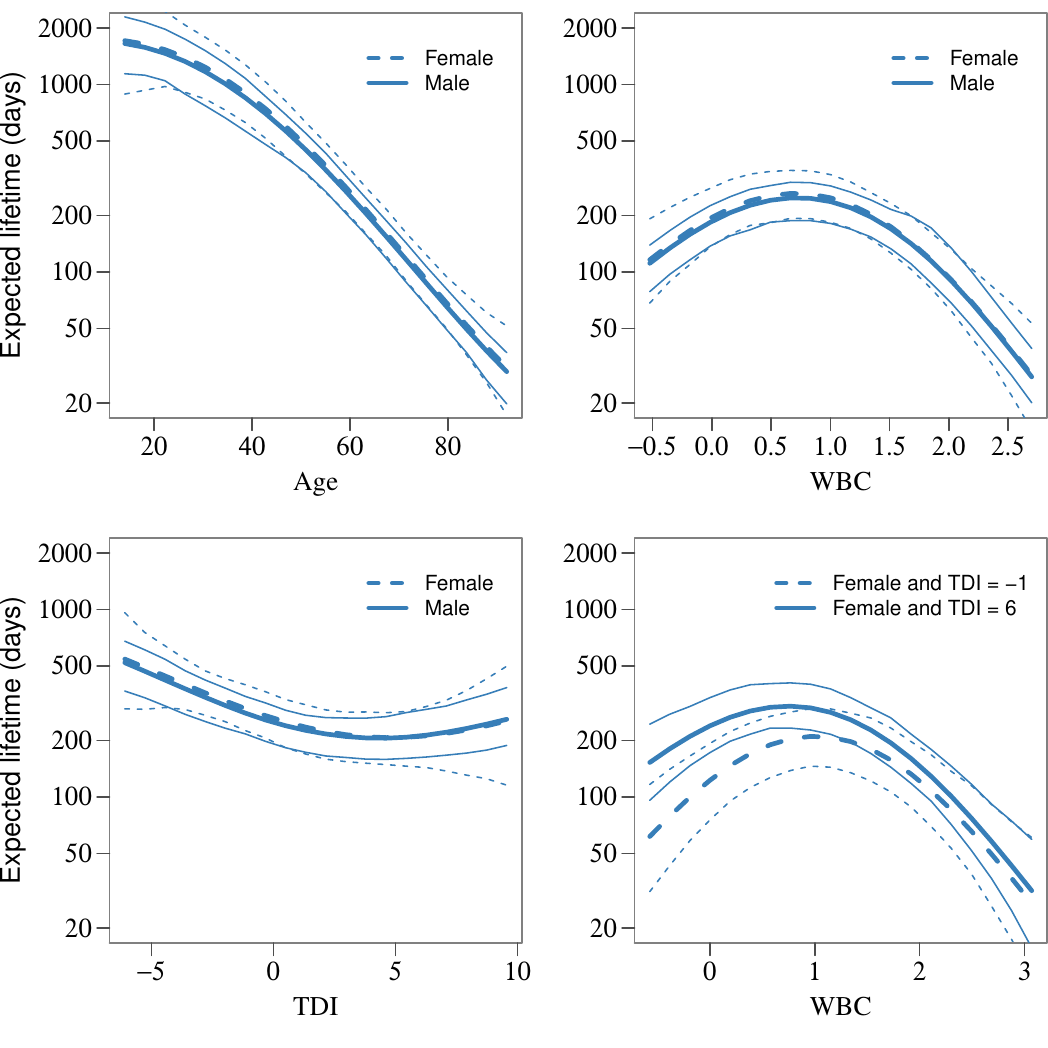}
\caption{Expected lifetime conditional comparison for each predictor with other predictors fixed to their mean values. The thick line in each graph is the posterior mean estimated using a HSGP model, and the thin lines represent pointwise 95\% credible intervals.}
  \label{fig16_posteriors_leukemia}
\end{figure*}

Using the HSGP approximation, the functions $h(\bm{x})$ and $g(\bm{x})$ are approximated as in eq.~(\ref{eq_approxf_multi}), with the $D$-dimensional (with a scalar length-scale) squared exponential spectral density $S$ as in eq.~(\ref{eq_specdens_inf}), and the multivariate eigenfunctions $\phi_j$ and the $D$-vector of eigenvalues $\bm{\lambda}_j$ as in equations (\ref{eq_eigenfunction_multi}) and  (\ref{eq_eigenvalue_multi}), respectively.

\begin{figure*}[h]
\centering
\subfigure{\includegraphics[scale=0.38, trim = 0mm 8mm 0mm 0mm, clip]{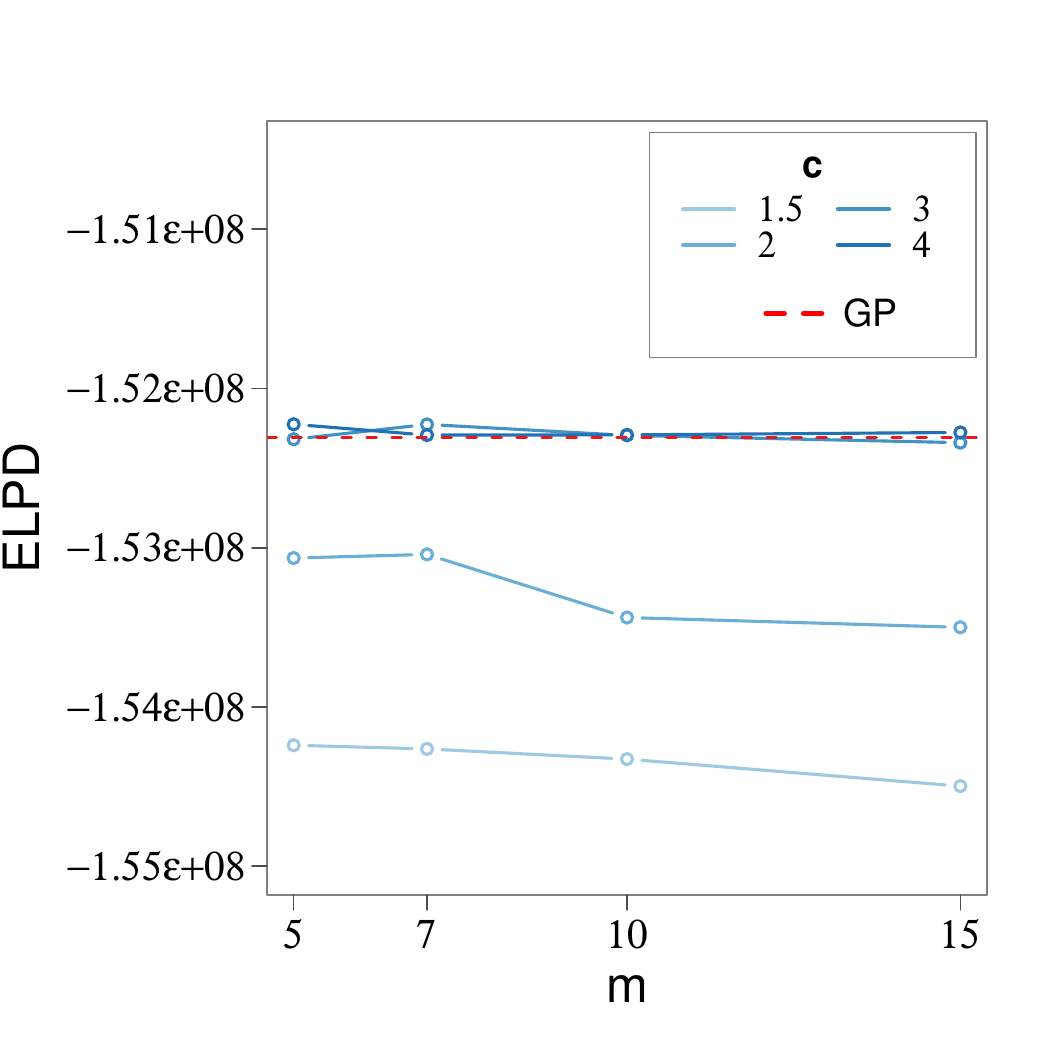}} 
\subfigure{\includegraphics[scale=0.38, trim = 0mm 8mm 0mm 0mm, clip]{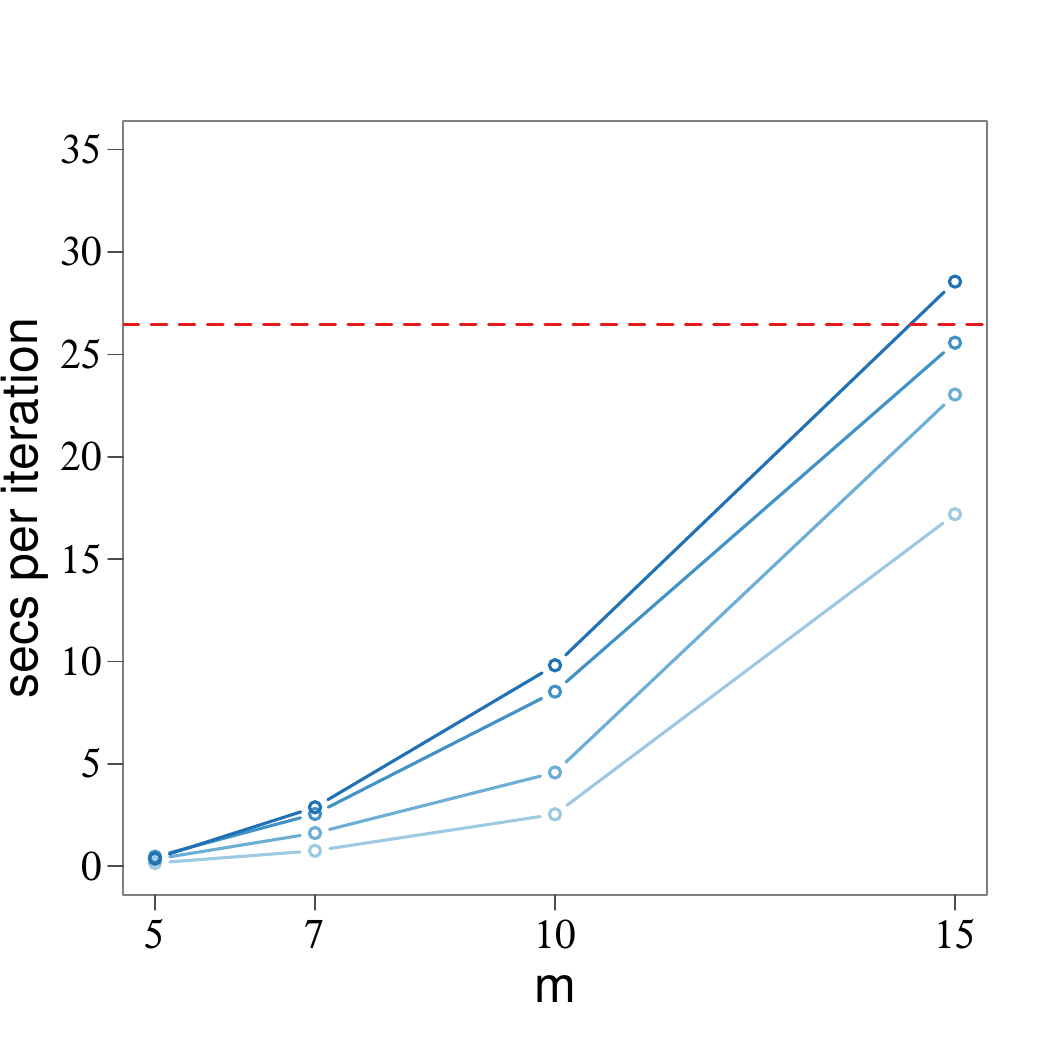}}
\caption{Expected log predictive density (ELPD; left) and time of computation in seconds per iteration (iteration of the HMC sampling method; right) as a function of the number of basis functions $m$ and boundary factor $c$.}
  \label{fig17_elpd_leukemia}
\end{figure*}

Figure \ref{fig16_posteriors_leukemia} shows estimated conditional functions of each predictor with all others fixed to their mean values. These posterior estimates correspond to the HSGP model with $m=10$ basis functions and $c=3$ boundary factor. There are clear  non-linear patterns and the right bottom subplot also shows that the conditional function associated with WBC has an interaction with TDI.
Figure~\ref{fig17_elpd_leukemia} shows the expected log predictive density \citep[ELPD; ][]{vehtari_2012,Vehtari+Gelman+Gabry:2017_practical} and time of computation as function of the number of univariate basis functions $m$ ($m^{\ast}=m^D$ in eq.~(\ref{eq_approxf_multi})) and boundary factor $c$. As the functions are smooth, a few number of basis functions and a large boundary factor are required to obtain a good approximation (Figure \ref{fig17_elpd_leukemia}-left); Small boundary factors are not appropriate for models for large length-scales, as can be seen in Figure \ref{fig6_relationships}. Increasing the boundary factor also significantly increases the time of computation (Figure \ref{fig17_elpd_leukemia}-right). With a moderate number of univariate basis functions ($m=15$), the HSGP model becomes slower than the exact GP model, in this specific application with $3$ input variables, as the total number of multivariate basis functions becomes $15^3 = 3375$ and is therefore quite high. 

R code and the Stan model code for the exact GP and the HSGP models of this case study can be found online at {\small \url{https://github.com/gabriuma/basis_functions_approach_to_GP/tree/master/Paper/Case-study_Leukemia-data}}\,.

\section{Conclusion}\label{sec_conclusion}

Modeling unknown functions using exact GPs is computationally intractable in many applications. This problem becomes especially severe when performing full Bayesian inference using sampling-based methods. In this paper, a recent approach for a low-rank representation of stationary GPs, originally proposed by \citet{solin2018hilbert}, has been analyzed in detail. The method is based on a basis function approximation via Laplace eigenfunctions. The method has an attractive computational cost as it effectively approximates GPs by linear models, which is also an attractive property in modular probabilistic programming programming frameworks. The dominating cost per log density evaluation (during sampling) is $O(nm+m)$, which is a big benefit in comparison to $O(n^3)$ of an exact GP model. The obtained design matrix is independent of hyperparameters and therefore only needs to be constructed once, at cost $O(nm)$. All dependencies on the kernel and the hyperparameters are through the prior distribution of the regression weights. The parameters' posterior distribution is $m$-dimensional, where $m$ is usually much smaller than the number of observations $n$. 

As one of the main contributions of this paper, we provided an in-depth analysis of the approximation's performance and accuracy in relation to the key factors of the method, that is, the number of basis functions, the boundary condition of the Laplace eigenfunctions, and the non-linearity of the function to be learned. On that basis, as our second main contribution, we developed practical diagnostics to assess the approximation's performance as well as an iterative procedure to obtain an accurate approximation with minimal computational costs. 

The developed approximate GPs can be easily applied as modular components in probabilistic programming frameworks such as Stan in both Gaussian and non-Gaussian observation models. Using several simulated and real datasets, we have demonstrated the practical applicability and improved sampling efficiency, as compared to exact GPs, of the developed method. The main drawback of the approach is that its computational complexity scales exponentially with the number of input dimensions. Hence, choosing optimal values for the number of basis functions and the boundary factor, using the recommendations and diagnostics provided in Figure~\ref{fig6_relationships}, is essential to avoid a excessive computational time especially in multivariate input spaces. However, in practice, input dimensionalities larger than three start to be computationally demanding even for moderately wiggly functions and few basis functions per input dimension. In these high dimensional cases, the proposed approximate GP methods may still be used for low-dimensional components in an additive modeling scheme but without modeling very high dimensional interactions, as complexity is linear with the number of additive components.

In this paper, the obtained functional relationships between the key factors influencing the approximation and corresponding diagnostics were studied primarily for univariate inputs. Accordingly, investigating the functional relationships more thoroughly for multivariate inputs remains a topic for future research.

\appendix

\numberwithin{equation}{section}
\numberwithin{figure}{section}

\section{Approximation of the covariance function using Hilbert space methods} \label{app_approx_covfun}

In this section, we briefly present a summary of the mathematical details of the approximation of a stationary covariance function as a series expansion of eigenvalues and eigenfunctions of the Laplacian operator. This statement is based on the work by \citet{solin2018hilbert}, who developed the mathematical theory behind the Hilbert Space approximation for stationary covariance functions.

Associated to each covariance function $k(\bm{x},\bm{x}')$ we can also define a covariance operator $\mathcal{K}$ over a function $f(\bm{x})$ as follows:
\begin{equation*}
\mathcal{K} f(\bm{x}) = \int k(\bm{x},\bm{x}') f(\bm{x}') \,\mathrm{d}\bm{x}'.
\end{equation*} 

From the Bochner’s and Wiener-Khintchine theorems, the spectral density of a stationary covariance function $k(\bm{x},\bm{x}') = k(\bm{\tau})$, $\bm{\tau}=(\bm{x}-\bm{x}')$, is the Fourier transform of the covariance function,
\begin{equation*}
S(\bm{w}) = \int k(\bm{\tau}) e^{-2\pi i \bm{w} \bm{\tau}} \,\mathrm{d}\bm{\tau}, \nonumber
\end{equation*}
where $\bm{w}$ is in the frequency domain. The operator $\mathcal{K}$ will be translation invariant if the covariance function is stationary. This allows for a Fourier representation of the operator $\mathcal{K}$ as a transfer function which is the spectral density of the Gaussian process. 
Thus, the spectral density $S(\bm{w})$ also gives the approximate eigenvalues of the operator $\mathcal{K}$.

In the isotropic case $S(\bm{w}) = S(||\bm{w}||)$ and assuming that the spectral density function $S(\cdot)$ is regular enough, then it can be represented as a polynomial expansion:
\begin{equation}\label{eq_S}
S(||\bm{w}||)=a_0+a_1||\bm{w}||^2+a_2(||\bm{w}||^2)^2+a_3(||\bm{w}||^2)^3+\cdots.
\end{equation}
The Fourier transform of the Laplace operator $\nabla^2$ is $-||\bm{w}||$, thus the Fourier transform of $S(||\bm{w}||)$ is
\begin{equation}\label{eq_K}
\mathcal{K}=a_0+a_1(-\nabla^2)+a_2(-\nabla^2)^2+a_3(-\nabla^2)^3+\cdots,
\end{equation}
defining a pseudo-differential operator as a series of Laplace operators.

If the negative Laplace operator $-\nabla^2$ is defined as the covariance operator of the formal kernel $l$,
\begin{equation*}
-\nabla^2 f(\bm{x}) = \int l(\bm{x},\bm{x}') f(\bm{x}') \,\mathrm{d}\bm{x}',
\end{equation*} 
then the formal kernel can be represented as 
\begin{equation*}
l(\bm{x},\bm{x}')= \sum_j \lambda_j \phi_j(\bm{x}) \phi_j(\bm{x}'),
\end{equation*}
where $\{\lambda_j\}_{j=1}^{\infty}$ and $\{\phi_j(\bm{x})\}_{j=1}^{\infty}$ are the set of eigenvalues and eigenvectors, respectively, of the Laplacian operator. Namely, they satisfy the following eigenvalue problem in the compact subset $\bm{x} \in \Omega \subset {\rm I\!R}^D$ and with the Dirichlet boundary condition (other boundary conditions could be used as well):
\begin{align*}
-\nabla^2 \phi_j(\bm{x})&=\lambda \phi_j(\bm{x}), \hspace{1cm}  x\in \Omega \nonumber \\ 
\phi_j(\bm{x})&=0, \hspace{1.85cm} x\notin \Omega.
\end{align*}  
Because $-\nabla^2$ is a positive definite Hermitian operator, the set of eigenfunctions $\phi_j(\cdot)$ are orthonormal with respect to the inner product
\begin{align*}
\langle f,g \rangle=\int_\Omega f(\bm{x}) g(\bm{x}) \,\mathrm{d}(\bm{x})
\end{align*} 
that is,
\begin{align*}
\int_\Omega \phi_i(\bm{x}) \phi_j(\bm{x}) \,\mathrm{d}(\bm{x}) = \delta_{ij},
\end{align*} 
and all the eigenvalues $\lambda_j$ are real and positive. 

Due to normality of the basis of the representation of the formal kernel $l(\bm{x},\bm{x}')$, its formal powers $s=1,2,\dots$ can be written as
\begin{eqnarray}\label{eq_formalkernel}
l(\bm{x},\bm{x}')^s= \sum_j \lambda_j^s \phi_j(\bm{x}) \phi_j(\bm{x}'),
\end{eqnarray} 
which are again to be interpreted to mean that
\begin{equation*}
(-\nabla^2)^s f(\bm{x}) = \int l^s(\bm{x},\bm{x}') f(\bm{x}') \,\mathrm{d}\bm{x}'.
\end{equation*} 
This implies that we also have
\begin{align*}
&[a_0+a_1(-\nabla^2)+a_2(-\nabla^2)^2+\cdots] f(\bm{x}) = \\
& \int [a_0+a_1l^1(\bm{x},\bm{x}')+a_2l^2(\bm{x},\bm{x}')+\cdots] f(\bm{x}')  \,\mathrm{d}\bm{x}'.
\end{align*} 

Then, looking at equations (\ref{eq_K}) and (\ref{eq_formalkernel}), it can be concluded 
\begin{equation}\label{eq_k}
k(\bm{x},\bm{x}')= \sum_j [a_0+a_1\lambda_j^1+a_2\lambda_j^2+\cdots] \phi_j(\bm{x}) \phi_j(\bm{x}').
\end{equation} 
By letting $||\bm{w}||^2=\lambda_j$ the spectral density in eq.~(\ref{eq_S}) becomes
\begin{equation*}
S(\sqrt{\lambda_j})=a_0+a_1\lambda_j+a_2\lambda_j^2+a_3\lambda_j^3+\cdots,
\end{equation*}
and substituting in eq.~(\ref{eq_k}) then leads to the final form
\begin{equation}\label{eq_k_2}
k(\bm{x},\bm{x}')= \sum_j S(\sqrt{\lambda_j}) \phi_j(\bm{x}) \phi_j(\bm{x}'),
\end{equation} 
where $S(\cdot)$ is the spectral density of the covariance function, $\lambda_j$ is the $j$th eigenvalue and $\phi_j(\cdot)$ the eigenfunction of the Laplace operator in a given domain.

\section{Low-rank Gaussian process with a periodic covariance function}\label{sec_periodic}

A GP model with a periodic covariance function does no fit in the framework of the HSGP approximation covered in this study, but it has also a low-rank representation. In this section, we first give a brief presentation of the results by \citet{solin2014explicit}, who obtain an approximate linear representation of a periodic squared exponential covariance function based on expanding the periodic covariance function into a series of stochastic resonators. Secondly, we analyze the accuracy of this approximation and, finally, we derive the GP model with this approximate periodic squared exponential covariance function.

The periodic squared exponential covariance function takes the form
\begin{equation} \label{eq_cov_periodic}
k(\tau)= \alpha \exp\left( -\frac{2\sin^{2}(\omega_0\frac{\tau}{2})}{\ell^2}\right),
\end{equation}
where $\alpha$ is the magnitude scale of the covariance, $\ell$ is the characteristic length-scale of the covariance, and $\omega_0$ is the angular frequency defining the periodicity. 

\citet{solin2014explicit} derive a cosine series expansion for the periodic covariance function (\ref{eq_cov_periodic}) as follows,
\begin{equation} \label{eq_cov_periodic_taylor_approx}
k(\tau)= \alpha \sum_{j=0}^{J} \tilde{q}_j^2 \cos(j\omega_0 \tau),
\end{equation}
which comes basically from a Taylor series representation of the periodic covariance function. The coefficients $\tilde{q}_j^2$ 
\begin{equation} \label{eq_q}
\tilde{q}_j^2= \frac{2}{\exp(\frac{1}{\ell^2})} \sum_{j=0}^{\lfloor \frac{J-j}{2} \rfloor} \frac{(2\ell^2)^{-j-2}}{(j+i)!i!},
\end{equation}
where $j=1,2,\ldots,J$, and $\lfloor \cdot \rfloor$ denotes the floor round-off operator. For the index $j=0$, the coefficient is 
\begin{equation} \label{eq_q0}
\tilde{q}_0^2= \frac{1}{2} \frac{2}{\exp(\frac{1}{\ell^2})} \sum_{j=0}^{\lfloor \frac{J-j}{2} \rfloor} \frac{(2\ell^2)^{-j-2}}{(j+i)!i!}.
\end{equation}
The covariance in eq.~(\ref{eq_cov_periodic_taylor_approx}) is a $J$th order truncation of a Taylor series representation. This approximation converges to eq.~(\ref{eq_cov_periodic}) when $J \rightarrow \infty$.

An upper bounded approximation to the coefficients $\tilde{q}_j^2$ and $\tilde{q}_0^2$ can be obtained by taking the limit $J \rightarrow \infty$ in the sub-sums in the corresponding equations (\ref{eq_q}) and (\ref{eq_q0}), and thus leading to the following variance coefficients:
\begin{equation}\label{eq_q_2}
\begin{split}
\tilde{q}_j^2=& \frac{2\mathrm{I}_j(\ell^{-2})}{\exp(\frac{1}{\ell^2})}, \\
\tilde{q}_0^2=& \frac{\mathrm{I}_0(\ell^{-2})}{\exp(\frac{1}{\ell^2})},
\end{split}
\end{equation} 
for $j=1,2,\ldots,J$, and where the $\mathrm{I}_{j}(z)$ is the modified Bessel function \citep{handbook1970m} of the first kind. This approximation implies that the requirement of a valid covariance function is relaxed and only an optimal series approximation is required. A more detailed explanation and mathematical proofs of this approximation of a periodic covariance function are provided by \citet{solin2014explicit}. 

\begin{figure*}
\centering
\subfigure{\includegraphics[scale=0.40, trim = 0mm 8mm 5mm 0mm, clip]{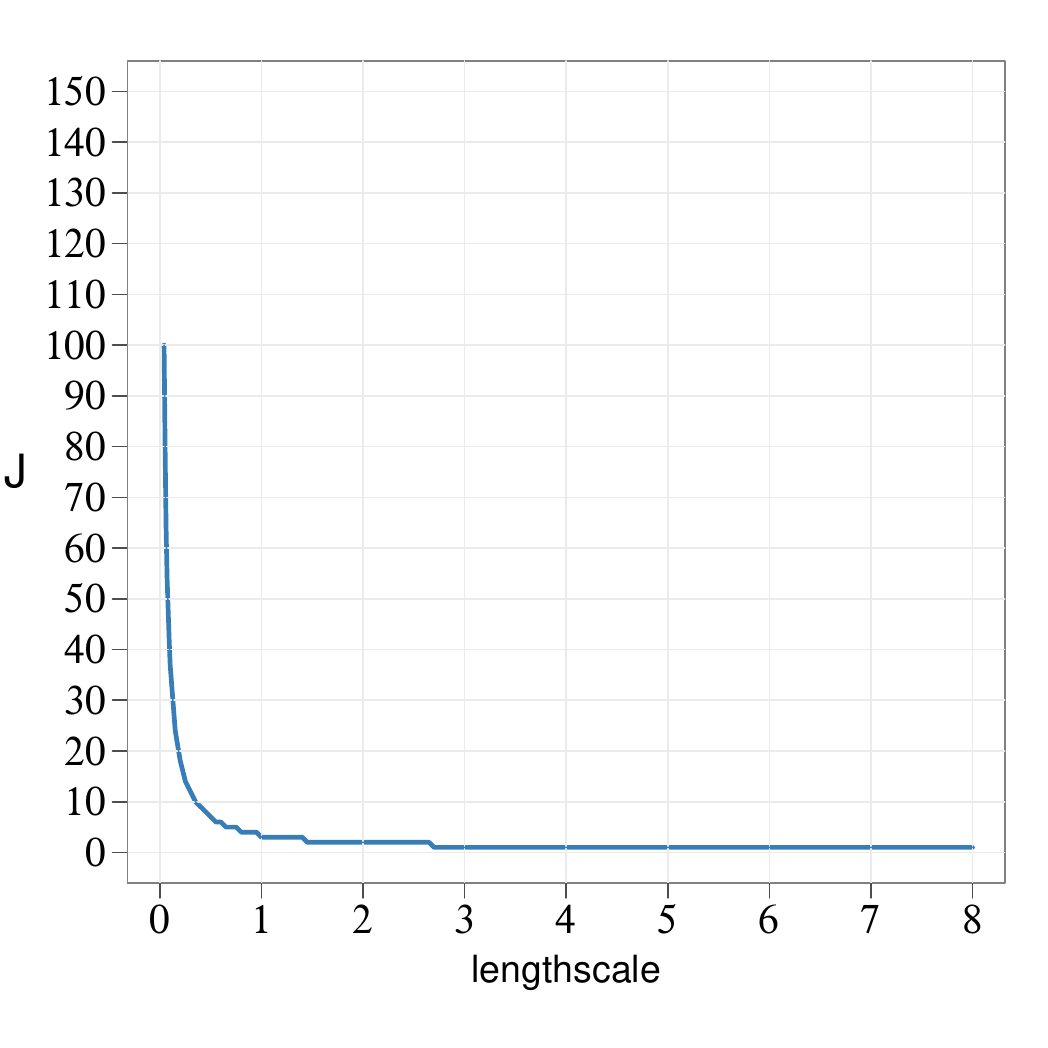}}
\hspace{3mm}
\subfigure{\includegraphics[scale=0.40, trim = 0mm 8mm 5mm 0mm, clip]{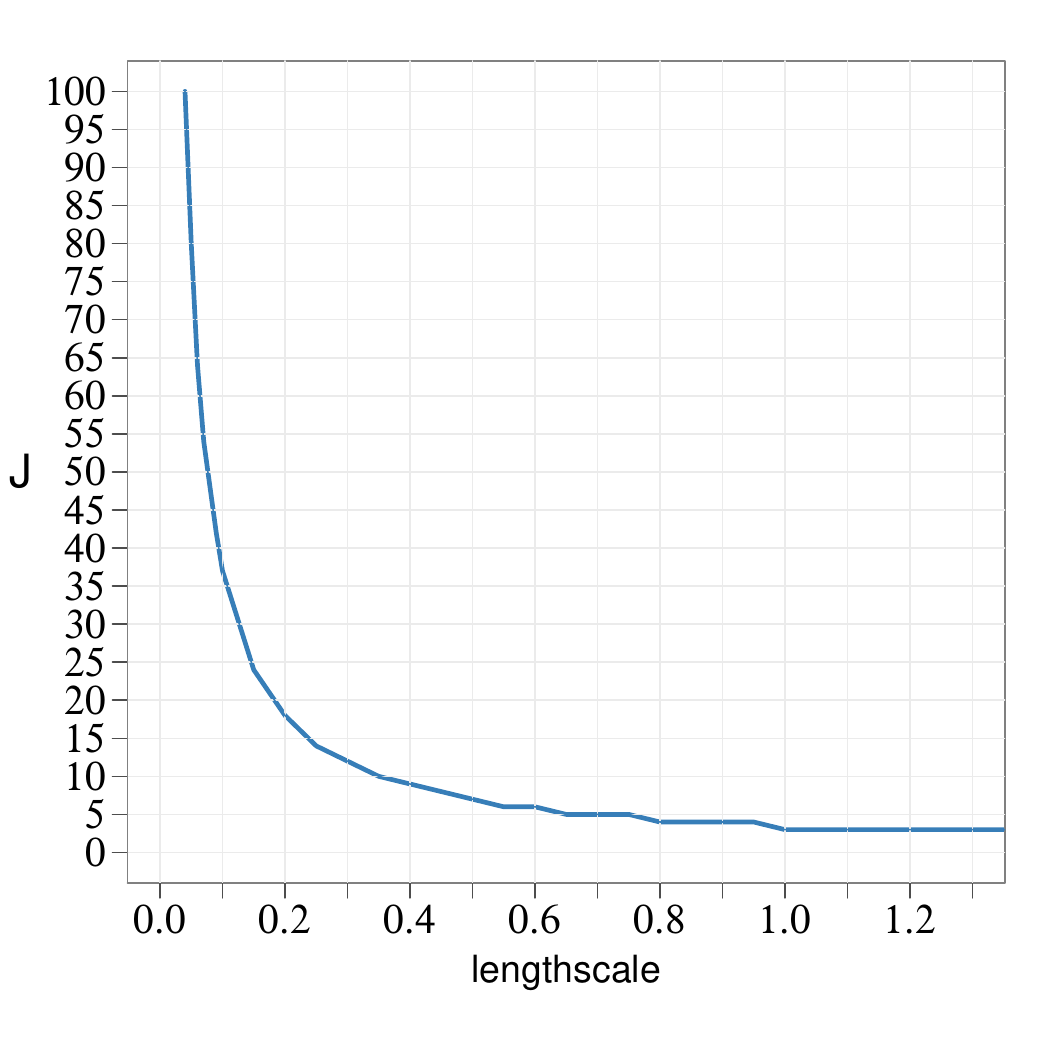}}
\caption{Relation among the minimum number of terms $J$ in the approximation and the length-scale ($\ell$) of the periodic squared exponential covariance function. The right-side plot is a zoom in of the left-side plot.}
  \label{figB1_m_lscale_periodic}
\end{figure*}

In order to assess the accuracy of this representation as a function of the number of cosine terms $J$ considered in the approximation, an empirical evaluation is carried out in a similar way than that in Section \ref{sec_accuracy} of this work. Thus, Figure \ref{figB1_m_lscale_periodic} shows the minimum number of terms $J$ required to achieve a close approximation to the exact periodic squared exponential kernel as a function of the length-scale of the kernel. We have considered an approximation to be close enough in terms of satisfying eq.~(\ref{eq_diff_covs}) with $\varepsilon=0.005 \int k(\tau) \,\mathrm{d}\tau$ (0.5$\%$ of the total area under the curve of the exact covariance function $k$). Since this is a series expansion of sinusoidal functions, the approximation does not depend on any boundary condition.

From the empirical observations, a numerical equation governing the relationships between $J$ and $\ell$ were estimated in equation \ref{eq_J_l_periodic}, which show linear proportionality between $J$ and $\ell$:
\begin{align}\label{eq_J_l_periodic}
&J \geq \frac{3.72}{\ell} \;\; \Leftrightarrow \;\; \ell \geq \frac{3.72}{J}.
\end{align}

The function values of a GP model with this low-rank representation of the periodic exponential covariance function can be easily derived. Considering the identity
\begin{equation*}
\cos(j\omega_0 (x-x'))=\cos(j\omega_0 x) \cos(j\omega_0 x') + \sin(j\omega_0 x) \sin(j\omega_0 x'),
\end{equation*}
the covariance $k(\tau)$ in eq.~(\ref{eq_cov_periodic_taylor_approx}) can be written as
\begin{align} \label{eq_cov_periodic_taylor_approx_2}
k(x,x')\approx \; & \alpha \bigg( \sum_{j=0}^{J} \tilde{q}_j^2 \cos(j\omega_0 x)  \cos(j\omega_0 x')  \nonumber\\
& + \sum_{j=1}^{J} \tilde{q}_j^2 \sin(j\omega_0 x) \sin(j\omega_0 x') \bigg).
\end{align}
With this approximation for the periodic squared exponential covariance function $k(x,x')$, the approximate GP model $f(x) \sim \GP\left(0,k(x,x')\right)$ equivalently leads to a linear representation of $f(\cdot)$ via
\begin{align} \label{eq_f_period}
f(x) \approx \; & \alpha^{1/2} \bigg( \sum_{j=0}^J  \tilde{q}_j \cos(j\omega_0 x) \beta_j \nonumber\\
& +  \sum_{j=1}^J \tilde{q}_j \sin(j\omega_0 x) \beta_{J+1+j} \bigg),
\end{align}
where $\beta_j \sim \Normal(0,1)$, with $j=1,\dots,2J+1$. The cosine $\cos(j\omega_0 x)$ and sinus $\sin(j\omega_0 x)$ terms do not depend on the covariance hyperparameters $\ell$. The only dependence on the hyperparameter $\ell$ is through the coefficients $\tilde{q}_j$, which are $J$-dimensional. The computational cost of this approximation scales as $O\big(n(2J+1) + (2J+1)\big)$, where $n$ is the number of observations and $J$ the number of cosine terms.
The parameterization in eq.~(\ref{eq_f_period}) is naturally in the non-centered form with independent prior distributions on
$\beta_j$, which makes posterior inference easier.

\section{Description of the iterative steps applied to perform diagnostic on two of the data sets in Figure~\ref{tab1_diagnostic}}\label{sec_ex_iterative_steps}

As concrete examples, the iterative steps applied to perform diagnostic on two of the data sets in Table~\ref{tab1_diagnostic} are described next:

\paragraph*{Data set with the squared exponential kernel and true length-scale $\ell_{\text{GP}} = 0.08$:} 
\begin{itemize}
\item[] Iteration 1:
	\begin{enumerate}
	\item Make the initial guess that the length-scale $\ell_1=0.5$.
	
	\item Compute the optimal $c_1$ as a function of $\ell_1$ by using eq.~(\ref{eq_c_vs_l_QE}).
	
	TODO: name the value of $c_1$ and all the other quantities in the following. Otherwise, what is the purpose of this example?
	
	\item Compute the minimum $m_1$ as a function of $c_1$ and $\ell_1$ by using eq.~(\ref{eq_m_l_QE}).
	
	\item Fit the HSGP model and obtain the estimated $\hat{\ell}_1$.
	
	TODO: provide at least a mean value for the posterior of  $\hat{\ell}_1$ here?
	
	\item The diagnostic $\hat{\ell}_1 + 0.01 \geq \ell_1$ gives FALSE.
	
	TODO: I think you wanted to use $\hat{\ell}_1 - 0.01 \geq \ell_1$ and I corrected it above. Can you verify whether you want + or -?
	\end{enumerate}
	
\item[] Iteration 2:
	\begin{enumerate}
	\item As the diagnostic in iteration 1 was FALSE, set $\ell_2 = \hat{\ell}_1$.
	
	\item Compute the optimal $c_2$ as a function of $\ell_2$ and $m_1$ as a function of $c_2$ and $\ell_2$ by using equations~(\ref{eq_c_vs_l_QE}) and eq.~(\ref{eq_m_l_QE}), respectively.
	
	TODO: why make this a single step here when it was two steps in Section 4.4.1?
	
	\item Fit the HSGP model using $c_2$ and $m_2$ and estimating $\hat{\ell}_2$.
	
	\item The diagnostic $\hat{\ell}_2 + 0.01 \geq \ell_2$ gives FALSE.
	\end{enumerate}
	
\item[] Iteration 3:
	\begin{enumerate}
	\item As the diagnostic in previous iteration 2 was FALSE, updating $\ell_3$ with $\hat{\ell}_2$ and repeat the process as in iteration 2.
	
	\item The diagnostic $\hat{\ell}_3 + 0.01 \geq \ell_3$ gives TRUE.
	\end{enumerate}
	
\item[] Iteration 4:
	\begin{enumerate}
	\item As the diagnostic in step 4 in previous iteration 3 was TRUE, set $m_4 = m_3 + 5$.
	
	\item Compute the optimal $c_4$ as a function of $\ell_3$ by using eq.~(\ref{eq_c_vs_l_QE}).
	
	\item Fit the HSGP model using $c_4$ and $m_4$ and estimating $\hat{\ell}_4$.
	
	\item The diagnostic $\hat{\ell}_4 + 0.01 \geq \ell_4$ gives TRUE.
	
	\item As RMSE, $R^2$, and ELPD are stable relative to previous iteration 4, the HSGP approximation is likely sufficiently accurate and the procedure ends here.
	\end{enumerate}
	
\end{itemize} 

\paragraph*{Data set with the squared exponential kernel and true length-scale $\ell_{\text{GP}} = 1.4$:} 
\begin{itemize}
\item[] Iteration 1:
	\begin{enumerate}
	\item Make the initial guess that the length-scale $\ell_1=1$.
	
	\item Compute the optimal $c_1$ as a function of $\ell_1$ by using eq.~(\ref{eq_c_vs_l_QE}).
	
	\item Compute the minimum $m_1$ as a function of $c_1$ and $\ell_1$ by using eq.~(\ref{eq_m_l_QE}).
	
	\item Fit the HSGP model and obtain the estimated $\hat{\ell}_1$.
	
	\item The diagnostic $\hat{\ell}_1 + 0.01 \geq \ell_1$ gives TRUE.
	\end{enumerate}
	
\item[] Iteration 2:
	\begin{enumerate}
	\item As the diagnostic in previous iteration 1 was TRUE, set $m_2 = m_1 + 5$.
	
	\item Compute the optimal $c_2$ as a function of $\ell_2$ by using eq.~(\ref{eq_c_vs_l_QE}).
	
	\item Fit the HSGP model using $c_2$ and $m_2$ and estimating $\hat{\ell}_2$.
	
	\item The diagnostic $\hat{\ell}_2 + 0.01 \geq \ell_2$ gives TRUE.
	
	\item As RMSE, $R^2$, and ELPD are stable relative to previous iteration 2, the HSGP approximation is likely sufficiently accurate and the procedure ends here.
	\end{enumerate}
	
\end{itemize}

\section*{Acknowledgements}
We thank Academy of Finland (grants 298742, 308640, and 313122), Finnish Center for Artificial Intelligence, and Technology Industries of Finland Centennial Foundation (grant 70007503; Artificial Intelligence for Research and Development) for partial support of this research. We also acknowledge the computational resources provided by the Aalto Science-IT project.

\bibliography{../references}

\end{document}